\documentclass[11pt]{article}
\pdfoutput=1
\usepackage{jheppub}
\usepackage[T1]{fontenc}
\usepackage{cancel,tensor,enumitem,fix-cm}
\usepackage{epsfig}
\usepackage{graphicx}
\usepackage{amsmath}
\usepackage{textcomp}
\usepackage{multirow}
\usepackage{booktabs}
\usepackage{tikz}
\usepackage{overpic}
\usepackage{parskip}
\usetikzlibrary{3d}
\newcommand{\bt}{{\mathbf{t}}}
\newcommand{\by}{{\mathbf{x}}}

\newcommand{\unit}{1\!\!1}

\usepackage{hyperref} 
\hypersetup
{
     colorlinks=true,  
     linkcolor=blue,  
     filecolor=cyan,   
     urlcolor=blue,   
     linktoc=page,  
     citecolor=red, 
}

\def\bw{\bar{w}}
\def\bz{\bar{z}}

\newcommand{\bea}{\begin{eqnarray}}
\newcommand{\eea}{\end{eqnarray}}
\newcommand{\be}{\begin{equation}}
\newcommand{\ee}{\end{equation}}

\def\Tr{\textrm{Tr}}

\def\({\left(}
\def\){\right)}
\def\[{\left[}
\def\]{\right]}

\newcommand{\bra}[1]{\langle #1|}
\newcommand{\ket}[1]{|#1\rangle}

\def\tr{\tilde{\rho}}

\def\tz0{\tilde{z}_0}

\def\CO{{\cal O}}

\definecolor{palatinatepurple}{rgb}{0.41, 0.16, 0.38}

\definecolor{uglybrown}{rgb}{0.8,  0.7,  0.5}

\def\cO{\mathcal{O}}

\def\tr{\text{tr}}

\subheader{\begin{flushright}
\texttt{YITP-20-09, BRX-TH-6663}
\end{flushright}}

\title{Local quenches, bulk entanglement entropy and a unitary Page curve}

\author[a]{Cesar A. Ag\'on,}
\author[b]{Sagar F. Lokhande}
\author[c,d]{and Juan F. Pedraza}
\affiliation[a]{C.~N. Yang Institute for Theoretical Physics, State University of New York,\\
Stony Brook NY 11794, USA}
\affiliation[b]{Department of Physics, University of Illinois, Urbana-Champaign, Urbana IL 61801, USA}
\affiliation[c]{Department of Physics and Astronomy, University College London, London WC1E 6BT, UK}
\affiliation[d]{Martin Fisher School of Physics, Brandeis University, Waltham MA 02453, USA}
\emailAdd{cesar.agon@stonybrook.edu}
\emailAdd{sagar.f.lokhande@gmail.com}
\emailAdd{j.pedraza@ucl.ac.uk}

\abstract{Quantum corrections to the entanglement entropy of matter fields interacting with dynamical gravity have proven to be very important in the study of the black hole information problem. We consider a one-particle excited state of a massive scalar field infalling in a pure AdS$_3$ geometry and compute these corrections for bulk subregions anchored on the AdS boundary. In the dual CFT$_2$, the state is given by the insertion of a local primary operator and its evolution thereafter. We calculate the area and bulk entanglement entropy corrections at order $\mathcal{O}(N^0)$, both in AdS and its CFT dual. The two calculations match, thus providing a non-trivial check of the FLM formula in a dynamical setting. Further, we observe that the bulk entanglement entropy follows a Page curve. We explain the precise sense in which our setup can be interpreted as a simple model of black hole evaporation and comment on the implications for the information problem.}

\begin{document}
\maketitle
\flushbottom

\section{Introduction}

\subsection{The big picture}

The presence of entanglement is an essential and ubiquitous feature of quantum systems. It is often quantified using entanglement entropy $S_A$. Given a quantum system in a state $\ket{\psi}$, one starts by partitioning its space into subsystems $A$ and $A^c$. The entanglement entropy between the subsystems is then defined as the von Neumann entropy $S_A \equiv - \text{tr} [ \rho_A \, \log \rho_A ]$, where $\rho_A \equiv \text{tr}_{A^c}[\ket{\psi} \bra{\psi}]$ is the \textit{reduced density matrix} associated with the region $A$. It is interesting to study this quantity for matter fields interacting with gravity, particularly, in situations where the state is fully dynamical, e.g., for evaporating black holes. Indeed, one of the observations that led to the black hole information paradox is that the semiclassical result for the entanglement entropy of matter fields outside an evaporating black hole initially in a pure state, keeps increasing with time \cite{Hawking:1974sw}, a behaviour that contradicts the expectations from unitarity \cite{Page:1993df}. This result could be interpreted in two ways: either, one should accept that the information is lost in black holes, challenging the rules of quantum mechanics; or one should understand the breakdown of unitarity as an artifact of the semiclassical approximation.

Recently, a resolution of this tension was put forward by suggesting a new prescription
for the computation of the entanglement entropy of a quantum gravitational
system in the semiclassical approximation. It amounts to adding a new term dubbed the \textit{island contribution} which accounts for the possibility of new saddle points in the semiclassical path integral \cite{Almheiri:2019hni}
\begin{equation}
\label{eq:islandform}
S_A(t) = \underset{\mathcal{I}_{\text{ext}}}{\text{min}} \, \underset{\mathcal{I}}{\text{ext}} \, \[ S^{\text{eff}}_{A \cup \mathcal{I}} + \frac{\text{Area}(\partial \mathcal{I})}{4 G}  \]\,.
\end{equation}
Here $\mathcal{I}$ denotes the so-called \textit{island}, $\partial \mathcal{I}$ is its boundary, $G$ is the Newton's constant of the gravity theory and $S^\text{eff}_A$ is the effective, or coarse-grained entropy, which is calculated using the semiclassical density matrix, i.e., the $\rho_A^{\text{eff}}$ that describes the quantum dynamics of matter fields living on a \textit{classical} geometry. On the other hand, the entropy in the left-hand side of (\ref{eq:islandform}) uses the fine-grained density matrix of quantum fields, including gravity. The problem is that there is no simple way to calculate the latter from a first principle calculation since this task would require starting from a complete and consistent quantum theory of gravity.

There is some evidence, owing to studies in low-dimensional models of conformal matter interacting with gravity, that such a formula may be correct \cite{Almheiri:2019qdq}. These models study gravity in AdS spacetime and make use of holography. One of the precise entries in holographic dictionary is the statement that the entanglement entropy of a subregion $A$ in the dual CFT is equal to the area of an extremal codimension-two surface $\gamma_A^{\text{ext}}$ anchored at the boundary of AdS and homologous to the subregion $A$ \cite{Ryu:2006bv}
\begin{equation}
S_A = \underset{\gamma_A^{\text{ext}}}{\text{min}}\, \underset{\gamma_A}{\text{ext}} \[ \frac{\text{Area}(\gamma_A)}{4 G}\]\, .
\end{equation}
This is known as the RT/HRT formula and the extremal surface $\gamma_A^\text{ext}$ is often referred to as the RT surface. However, this formula gives only the leading $\mathcal{O}(1/G)$ contribution to the entanglement entropy, and ignores corrections coming from quantum fields in the bulk. To obtain a more accurate answer one must naturally include the entanglement entropy of these fields, an idea first advocated in \cite{Faulkner:2013ana}  and now called the FLM formula
\begin{equation}
\label{eq:flmform}
S_A = \underset{\gamma_A^{\text{ext}}}{\text{min}}\,\underset{\gamma_A}{\text{ext}} \,  \[ \frac{\text{Area}(\gamma_{A})}{4 G} \] + S_{\text{bulk}}(\Sigma_A)\, .
\end{equation}
Here $\Sigma_A$ is the codimension-one region in the bulk between the RT surface and the boundary region $A$ in a time-slice that contains the two, and $S_\text{bulk}$ is the von Neumann entropy of the semiclassical bulk density matrix of quantum fields in $\Sigma_A$.

There is a further proposal \cite{Engelhardt:2014gca} that extends the FLM formula and is analogous to the \text{island formula} in equation \eqref{eq:islandform} \cite{Penington:2019npb,Almheiri:2019psf}. According to this proposal, the entanglement entropy is given in terms of a different extremal surface $\tilde{\gamma}_A^{\text{ext}}$
called {\it quantum extremal surface} so that
\begin{equation}\label{QES}
S_A = \underset{\tilde{\gamma}_A^{\text{ext}}}{\text{min}}\,\underset{\gamma_A}{\text{ext}} \,  \[  \frac{\text{Area}(\gamma_A)}{4 G} +S_{\text{bulk}}(\Sigma_{A^c})\]\,.
\end{equation}
This new prescription agrees with FLM at order $\mathcal{O}(1)$, but generalizes it to all orders in $G$. However, the implementation of (\ref{QES}) would require understanding of the bulk entanglement entropy for general regions, a task that is very difficult to accomplish in practice, in higher than two bulk dimensions.

These proposals highlight the utility of holography in the study of non-perturbative aspects of the black hole information paradox. In this paper we will use this framework to start a systematic study of entanglement entropy in cases where the bulk state is fully time dependent and completely under control, and interpret the results in terms of the microscopic CFT description. As a first step, we will focus on the leading $\mathcal{O}(1)$ corrections to the RT/HRT prescription which are given by the FLM formula (\ref{eq:flmform}). In particular, our goal is to discuss
\begin{enumerate}
\item Evidence that the FLM formula works in dynamical settings.
\item Implications of unitarity for the different terms in equation \eqref{eq:flmform}.
\item Properties of bulk entanglement entropy $S_\text{bulk}$.
\end{enumerate}
We will do so in a simple model of a quantum field theory interacting with dynamical gravity in AdS$_3$. We will also check all our calculations in the dual CFT$_2$ and explain how the different terms in the left-hand side and right-hand side of equation (\ref{eq:flmform}) match. As a byproduct, we will show that we can interpret our setup as a simple toy model for black hole evaporation, thus, providing fresh insight on the black hole information problem.

\subsection{Road map and summary}

Dynamical gravity backgrounds can be obtained by starting with a time-independent geometry and then perturbing it globally or locally. These backgrounds are holographically dual to time-dependent states in the dual CFT where these perturbations are also known as \textit{quantum quenches}. The two categories of global or local perturbations are dubbed \textit{global} and \textit{local} quenches, respectively. Both types of quenches have been widely studied in the context of holography \cite{Basu:2011ft,Das:2011nk,Basu:2012gg,Buchel:2012gw,Buchel:2013lla,Nozaki:2013wia,Caputa:2014eta,Arefeva:2017pho,Caputa:2019avh,Ageev:2020acl}. They have been used to study the problem of thermalization of closed quantum systems \cite{Das:2010yw,Ebrahim:2010ra,Balasubramanian:2010ce,Balasubramanian:2011ur,Caceres:2012em,Galante:2012pv,Fischler:2013fba} and to describe the process of black hole formation \cite{Bhattacharyya:2009uu,Garfinkle:2011hm,Garfinkle:2011tc,Wu:2012rib,Balasubramanian:2013rva,Balasubramanian:2013oga,Caceres:2014pda,Anous:2016kss}, an important problem in gravity. In this paper, we will however focus only on local quantum quenches since they provide more realistic models of perturbations as well as observers.

Local quenches have been used to describe the spread of local perturbations in QFTs and many-body systems \cite{lieb1972, hastings2010locality}. For holographic field theories, the holographic description of local quenches is useful to study quantum chaos \cite{larkin1969quasiclassical,Shenker2014BlackHA}. They also serve as models for generation of quantum entanglement  \cite{Calabrese:2007mtj,Eisler_2007,Stephan_2011}, a resource for quantum computations. In fact, local quenches are approximately equivalent to a quantum gate \cite{Shimaji:2018czt} and sequence of quantum gates are ubiquitous in quantum computations \cite{nielsen2001quantum}. The study of local quenches is exciting also because there is hope that local quenches could be simulated in condensed-matter systems like cold atoms \cite{Hofferberth_2007,Langen_2013,Meinert_2013}. From a holographic perspective, local quantum quenches provide models for the dynamics of localized perturbations in the dual gravity theory \cite{Nozaki:2013wia}. One can use these states to study the real time dynamics of bulk fields coupled to gravity. For example, if the unperturbed background is that of a black hole, one can compute quantities such as quasi-normal modes and other time-dependent bulk observables of interest. This is the setup that we will mostly work with. In particular, we will be interested in understanding entanglement structure of matter fields interacting with dynamical gravity with local perturbations.

In Section \ref{sec:CFTEE}, we start by describing local quench states in $2d$ CFTs. We calculate the time-dependent entanglement entropy for a single interval of length $2 R$ centered at $x=x_c$, using the replica trick. The general formula for the change of entanglement entropy in the excited state is given by equation \eqref{regEE-3}, and can be written as a sum of two contributions,
\begin{equation}\nonumber
\delta S_A = \delta S^{uni}_A + \delta S^{dyn}_A\,.
\end{equation}
defined by equations \eqref{Suni} and \eqref{Sdyn}, respectively. The final result for the two terms are:
\begin{align*}
\delta S^{uni}_A &=\Delta \left[2-\( \frac{1}{\eta_+}{\rm arctan}\, \eta_+ + \frac{1}{\eta_-}{\rm arctan}\, \eta_- \) \right], \qquad \eta_{\pm} \equiv \frac{2\alpha R}{R^2-(x_0-x_c\pm t)^2-\alpha^2}\,, \\
\delta S^{dyn}_A &= -\frac{\Gamma\(\frac{3}{2}\)\Gamma(2\Delta+1)}{\Gamma(2\Delta+\frac{3}{2})}\!\[\frac{2\alpha  R }{\left|(x_0-x_c-t)^2-R^2+\alpha^2\right|}\]^{2\Delta}
\!\!\[\frac{2\alpha  R }{\left|(x_0-x_c+t)^2-R^2+\alpha^2\right|}\]^{2\Delta} ,
\end{align*}
given by \eqref{deltaS+-} and \eqref{realSdyn}, respectively. Here, $\Delta$ is the conformal dimension of the perturbing operator, which is inserted at $x=x_0$ and spatially smeared over a length-scale of order $\alpha$. Later in the paper, we set $x_0=0$ which can be done without any loss of generality, given the translation invariance of the vacuum state. The $\delta S^{uni}_A$ term is fixed by kinematics and is a \emph{universal}, i.e., valid for any CFT. The $\delta S^{dyn}_A$ term depends on \emph{dynamical} data, in particular, on a specific correlation function that is not fixed by symmetries. For the evaluation of this term we use an OPE expansion and keep only the first non-vanishing contribution, therefore the final result is only valid in the small $R$ limit. Further, we use OPE data for large-$c$ CFTs and include all corrections from single- and multiple-trace operators at order $\mathcal{O}(c^0)$.

In Section \ref{sec:gravity}, we discuss the holographic time-dependent geometry dual to the local quench state. We argue that the time-dependent metric is given by the action of a specific large diffeomorphism, given by equations \eqref{eq:cotrans}-\eqref{eq:cotrans3}, on the backreacted metric of a one-particle state in global AdS, given in \eqref{backMetric}.  We also compute the expectation value of CFT operators dual to light bulk fields, with particular emphasis on the stress-energy tensor and its time-evolution following the quench. The result of this calculation is given in equation \eqref{eq:Tmunu}.

In Section \ref{sec:HEE}, we discuss the holographic computation of entanglement entropy at order $\mathcal{O}(G^0)$ in the excited state. The calculation naturally splits in two contributions, corresponding to the two terms in the FLM formula (\ref{eq:flmform}). Firstly, there is the correction to the area term due to change in geometry, dubbed geometric correction $\delta S^\text{geom}_A$, which amounts to compute \eqref{deltaArea}. For intervals centered at $x_c=x_0=0$, we show in equation \eqref{eq:finalcentered0} that it is equal to
\begin{equation}
\delta S^\text{geom}_A\big|_{x_c=0}=2\Delta\left[1-\arcsin\big(\tfrac{R}{a(t)}\big)\sqrt{\tfrac{a(t)^2}{R^2}-1}\right]
-\frac{\Gamma(\tfrac{3}{2})\Gamma(\Delta+1)R^{2\Delta}}{\Gamma(\Delta+\tfrac{3}{2})a(t)^{2\Delta}}\,\!_2F_1\left[1,\Delta,\Delta+\tfrac{3}{2},\tfrac{R^2}{a(t)^2}\right],\nonumber
\end{equation}
with $a(t)$ given in (\ref{def:aoft}). We observe that the first term above matches exactly the universal correction $\delta S^{uni}_A$ obtained from CFT calculation.
For non-centered intervals, we show that the geometric correction can be written as a simple integral expression, given in equation \eqref{intSgeneral}, which can be explicitly integrated for integer values of $\Delta$. For arbitrary values of $\Delta$, we show that it can generically be expressed as a sum of two series expansions,
\begin{align}
\begin{split}
\delta S^\text{geom}_A &=\frac{8 \Delta \alpha ^2 R^2}{3} \frac{(t^2+x_c^2+\alpha ^2)^2+4 t^2 x_c^2}{[(t^2-x_c^2+\alpha ^2)^2+4 \alpha ^2 x_c^2]^2}\Big[1+\sum_{i=1}^{\infty}\mathcal{P}_i(t,x_c,\alpha)R^{2i}\Big]\\
&\quad-\frac{\Gamma(\tfrac{3}{2}) \Gamma (\Delta +1)}{\Gamma(\Delta +\tfrac{3}{2})}\frac{ (2 \alpha R)^{2 \Delta }}{ [(t^2-x_c^2+\alpha ^2)^2+4 \alpha ^2 x_c^2]^{\Delta }}\Big[1+\sum_{i=1}^{\infty}\mathcal{Q}_i(t,x_c,\alpha)R^{2i}\Big]\,.\nonumber
\end{split}
\end{align}
Once again, the first series matches the universal correction $\delta S^{uni}_A$ obtained from CFT calculation. We further show explicitly that this series follows from an application of first law of entanglement in the CFT, equation \eqref{eq:deltaSfirstL}, which gives the result at linear order in the change of the density matrix $\delta\rho$. The appearance of the second series, with terms $\propto R^{2\Delta+2i}$ not seen in the CFT expressions, raises a puzzle which we resolve in Subsection \ref{sec:quanC}.

In Subsection \ref{sec:quanC}, we compute the corrections of order $\mathcal{O}(G^0)$ coming from bulk entanglement entropy $\delta S_\text{bulk}$ in the one-particle excited state. We show that the change in the bulk entanglement entropy up to linear order in the bulk density matrix $\delta \rho_\text{bulk}$ is
\begin{equation}
\delta S_{\text{bulk}}^{(\delta\rho)}=\frac{\Gamma(\tfrac{3}{2})  \Gamma(\Delta+1)}{\Gamma (\Delta+\tfrac{3}{2} )}\frac{(2 \alpha R)^{2\Delta} }{[(t^2-x_c^2+\alpha ^2)^2+4 \alpha ^2 x_c^2]^{\Delta }}\Big[1+\sum_{i=1}^{\infty}\mathcal{Q}_i(t,x_c,\alpha)R^{2i}\Big]\,,\nonumber
\end{equation}
as shown in equation \eqref{linbulkEE}. This clearly cancels the second series in $\delta S^\text{geom}_A$, thus removing the terms that do not appear in the CFT calculation. We explain that this cancelation is, in fact, expected from the precise relation between the CFT and bulk modular Hamiltonians, given in equation (\ref{modularHs}). Then, we use the \textit{bulk} replica trick to calculate the bulk entanglement entropy at quadratic order in $\delta \rho_\text{bulk}$. This calculation involves the Bogoliubov coefficients that bring the global state to the Rindler state associated with the entanglement wedge of $A$, which we only obtain in the small $R$ limit. The final result, given in equation \eqref{quadbulkEE}, reads
\begin{equation}
\delta S^{(\delta\rho^2)}_{\text{bulk}}=-\frac{\Gamma(\tfrac{3}{2})\Gamma(2\Delta+1)}{\Gamma(2\Delta+\tfrac{3}{2})}\[\frac{2 \alpha R }{\sqrt{4\alpha^2x_c^2+\left(\alpha^2+t_c^2-x_c^2\right)^2}}\]^{4\Delta}\,.\nonumber
\end{equation}
This term is in perfect agreement with the dynamical correction $\delta S^{dyn}_A$ obtained in the CFT, thus providing a explicit check of the FLM formula in our dynamical setting.

In Subsection \ref{bulkInterp}, we argue that our setup can be interpreted as a simple toy model for black hole evaporation. In order to reach this conclusion, we use the CHM map \cite{Casini:2011kv} and specialize to a Rindler observer adapted to the entanglement wedge of the region $A$, as shown in Figure \ref{fig:globalfigs2}. Indeed, for such an observer the bulk geometry coincides with that of a planar BTZ black hole, with metric given in \eqref{btzmetric}, so that $\delta S_{\text{bulk}}$ computes the entropy of matter fields outside the black hole. We further observe that, as seen from the global perspective, the RT surface and hence the black hole horizon generically vary in time. For the particular case of centered intervals, the entanglement wedge corresponds to that of an interval with $\delta \tau\equiv|\tau_2-\tau_1|=0$ and opening angle $\delta \theta\equiv |\theta_2-\theta_1|$ given by equation (\ref{deltatheta}),
\begin{align}\nonumber
\delta \theta&=\begin{cases}
  \displaystyle 2\pi - \left|2 \arctan\left(\tfrac{2\alpha R}{R^2-\alpha^2-t^2}\right)\right| \, , & \displaystyle 0<t<t_{\text{Page}} \ ,\\[3ex]
  \displaystyle  \left|2 \arctan\left(\tfrac{2\alpha R}{R^2-\alpha^2-t^2}\right)\right| \, , & \displaystyle t>t_{\text{Page}} \ ,
 \end{cases}
\end{align}
with $t_{\text{Page}}=\sqrt{R^2-\alpha ^2}$. Thus the horizon decreases monotonically in time as time goes from $t=0$ to $t=\infty$. In this way, we can interpret $\delta S_{\text{bulk}}$ as the entropy of Hawking radiation in an \textit{eternally evaporating} black hole geometry. In fact, we show that $\delta S_{\text{bulk}}$ in this case follows the expected behavior for a unitary \emph{Page curve}, as seen in Figure \ref{fig:bulkEE}, i.e., increasing from $t=0$ up to $t=t_{\text{Page}}$ and then decreasing as the black hole fully evaporates. We explain that this behavior follows directly from the purity of the quantum state in combination with the Araki-Lieb inequality \eqref{araki-lieb}.
This shows that unitarity \emph{can} be preserved in a semiclassical analysis, and implies that in more realistic models of black hole evaporation, the information loss or the lack thereof should \emph{not} be interpreted as an artifact of the approximation.

In Section \ref{sec:conclusions}, we conclude with some important remarks and a list of open questions.

\section{CFT entanglement entropy\label{sec:CFTEE}}

In this section we will discuss the CFT calculation of the entanglement entropy of a single interval after a local quench.

\subsection{Reduced density matrix for the local quench}

\subsubsection*{Regularized quenched state}

Consider the vacuum state of a two-dimensional, large-$c$ CFT: $\ket{0}$. We are interested in a special class of excited states produced by locally quenching the CFT at time $t=0$ by inserting an operator $\cO(0,x_0)$ at a point $x=x_0$. Due to the quench, the state $\ket{0}$ changes to the excited state
\begin{equation}\label{quenchstate}
\ket{\psi} = \cO(0,x_0) \ket{0}\,.
\end{equation}
Exactly localized states of the form (\ref{quenchstate}) contain modes of unbounded frequency, which is problematic if we want quantities like energy (density) and entanglement entropy in the state $\ket{\psi}$ to be finite. Hence, we need to regularize the state. A convenient way to do this is to give the time coordinate of $\cO$ a small imaginary part $\alpha$. We will thus take the state to be
\begin{equation}
\label{eq:normstate}
\ket{\psi} = \sqrt{\mathcal{N}} \, e^{-\alpha H} \, \cO(0,x_0)  \ket{0}\,.
\end{equation}
Notice that in this state $\alpha$ acts as a UV regulator, so high energy modes are effectively suppressed by the factor $e^{-\alpha H}$. The constant $\sqrt{\mathcal{N}}$ is an appropriate normalization that ensures that the state $\ket{\psi}$ as written above has norm one. We will omit this factor from here on. We can recover the proper normalization of various physical quantities whenever necessary.

In the Schr\"odinger picture, this excited state under consideration evolves under the CFT Hamiltonian $H$ for $t>0$, leading to the spread of energy and entanglement. Specifically, at time $t>0$ the state can be written as
\begin{equation}
\ket{\psi(t)} = e^{-i H t} \, e^{-\alpha H} \, \cO(0,x_0) \ket{0}\,.
\end{equation}
Consequently, the density matrix of the CFT is given by
\begin{equation}\label{densityop}
\rho(t) = e^{-i H t} \, e^{-\alpha H} \, \cO(0,x_0) \ket{0} \bra{0} \cO^\dagger(0,x_0)\,  e^{-\alpha H} \, e^{i H t}\,.
\end{equation}
One could carry out many of the computations in the real time formalism; however, we find it convenient to work in Euclidean time, by doing the Wick rotation
\bea\label{wick1}
t \to -i\tau\,,
\eea
and taking $\tau$ as a real variable. At the end of the calculation, we would need to Wick rotate back to obtain the final results in real time. The Euclidean density operator (\ref{densityop}) is
\bea\label{rho-tau}
\rho(\tau)
&\equiv &\cO(\tau_2,x_0) \, \ket{0} \bra{0} \, \cO^\dagger(\tau_1,x_0)\,,
\eea
where the operator $\cO(\tau,x)$ is now in the (Euclidean) Heisenberg picture, i.e.,
\be\label{HeisOps}
\cO(\tau_i,x)=e^{H\tau_i} \cO(0,x_0) e^{-H\tau_i}\,,
\ee
and we have defined the Euclidean times $\tau_1=-\tau+\alpha$ and $\tau_2=-\tau-\alpha$. This object has a well defined Euclidean path integral representation, namely its matrix elements are given in terms of an Euclidean path integral over $\mathbb{R}^2$ with open cuts on the $\tau=0^+$ and $\tau=0^-$ surfaces respectively. Schematically this is
\bea\label{PI-rep-A}
\langle \psi|\rho(\tau)|\psi' \rangle=\int [d\phi] \, \cO^\dagger(\tau_1,x_0)\,\cO(\tau_2,x_0) \delta[\phi(0^-,x)-\psi] \delta[\phi(0^+,x)-\psi'] e^{-S_E[\phi]}\,,
\eea
where $\phi$ represents the field content of the theory, $S_E[\phi]$ the associated Euclidean action, and $[d\phi]$ the appropriate path integral measure.

\subsubsection*{Reduced density matrix and its moments }

Let us now consider a subsystem  $A \equiv \{ \, x  \, | \,  x \in [x_L, x_R] \, \}$. Starting from (\ref{PI-rep-A}), we can arrive to a formula for the reduced density matrix on $A$, defined as the partial trace
\begin{equation}
\rho_A(\tau) \equiv \text{tr}_{A^c}\rho(\tau)\,,
\end{equation}
with $A^c\equiv \{ \, x  \, |x\notin [x_L,x_R]\, \}$. In practice, this is implemented by identifying the open cuts at $\tau=0^+$ with the one at $\tau=0^-$ on $A^c$ and summing over all field configurations there. The result is a path integral over $\mathbb{R}^2$ with open cuts along $A$ at $\tau=0$,
\be\label{PI-rep-A2}
\langle \psi_A|\rho_A(\tau)|\psi_A' \rangle=\int [d\phi] \, \cO^\dagger(\tau_1,x_0)\,\cO(\tau_2,x_0) \delta[\phi(0^-)-\psi_A] \delta[\phi(0^+)-\psi_A'] e^{-S_E[\phi]}\,,
\ee
where the field configurations $\phi$ and boundary conditions $\psi_A$ and $\psi'_A$ are defined on $A$.

The path integral representation is quite convenient since it relates the calculation of the moments of the reduced density matrix  $\tr \rho_A^n$ with a path integral computation over a multi-sheeted Riemann surface $\Sigma^A_n$, obtained by taking $n$ copies of the QFT on $\(\mathbb{R}^{2}\)^{\otimes n}$ and sewing them together along the region $A$.  We denote that path integral on this state by $Z_n(A)$, so that the above relation can be written as
\bea
\tr \rho_A^n=Z_n(A)\,.
\eea
For the normalized density matrix $\hat{\rho}_A=\rho_A/\tr{\rho}_A$ one has the formula
\bea\label{moments}
\tr \hat{\rho}_A^n=\frac{Z_n(A)}{Z^n_1}\,,
\eea
where $Z_1$  is the Euclidean path integral evaluated on a single copy of $\mathbb{R}^{2}$, with the operator insertions that create the state. Specifically, taking the trace of (\ref{PI-rep-A2}), we arrive at
\bea
Z_1 \equiv \tr\rho_A(\tau)=\int [d\phi] \, \cO^\dagger(\tau_1,x_0)\,\cO(\tau_2,x_0) e^{-S_E[\phi]} \, .
\eea
It is easy to see that $Z_1$ can be further related to the two-point function of Euclidean operators,
\bea\label{correlator-OO}
\langle  \cO^\dagger(\tau_1,x_0)  \cO(\tau_2,x_0) \rangle\equiv \frac{\int [d\phi] \, \cO^\dagger(\tau_1,x_0)\,\cO(\tau_2,x_0) e^{-S_E[\phi]}}{\int [d\phi]  e^{-S_E[\phi]}}\,.
\eea
Identifying the denominator in (\ref{correlator-OO}) as the trace of the density operator in the ground state, $\tr\,\ket{0} \bra{0} =Z_1^{gs}$, we obtain that
\bea\label{Z1Z1gs}
Z_1=\langle  \cO^\dagger(\tau_1,x_0)  \cO(\tau_2,x_0) \rangle\, Z_1^{gs}\,.
\eea
A similar analysis leads to an expression on the multi-sheeted surface that generalizes (\ref{Z1Z1gs}):
\bea\label{mom-corr}
Z_n(A)= \big \langle \prod_{k=1}^n \cO_{(k)}^{\dagger}(\tau_1,x_0)  \cO_{(k)}(\tau_2,x_0) \big \rangle_{\Sigma_n^A} \,\, Z^{gs}_n(A)\,,
\eea
where the index  $k$ represents the fact that the operator is inserted in the $k^{\text{th}}$ sheet of $\Sigma_n^A$ and $Z_n^{gs}(A)$ represents the ground state partition function on the replicated manifold $\Sigma_n^A$, i.e. one without any operator insertions.

\subsection{The replica trick}

We are now interested in calculating the von Neumann entropy associated with $\rho_A$,
\begin{equation}
S_A = - \tr(\hat{\rho}_A \, \log \hat{\rho}_A)\,.
\end{equation}
However, in QFT the Hilbert space is infinite dimensional and it is hard to calculate the logarithm of $\hat{\rho}_A$. Instead, one often uses the so-called \textit{replica trick}, where one first computes the R\'enyi entropies, defined as
\begin{equation}
S_n \equiv  \frac{\log \tr \hat{\rho}_A^n}{1-n}\,,
\end{equation}
and then extract the entanglement entropy as the formal limit
\begin{equation}\label{renyiA}
S_A = \lim_{n \to 1} \, S_n\,.
\end{equation}
In practice, most methods allow us to calculate $S_n$ for $n \in \mathbb{Z}$. To take the limit in $(\ref{renyiA})$ one analytically continues $n$ to real numbers and defines $S_n$ in the neighbourhood of $n=1$. Such a continuation exists and is unique provided $S_n$ has proper asymptotics.\footnote{The exact conditions are stated in what is known as Carlson's theorem.}

Using the path integral representation of the reduced density operator and its moments, one can write down the following formula for the R\'enyi entropies
\bea
S_n&=&\frac{1}{1-n}\log\(\frac{Z_n(A)}{Z_1^n}\)\,,
\eea
where $Z_1$ and $Z_n(A)$ are given by (\ref{Z1Z1gs}) and (\ref{mom-corr}), respectively. This expression can be massaged into the following form
\bea \label{Renyi}
S_n&=&\frac{1}{1-n}\log\left[\frac{ \big\langle \, \prod_{k=1}^n \cO_{(k)}^{ \dagger}(\tau_1,x_0)  \cO_{(k)}(\tau_2,x_0) \, \big\rangle_{\Sigma_n^A}}{ \langle  \cO^\dagger(\tau_1,x_0)  \cO(\tau_2,x_0) \rangle^n} \frac{Z^{gs}_n(A)}{\(Z^{gs}_1\)^n}\right]\,,\nonumber \\
&=&\frac{1}{1-n}\log\left[\frac{ \big \langle \, \prod_{k=1}^n \cO_{(k)}^{ \dagger}(\tau_1,x_0)  \cO_{(k)}(\tau_2,x_0) \,  \big \rangle_{\Sigma_n^A}}{ \langle  \cO^\dagger(\tau_1,x_0)  \cO(\tau_2,x_0) \rangle^n}\right]+ \frac{1}{1-n}\log\left[ \frac{Z^{gs}_n(A)}{\(Z^{gs}_1\)^n}\right]\,.\
\eea
Identifying the second term in  (\ref{Renyi}) as the ground state R\'enyi entropy $S_n^{gs}$, one can then write down a simple formula for the regularized R\'enyi entropy, $\delta S_n\equiv S_n-S^{gs}_n$,
\bea
\delta S_n=\frac{1}{1-n}\log\left[\frac{ \big\langle \, \prod_{k=1}^n \cO_{(k)}^{ \dagger}(\tau_1,x_0)  \cO_{(k)}(\tau_2,x_0) \, \big \rangle_{\Sigma_n^A}}{ \langle  \cO^\dagger(\tau_1,x_0)  \cO(\tau_2,x_0) \rangle^n}\right]\,.
\eea
Provided one can compute this quantity and find its analytic continuation for $n\approx 1$, the replica trick gives us the regularized von Neumann entropy $\delta S_A$,
\bea\label{regEE}
\delta S_A=\lim_{n\to 1} \frac{1}{1-n}\log\left[\frac{ \big \langle \, \prod_{k=1}^n \cO_{(k)}^{ \dagger}(\tau_1,x_0)  \cO_{(k)}(\tau_2,x_0) \,  \big  \rangle_{\Sigma_n^A}}{ \langle  \cO^\dagger(\tau_1,x_0)  \cO(\tau_2,x_0) \rangle^n}\right]\,.
\eea
We will now proceed to evaluate this quantity and study it in detail.

\subsubsection{Conformal mapping}

In order to exploit the full power of conformal invariance in two dimensions, it is convenient to use complex coordinates $w$ and $\bw$ to label our space-time points, where
\bea
w=x+i\tau,\qquad \bw =x-i\tau\,.
\eea
In these coordinates the full density operator becomes
\begin{equation}
\label{eq:fullrho}
\rho = \cO(w_2,\bar{w}_2) \ket{0} \bra{0} \cO^\dagger(w_1,\bar{w}_1)
\end{equation}
where $w_i \equiv x_0+i\tau_i$, $\bar{w}_i \equiv x_0-i\tau_i$,  and with the
$\tau_{i}$ given below (\ref{HeisOps}). Expanding it out, these coordinates are explicitly given by
\be
\begin{split}\label{w12}
&w_1=x_0-i(\tau-\alpha),\quad \bar{w}_1=x_0+i(\tau-\alpha)\,,  \\
&w_2=x_0-i(\tau+\alpha),\quad \bar{w}_2=x_0+i(\tau+\alpha) \,.
\end{split}
\ee
Next, we need to express (\ref{regEE}) in complex coordinates. For that purpose we adopt the following prescription. We label the coordinates on the $k^{\text{th}}$ sheet of $\Sigma_n^A$ as $(w^{(k)}, \bw^{(k)})$ and leave the operators $\cO$ and $\cO^\dagger$ unlabeled. We further define the insertion points of operators on the $k^{\text{th}}$ sheet to be
\bea\label{eq:defws}
w_1^{(k)}=w_{2k-1}\,,\quad \bw_1^{(k)}=\bw_{2k-1}\,,\quad w_2^{(k)}=w_{2k}\,,\quad  \bw_2^{(k)}=\bw_{2k}\,  \, .
\eea
With these changes in mind, the entanglement entropy (\ref{regEE}) becomes
\bea\label{regEE-2}
\delta S_A=\lim_{n\to 1} \frac{1}{1-n}\log\left[\frac{ \big \langle \,  \prod_{k=0}^{n-1}  \cO^\dagger(w_{2k+1}, \bar{w}_{2k+1}) \, \cO(w_{2k+2},\bar{w}_{2k+2} ) \,  \big \rangle_{\Sigma_n^A}}{ \langle   \cO^\dagger(w_1, \bar{w}_1) \, \cO(w_2,\bar{w}_2) \,  \rangle^n}\right]\,.
\eea

\subsection*{The uniformization map}

\begin{figure}[t!]
\begin{center}
  \includegraphics[angle=0,width=0.4\textwidth]{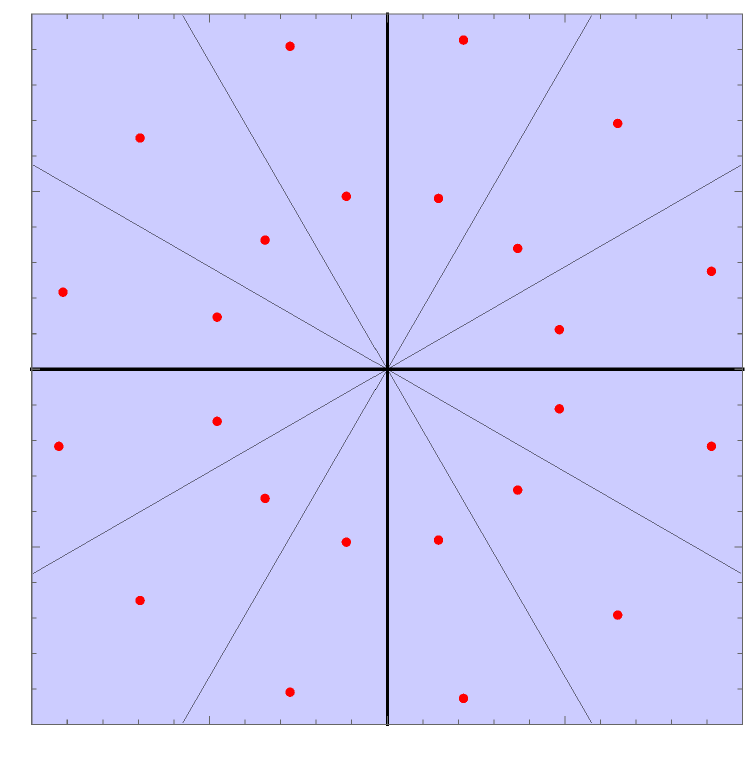}
\begin{picture}(0,0)
\put(-17,120){{\small $z_2$}}
\put(-53,106){{\small $z_1$}}
\put(-39,154){{\small $z_4$}}
\put(-63,125){{\small $z_3$}}
\qbezier(-55,141)(-57,146)(-65,147)
\put(-67,146.7){\vector(-1,0){0.5}}
\put(-18,67){{\small $z_{24}$}}
\put(-54,76){{\small $z_{23}$}}
\put(-17,166){{\small $\mathbb{C}$}}
\end{picture}
\end{center}
\vspace{-0.5cm}
\caption{\small Schematic representation of the uniformization map (\ref{eq:zofw}). After this transformation is implemented, each sheet of the original Riemann surface $\Sigma_n^A$ maps to a wedge in the complex plane $\mathbb{C}$ with $2\pi (k-1)/n \leq \theta \leq 2\pi k/n$. The insertion points are mapped according to (\ref{eq:oddz}) and (\ref{eq:z1z2map}) which means that $i)$ only two of these points are inserted in each wedge and $ii)$ points with $k\geq1$ differ only by a phase to those with $k=0$. For this plot we have set $n=12$ as an illustration, but for our particular calculation we are interested in smaller values of $n$, specifically, in the vicinity of $n\approx 1$. }
\label{fig:UniMap}
\end{figure}
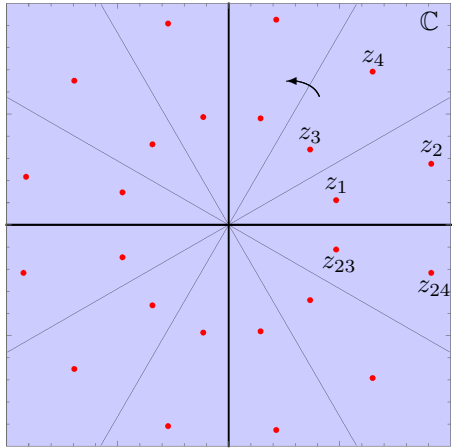

In general, computing a $2n$-point correlation function on a non-trivial Riemann surface such as $\Sigma_n^A$ is a very complicated task. Fortunately, it is simplified in two dimensions due to large conformal symmetry. The group of conformal transformations coincides with the analytic coordinate transformations:
\bea
z=f(w)\,, \qquad \bz=\bar{f}(\bw)\,.
\eea
By using a suitable function $f$ such that $f:\Sigma_n^A \to \mathbb{C}$, one can map the $2n$-point correlation function on $\Sigma_n^A$ in (\ref{regEE-2}) to a $2n$-point correlation function on $ \mathbb{C}$, which is simpler by the virtue of the analytic transformations mentioned above. Such a map exists and is known as the \textit{uniformization map}
\begin{equation}
\label{eq:zofw}
z = \bigg( \frac{w - x_L}{w-x_R} \bigg)^{1/n} \,, \quad \bar{z} = \bigg( \frac{\bar{w} - x_L}{\bar{w}-x_R} \bigg)^{1/n}\,,
\end{equation}
where $w\in \Sigma_n^A$, and $z\in \mathbb{C}$. Under its action, the left end-point $x_L$ of the open cut on $\Sigma_n^A$ is mapped to the origin of the complex plane $\mathbb{C}$, whereas the right end-point $x_R$ is mapped to complex infinity along a direction that differs from one sheet to another. In particular, on the $k^{\text{\text{th}}}$ sheet, $x_R$ is mapped to complex infinity along the angle $\theta_k=2\pi (k-1)/n$ wth respect to to the real axis. This is easy to prove by studying the $w^{(k)} \to x_L$ and $w^{(k)}\to x_R$ limits of (\ref{eq:zofw}). Near the left end-point on the $k^{\text{th}}$ sheet, $w^{(k)}$ is obtained by circling counter-clockwise $(k-1)$ times around $x_L$, at a fixed but infinitesimal distance $\epsilon$ away from it. That is, $w^{(k)}=x_L+\epsilon\, e^{2\pi i (k-1)}$. Similarly, near the right end-point $x_R$ $(k-1)$ on the $k^{\text{th}}$ sheet, $w^{(k)}$ is given by circling clockwise around $x_R$ at a fixed distance $\epsilon$, i.e. $w^{(k)}=x_R-\epsilon \, e^{-2\pi i (k-1)}$. In the limit, this gives us
\bea
\label{limit0}
\lim_{w^{(k)}\to x_L } z(w^{(k)})&=&e^{2\pi i (k-1) /n}\lim_{\epsilon \to 0} \(\frac{\epsilon}{\ell}\)^{1/n} \,,\\
\label{limitinfty}
\lim_{w^{(k)}\to x_R } z(w^{(k)})&=&e^{2\pi i (k-1) /n}\lim_{\epsilon \to 0} \(\frac{\ell}{\epsilon} \)^{1/n}\,,
\eea
where $\ell\equiv x_R - x_L$ is the length of the interval. Moreover, under this map the $k^{\text{th}}$ sheet of $\Sigma_n^A$ is mapped to a sector of the complex plane $\mathbb{C}$ defined within angles $2\pi (k-1)/n \leq \theta \leq 2\pi k/n$, as shown in Figure \ref{fig:UniMap}. Finally, it is easy to see that the insertion points $w_{2k+1}$, $\bw_{2k+1}$, $w_{2k+2}$ and  $\bw_{2k+2}$ in $\Sigma_n^A$ are mapped to
\begin{align}
\begin{split}
\label{eq:oddz}
& z_{2k+1} = e^{i 2 \pi k/n} \, z_1 \, , \qquad \, z_{2k+2} = e^{i 2 \pi k/n} z_2\,, \\
& \bz_{2k+1} = e^{-i 2 \pi k/n} \, \bz_1 \, , \qquad \bar{z}_{2k+1} = e^{-i 2 \pi k/n} \, \bar{z}_2\,,
\end{split}
\end{align}
on $\mathbb{C}$, respectively, where $z_1$, $\bz_1$, $z_2$, and $\bz_2$ are the principal roots in
\begin{align}
\begin{split}
\label{eq:z1z2map}
z_1 \equiv \bigg( \frac{w_1 - x_L}{w_1-x_R} \bigg)^{1/n} , \, \qquad \bar{z}_1 \equiv \bigg( \frac{\bar{w}_1 - x_L}{\bar{w}_1-x_R} \bigg)^{1/n}\,, \\
z_2 \equiv \bigg( \frac{w_2 - x_L}{w_2-x_R} \bigg)^{1/n} , \, \qquad \bar{z}_2 \equiv \bigg( \frac{\bar{w}_2 - x_L}{\bar{w}_2-x_R} \bigg)^{1/n}\,.
\end{split}
\end{align}
In Figure \ref{fig:UniMap} we show an illustrative example of the transformation.

Under the map \eqref{eq:zofw}, a primary operator $\cO$ without spin transforms as
\bea\label{map}
\cO(w,\bw)= \bigg(\frac{d z}{d w} \bigg)^{h} \bigg(\frac{d \bz}{d \bw} \bigg)^{\bar{h}}\cO(z,\bz)\,,\qquad \bar{h}=h\,,
\eea
and similarly for $\cO^\dagger$.\footnote{We will continue using $h$ for the time being but at the end of the calculation we will express the final answers in terms of the conformal dimension $\Delta=h+\bar{h}=2h$.} The $2n$-point correlator on $\Sigma_n^A$ appearing in (\ref{regEE-2}) then transforms to the following $2n$-point correlator on $\mathbb{C}$
\be\label{corr-rels}
\big \langle \, \prod_{k=0}^{n-1}  \cO^\dagger(w_{2k+1}, \bar{w}_{2k+1}) \, \cO(w_{2k+2},\bar{w}_{2k+2}) \, \big \rangle_{\Sigma_n^A}= \mathcal{J}^n_{\cO^\dagger}\mathcal{J}^n_{\cO} \, \big \langle  \, \prod_{k=0}^{n-1}  \cO^\dagger(z_{2k+1}, \bar{z}_{2k+1}) \, \cO(z_{2k+2},\bar{z}_{2k+2}) \, \big \rangle_{\mathbb{C}}\,,
\ee
where $\mathcal{J}^n_{\cO}$ and $\mathcal{J}^n_{\cO^\dagger} $ combine all the Jacobian factors coming from the map \eqref{map} for the $2n$ operators and are given by
\begin{align}
\label{eq:splitJac1}
\mathcal{J}^n_{\cO^\dagger} &\equiv  \prod\limits_{k=0}^{n-1} \, \bigg(\frac{d z}{d w} \bigg)^{h} \bigg|_{z_{2k+1}} \,  \bigg(\frac{d \bar{z}}{d \bar{w}} \bigg)^{\bar{h}} \bigg|_{\bar{z}_{2k+1}} \,, \\
\label{eq:splitJac2}
\mathcal{J}^n_{\cO} &\equiv \prod\limits_{k=0}^{n-1} \, \bigg(\frac{d z}{d w} \bigg)^{h} \bigg|_{z_{2k+2}} \,  \bigg(\frac{d \bar{z}}{d \bar{w}} \bigg)^{\bar{h}} \bigg|_{\bar{z}_{2k+2}} \,.
\end{align}
It is also convenient to define the factor $\mathcal{J}^1_{\cO^\dagger}\mathcal{J}^1_{\cO}$ such that\footnote{Notice that this factor do \emph{not} follow from setting $n=1$ in (\ref{eq:splitJac1}) and (\ref{eq:splitJac2}). In fact, both correlators in (\ref{corr-rels-2}) are computed in $\mathbb{C}$.}
\bea \label{corr-rels-2}
\langle \cO^\dagger(w_{1}, \bar{w}_{1}) \, \cO(w_{2},\bar{w}_{2}) \rangle_{\mathbb{C}}= \mathcal{J}^1_{\cO^\dagger}\mathcal{J}^1_{\cO}  \, \big \langle \,   \cO^\dagger(z_{1}, \bar{z}_{1}) \, \cO(z_{2},\bar{z}_{2}) \, \big  \rangle_{\mathbb{C}}\,.
\eea
With these notations in mind, we can now rewrite the expression for the regularized entanglement entropy (\ref{regEE-2}) as\footnote{One must be careful when separating logarithms inside a limit. This particular separation is possible because the arguments of both logarithms approach unity as $n\to 1$ in such a way that the limits remain finite.}
\bea\label{regEE-3}
\delta S_A&=&\lim_{n\to 1} \frac{1}{1-n}\log\[\frac{\mathcal{J}^n_{\cO^\dagger}\mathcal{J}^n_{\cO} }{\(\mathcal{J}^1_{\cO^\dagger}\)^n\(\mathcal{J}^1_{\cO}\)^n }\] \nonumber \\
&& \,\, +  \lim_{n\to 1} \frac{1}{1-n}\log \left[\frac{ \big \langle \, \prod_{k=0}^{n-1}  \cO^\dagger (z_{2k+1}, \bar{z}_{2k+1}) \, \cO(z_{2k+2},\bar{z}_{2k+2}) \, \big \rangle_\mathbb{C} }{\big( \langle  \cO^\dagger(z_1, \bar{z}_1) \, \cO(z_2,\bar{z}_2) \rangle_{\mathbb{C}} \big)^n } \right] \,.
\eea
which is naturally separated into two contributions: one which we call the universal part,
\bea\label{Suni}
\delta S_A^{uni}\equiv \lim_{n\to 1} \frac{1}{1-n}\log\[\frac{\mathcal{J}^n_{\cO^\dagger}\mathcal{J}^n_{\cO} }{\(\mathcal{J}^1_{\cO^\dagger}\)^n\(\mathcal{J}^1_{\cO}\)^n }\]\,,
\eea
and a dynamical part, defined as
\bea\label{Sdyn}
\delta S_A^{dyn}\equiv \lim_{n\to 1} \frac{1}{1-n}\log \left[\frac{ \big \langle  \, \prod_{k=0}^{n-1}  \cO^\dagger (z_{2k+1}, \bar{z}_{2k+1}) \, \cO(z_{2k+2},\bar{z}_{2k+2}) \, \big \rangle_{\mathbb{C}}  }{ \(  \langle  \cO^\dagger(z_1, \bar{z}_1) \, \cO(z_2,\bar{z}_2) \rangle_{\mathbb{C}} \)^n } \right]\,.
\eea
Notice that the former one depends only on the dimension of the quench operator $\cO$ and the underlying geometry of $\Sigma_n^A$, but is otherwise
independent of the coupling constants in the CFT under consideration. The latter one depends on a higher-point function, which is \emph{not} fixed by conformal symmetry, and it is therefore not universal.

\subsection{Regularized entanglement entropy}
In the previous subsection we showed that the regularized entanglement entropy after the local quench can be naturally separated into two contributions,
\bea
\delta S_A=\delta S_A^{uni}+\delta S_A^{dyn}\,,
\eea
with universal and dynamical parts, $\delta S_A^{uni}$ and $\delta S_A^{dyn}$, given in (\ref{Suni}) and (\ref{Sdyn}), respectively. We will now compute each of these contributions.

\subsubsection{Universal contribution}
The calculation of the universal part (\ref{Suni}) can be split in two steps: first, we need to compute the Jacobian factors $\mathcal{J}^n_{\cO^\dagger}$, $\mathcal{J}^n_{\cO}$,  $\mathcal{J}^1_{\cO^\dagger}$ and $\mathcal{J}^1_{\cO}$, and then we need to analytically continue the result for $n\in\mathbb{R}$ and take the limit $n\to1$. For the computation of the Jacobians we need the derivatives of $z$ and $\bar{z}$,
\bea\label{derivatives}
\frac{dz}{dw}=-\frac{z^{1-n}}{n \ell }(z^n-1)^2\,, \qquad \frac{d\bz}{d\bw}=-\frac{\bz^{1-n}}{n \ell }(\bz^n-1)^2\,,
\eea
with $\ell\equiv x_R-x_L$, evaluated at the points $z_{2k+1}$, $z_{2k+2}$, $\bz_{2k+1}$ and $\bz_{2k+2}$ respectively. Notice that we have expressed the answers directly in terms of the $z$-coordinates so the evaluation is now straightforward. To get a closed expression for  $\mathcal{J}^n_{\cO^\dagger} \, \mathcal{J}^n_{\cO} $ notice that the first factor of
(\ref{eq:splitJac1}) can be expressed as
\bea\label{factor-1}
 \!\!\prod\limits_{k=0}^{n-1} \bigg(\frac{d z}{d w} \bigg)^{\!h} \bigg|_{z_{2k+1}}\!\!=\prod_{k=0}^{n-1} \left(\frac{e^{\frac{2\pi i k(1-n)}{n}}z_1^{1-n} (z_1^n-1)^2 }{n\ell } \right)^{\!h}\!\! =\frac{e^{-i\pi (n-1)^2 h} z_1^{n (1-n) h} (z_1^n-1)^{2 nh } }{n^{n h}\ell ^{nh}}\,,
\eea
where we have used (\ref{eq:oddz}) and the fact that $e^{2\pi i k}=1$ for $k \in \mathbb{Z}$.  The second factor of (\ref{eq:splitJac1}) is just the complex conjugate of (\ref{factor-1}) so the product of the two cancels their phases. Repeating the same analysis for (\ref{eq:splitJac2}) and combining the two results, we obtain
\bea
\label{JJO}
\mathcal{J}^n_{\cO^\dagger} \, \mathcal{J}^n_{\cO} =\frac{\( |z_1| |z_2| \)^{2n(1-n)h} (z^n_1-1)^{2nh}(\bar{z}^n_1-1)^{2nh} (z^n_2-1)^{2nh}(\bar{z}^n_2-1)^{2nh}}{n^{4nh}\ell ^{4nh} }\,.
\eea
We also need the factor $\mathcal{J}^1_{\cO^\dagger} \, \mathcal{J}^1_{\cO} $. Writing down the explicit form of the two-point correlators in (\ref{corr-rels-2}), it follows that
\bea\label{JOO-2}
\mathcal{J}^1_{\cO^\dagger}\mathcal{J}^1_{\cO}=\frac{|z_1^n-1|^{4h}|z_2^n-1|^{4h} |z_1-z_2|^{4h}}{\ell^{4h} |z^n_1-z^n_2|^{4h}}\,,
\eea
which, in combination with (\ref{JJO}), leads to a closed expression for $\delta S^{uni}_A$ solely in terms of the insertion points $z_1$ and $z_2$,
\bea \label{Suni-z}
\delta S^{uni}_A=\lim_{n\to 1} \frac{1}{1-n} \, \log \left[ \frac{\(|z_1||z_2|\)^{2n(1-n)h}}{n^{4nh} }\(\frac{|z_1^n-z_2^n|}{|z_1-z_2|}\)^{4nh}  \right]\,.
\eea

It only remains to carry out the analytic continuation for $n\in\mathbb{Z}$ and take the explicit limit $n\to1$. We will merely transcribe the outcome of the calculation and relegate the details of this analysis to Appendix \ref{app:univ}. The final result yields:
\bea \label{Suni-1}
\delta S^{uni}_A=\Delta\left[2+\frac{1}{2}\(\frac{z_2+z_1}{z_2-z_1}\)\log\(\frac{z_1}{z_2}\)+\frac{1}{2}\(\frac{\bar{z}_2+\bar{z}_1}{\bar{z}_2-\bar{z}_1}\)\log\(\frac{\bar{z}_1}{\bar{z}_2} \) \right]\,,
\eea
where $\Delta=2h$ is the dimension of the quench operator $\mathcal{O}$. Here, the insertion points $z_1$, $\bz_1$, $z_2$ and $\bz_2$ are given by setting $n=1$ in the general expressions (\ref{eq:z1z2map}), i.e.,
\bea \label{zw}
z_1 \equiv \bigg( \frac{w_1 - x_L}{w_1-x_R} \bigg)\,, \qquad \bz_1 \equiv  \bigg( \frac{\bar{w}_1 - x_L}{\bar{w}_1-x_R} \bigg)\,,
\eea
and similarly for $z_2$ and $\bz_2$. Finally, we notice that by using the identity
\bea
\log(z)=2 \,{\rm arctanh}\(\frac{z-1}{z+1}\)\,,
\eea
we can rewrite (\ref{Suni-1}) in a slightly more useful form:
\bea \label{Suni-2}
\delta S^{uni}_A=\Delta\left[2-\(\frac{z_2+z_1}{z_1-z_2}\){\rm arctanh}\(\frac{z_1-z_2}{z_1+z_2}\) - \(\frac{\bz_2+\bz_1}{\bz_1-\bz_2}\){\rm arctanh}\(\frac{\bz_1-\bz_2}{\bz_1+\bz_2}\) \right]\,,
\eea
which is conveniently  written in terms of $(z_1-z_2)/(z_1+z_2)$ and its complex conjugate. It also has the advantage that its small interval expansion converges much faster than the one obtained from (\ref{Suni-1}), a property that will prove useful later.

\subsubsection{Dynamical contribution}

Next, let us consider the dynamical contribution (\ref{Sdyn}). To calculate the $2n$-point correlator in the numerator we will work in the special class of \textit{holographic CFTs}. We will further simplify the calculation by taking the limit of small intervals.

\begin{figure}[t!]
\begin{center}
  \includegraphics[angle=0,width=0.4\textwidth]{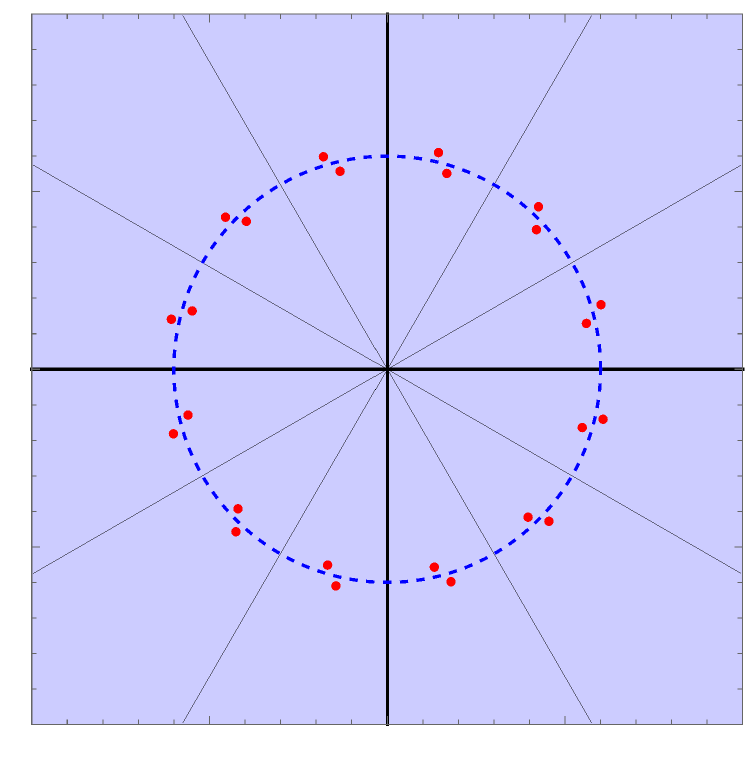}
\begin{picture}(0,0)
\put(-43,120){{\small $|z|=1$}}
\put(-37,105){{\small $z_2$}}
\put(-53,102){{\small $z_1$}}
\put(-17,166){{\small $\mathbb{C}$}}
\end{picture}
\end{center}
\vspace{-0.5cm}
\caption{\small Small interval limit of the $2n$-correlator in the complex plane $\mathbb{C}$. Assuming $R\to0$ all pairs of insertion points with equal $k$ (i.e., in the same sheet of the original Riemann surface $\Sigma_n^A$) approach to each other and one can carry out an OPE expansion between the two. This is possible provided that $n$ is not too large, so points with different $k$ are still a finite distance away from each other as $R\to0$. For this plot we have set $n=12$ as an illustration, but for our particular calculation we are interested in smaller values of $n$, specifically, in the vicinity of $n\approx 1$. }
\label{fig:OPElimit}
\end{figure}
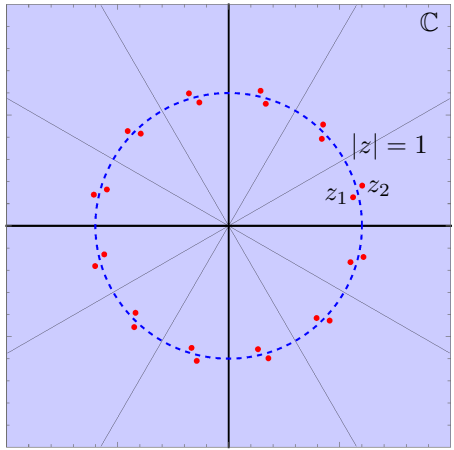

First, notice that in the small interval limit $x_R\to x_L$ or $\ell \to 0$, all the insertion points localize around the unit circle $|z|=1$, as shown in Figure \ref{fig:OPElimit}. This behavior can be traced back to the uniformization map (\ref{eq:zofw}), and can be seen more directly from the expressions (\ref{eq:oddz}) and (\ref{eq:z1z2map}). In this limit we obtain
\bea
\left|z_1-z_2\right|= \frac{4 R \alpha }{n \sqrt{(x_0-x_c)^2+\alpha^2-\tau^2)^2+4\tau^2(x_0-x_c)^2} }+\mathcal{O}(R^2)\,,
\eea
where we have used $x_c$ and $R$ to label the center and the half-length of the interval, respectively
\be
x_c\equiv \frac{x_R+x_L}{2}\,,\qquad R\equiv \frac{x_R-x_L}{2} = \frac{\ell}{2}\,.
\ee
Similarly, it is easy to see that in this limit $|z_{2k+1}-z_{2k+2}|\sim\mathcal{O}(R)$ for all other $k$ since the insertion points with $k\geq1$ differ only by a phase to those with $k=0$. This implies that \emph{all} pairs of operators with equal $k$ in the $2n$-point correlator in (\ref{Sdyn}) become arbitrarily close in the limit $R\to0$. See Figure \ref{fig:OPElimit} for an illustrative representation of this limit.
Assuming that $n$ is not too large, points with different $k$ will be a finite distance apart in this limit, and we can replace each product by an OPE expansion of the form
\bea
\!\!\!\!\!\!\!\!\cO^\dagger(z_{2k+1}, \bar{z}_{2k+1})   \cO(z_{2k+2},\bar{z}_{2k+2}) =
\frac{1}{|z_1-z_2|^{2\Delta }}\Big[1+ \sum_\Phi |z_1-z_2|^{\Delta_\Phi} C_{\mathcal{O} \mathcal{O} \Phi }\, \Phi(z_k^c, \bar{z}_k^c)\Big]+\cdots. \label{OPE-k}
\eea
In such an expansion, the sum runs over all primary operators $\Phi$ while the dots represent contributions from their descendents. Notice, however, that we have isolated the contribution from the identity operator $\Phi=\unit$ (the ``$1$'' outside the sum), whose coefficient is fixed by the normalization of the two-point correlator. Finally, the operators $\Phi$
inside the sum must be evaluated at an arbitrary point $z_k^c$ in $\mathcal{C}$ within the radius of convergence of the expansion, which we can choose to place the fused operators. A simple and convenient
choice would be to pick the ``center point'' (rotated to the appropriate sector corresponding to a given sheet), i.e.,
\be\label{eq:centpoints}
z_k^c=e^{2\pi i k/n}\(\frac{z_1+z_2}{2}\)\,,\qquad \bz_k^c=e^{-2\pi i k/n}\(\frac{\bz_1+\bz_2}{2}\)\,.
\ee

At this point, and in order to proceed with the calculation, we need to specify more data about the CFT of interest. Since we want to compare our results with a bulk calculation using
the AdS/CFT correspondence, we will focus on CFTs with holographic duals or, in other words, \emph{holographic CFTs}. Generically, these are theories with
large central charge $c$ and a sparse spectrum of low-dimension operators \cite{Heemskerk:2009pn,Heemskerk:2010ty,Fitzpatrick:2010zm,Fitzpatrick:2013twa}. One of the important features that follow from these properties
is large-$c$ factorization \cite{ElShowk:2011ag}. Large-$c$ factorization is the statement that for single-trace operators $\Phi$ the OPE coefficients $\mathcal{C}_{\mathcal{O} \mathcal{O} \Phi }$ are
all suppressed, i.e.,
\be
\mathcal{C}_{\mathcal{O} \mathcal{O} \Phi }\sim \frac{1}{\sqrt{c}}\,.
\ee
Since we are interested in the contribution to the entanglement entropy at order $\mathcal{O}(1)$, we can safely ignore these contributions. On the other hand, multi-trace operators
can indeed have OPE coefficients at $\mathcal{O}(1)$ so they must be considered in the sum of (\ref{OPE-k}). In particular, in the limit of small $R$, the leading contribution
coming from the sum in \eqref{OPE-k} is given by the lightest multi-trace operator that can appear in the OPE, while all others are suppressed by a higher power of $R$. This operator
is $\Phi=\,\,:\!\!\cO^2\!\!:$, a double-trace, and has conformal dimension $\Delta_\Phi=2\Delta$. In holographic CFTs, its OPE coefficient is given by \cite{Belin:2017nze}
\bea\label{OO2O}
C_{\cO \cO \, \cO^2 }=\sqrt{2}\,.
\eea
Next, we insert the OPE expansions in equation \eqref{OPE-k} for all pairs of operators with the same index $k$ in equation (\ref{Sdyn}). At leading order, the result is given by all pair of operators replaced by the identity contribution. This gives rise to
\bea\label{OPE-corrID}
\Big\langle \prod_{k=0}^{n-1}  \cO^\dagger(z_{2k+1}, \bar{z}_{2k+1}) \, \cO(z_{2k+2},\bar{z}_{2k+2}) \Big\rangle=\frac{1}{|z_1-z_2|^{2n \Delta}}\,.
\eea
If we plug this into (\ref{Sdyn}) we obtain exactly zero, but it is easy to understand why. The reason is that we have already factored out the Jacobians of the correlators
in the numerator and denominator of (\ref{regEE}), in what we have called the \emph{universal contribution} (\ref{Suni}). This means that we can in fact interpret the universal term
as the contribution coming from the identity operator.

Similarly, we can analyze the sub-leading corrections by imagining the case when only a few of the pairs are replaced by the operator $\cO^2$, while the remaining ones are replaced by the identity. Now if only one pair is replaced by $\cO^2$, this sub-leading correction to \eqref{Sdyn} vanishes. This is because all one-point functions of local operators are exactly zero in the vacuum due to conformal invariance. Therefore, the first non-trivial contribution to the sub-leading term appears when \emph{two} pairs are replaced by the operator $\cO^2$. Considering all possible Wick contractions, and using (\ref{OO2O}), this yields
\bea\label{OPE-corr}
\!\!\!\!\!\!\!\!\!\Big\langle \prod_{k=0}^{n-1}  \cO^\dagger(z_{2k+1}, \bar{z}_{2k+1}) \, \cO(z_{2k+2},\bar{z}_{2k+2}) \Big\rangle=
\frac{1}{|z_1-z_2|^{2n \Delta}}\Big[1+2|z_1-z_2|^{4\Delta} \sum_{k\neq l=0}^{n-1} \langle \cO^2_k \cO^2_l \rangle\Big],
\eea
where we have used the short hand notation $\cO^2_k\equiv\cO^2(z_k^c,\bz_k^c)$, with $z_k^c$ and $\bz_k^c$ given in (\ref{eq:centpoints}). Using the replica symmetry one can fix
 the location of one of the operators appearing in the double sum of (\ref{OPE-corr}), say to the location of the operator with $\ell=0$, and multiply the result by $n$. This leads to:
\bea\label{O2O2}
 \sum_{k\neq l=0}^{n-1} \langle \cO^2_k \cO^2_l \rangle=\frac{n}{2}\sum_{k=1}^{n-1}\langle \cO^2_k \cO^2_l \rangle
  =\frac{n}{2}\frac{1}{|z_1+z_2|^{4\Delta}}\sum_{k=1}^{n-1}\frac{1}{\left| \sin\(\frac{\pi k}{n} \)\right|^{4\Delta}}\,,
\eea
where the extra factor of $1/2$ is introduced to avoid double counting. The sum in (\ref{O2O2}) can be carried out for $n\approx 1$ as was done in \cite{Calabrese:2010he} and further generalized to generic thermal green functions in \cite{Agon:2015ftl}. We will merely write the answer here, and relegate the
details of this analysis to Appendix \ref{app:dyn}. At the end, the outcome of the calculation yields:
\bea\label{sum}
\sum_{k=1}^{n-1}\frac{1}{\left| \sin\(\frac{\pi k}{n} \)\right|^{4\Delta}}\approx (n-1)\frac{\Gamma\(\frac{3}{2}\)\Gamma(2\Delta+1)}{ \Gamma(2\Delta+\frac{3}{2})}\,.
\eea
Combining (\ref{OPE-corr})-(\ref{sum}) leads to an expression for the leading term in the $n\to1$ expansion of the logarithm in (\ref{Sdyn}), i.e.,
\bea
\!\!\!\!\!\log \left[\frac{ \langle \prod_{k=0}^{n-1}  \cO^\dagger (z_{2k+1}, \bar{z}_{2k+1}) \, \cO(z_{2k+2},\bar{z}_{2k+2})\rangle  }{ \(  \langle  \cO^\dagger(z_1, \bar{z}_1) \, \cO(z_2,\bar{z}_2) \rangle \)^n } \right]\approx (n-1)\frac{\Gamma\(\frac{3}{2}\)\Gamma(2\Delta+1)}{ \Gamma(2\Delta+\frac{3}{2})}\left|\frac{z_1-z_2}{z_1+z_2}\right|^{4\Delta}\!\!\!.
\eea
Finally, taking the $n\to 1$ limit in (\ref{Sdyn}) we obtain
\bea\label{Sdyn-2}
\delta S_A^{dyn}=-\frac{\Gamma\(\frac{3}{2}\)\Gamma(2\Delta+1)}{ \Gamma(2\Delta+\frac{3}{2})}\left|\frac{z_1-z_2}{z_1+z_2}\right|^{4\Delta}\,.
\eea
We emphasize that this is only the first term in the small $R$ expansion of $\delta S_A^{dyn}$. On the other hand, our result for $\delta S_A^{uni}$
 given by equation (\ref{Suni-2}), is valid for any $R$.

\subsection{Analytic continuation and real-time interpretation\label{sec:realtime}}

Given our final expressions for $\delta S_A^{uni}$ and $\delta S_A^{dyn}$ given in (\ref{Suni-2}) and (\ref{Sdyn-2}), the final step is to analytically continue these results to real time. We will do so in this subsection. Along the way, we will uncover a clear physical picture for the spread of entanglement after local quenches that will allow us interpret our results in a transparent way.

\subsubsection*{Universal contribution}

Consider our result for $\delta S_A^{uni}$ given by equation (\ref{Suni-2}). As explained before, this is conveniently written in terms of $(z_1-z_2)/(z_1+z_2)$ and its complex conjugate so we will start by writing this combination in terms of the parameters $\{\tau, x_0, x_c, R, \alpha\}$ using (\ref{w12}) and (\ref{zw}):
\bea
\frac{z_1-z_2}{z_1+z_2}=\frac{-2i\alpha R}{(x_0-x_c)^2-R^2+\alpha^2-\tau^2-2i\tau(x_0-x_c)}\,.
\label{parameter-2}
\eea
Performing the analytic continuation to Lorentzian time, $\tau \to it$, leads to
\bea
\frac{z_1-z_2}{z_1+z_2}=\frac{-2i\alpha R}{(x_0-x_c)^2-R^2+\alpha^2+t^2+2t(x_0-x_c)}\,,
\label{parameter-3}
\eea
and similarly
\bea
\frac{\bar{z}_1-\bar{z}_2}{\bar{z}_1+\bar{z}_2}=\frac{2i\alpha R}{(x_0-x_c)^2-R^2+\alpha^2+t^2-2t(x_0-x_c)} \label{parameter-3b}\,.
\eea
Notice that after the analytic continuation, both parameters become purely imaginary and are no longer the complex conjugate of each other. Defining the parameters
\bea\label{chi}
\eta_{\pm}=\frac{2\alpha R}{R^2-(x_0-x_c\pm t)^2-\alpha^2}\,,
\eea
and using the identity $\text{arctanh}(ix)=i \,\text{arctan}(x)$, we can rewrite $\delta S_A^{uni}$ as
\bea\label{deltaS+-}
\delta S_A^{uni}=\Delta \left[2-\( \frac{1}{\eta_+}{\rm arctan}\, \eta_+ + \frac{1}{\eta_-}{\rm arctan}\, \eta_- \) \right]\,.
\eea

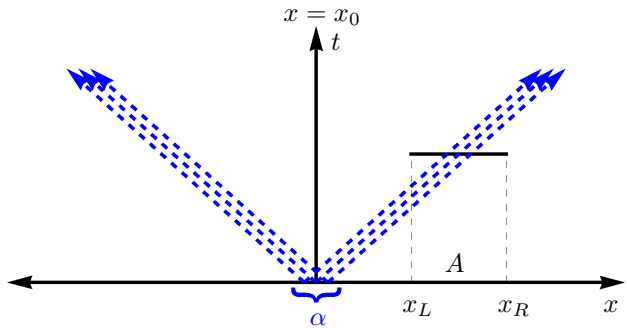
\begin{figure}[t!]
\begin{center}
  \includegraphics[angle=0,width=0.55\textwidth]{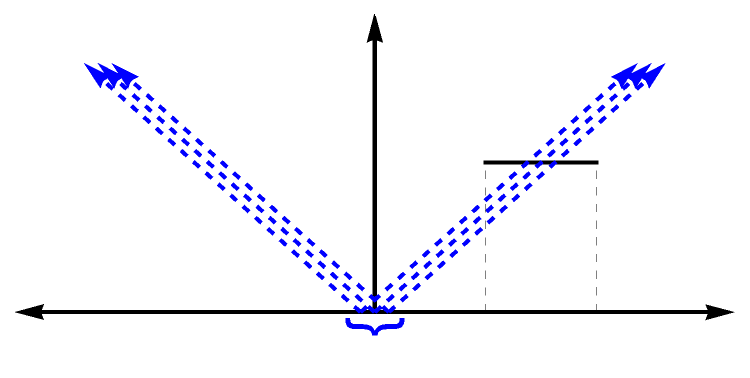}
\begin{picture}(0,0)
\put(-120,109){{\small $t$}}
\put(-138,120){{\small $x=x_0$}}
\put(-128,5){{\small{\color{blue}$\alpha$}}}
\put(-17,10){{\small $x$}}
\put(-77,25){{\small $A$}}
\put(-93,10){{\small $x_L$}}
\put(-57,10){{\small $x_R$}}
\end{picture}
\end{center}
\vspace{-0.5cm}
\caption{\small Pictorial representation of the universal contribution to the spread of entanglement entropy after a local quantum quench in 2d CFTs. The state at $t=0$ is the vacuum state perturbed by a local operator smeared over a region of compact support $\sim \alpha$. The state
evolves under the CFT Hamiltonian $H$ for $t > 0$, generating an entangled pair of wave packets that move in opposite directions at the speed of light. The wave packets eventually increase the entanglement entropy of region $A=\{x|x\in[x_L,x_R]\}$ ($x_L\equiv x_c-R$, $x_R\equiv x_c+R$),
and then decrease it as they disperse to infinity.}
\label{fig:bdyquench}
\end{figure}

Equation (\ref{deltaS+-}) allows an interesting interpretation. The entanglement generated by the perturbation can be interpreted as coming from two independent sources (corresponding to the two independent components $\eta_{\pm}$ moving in opposite directions and at the speed of light.
Each of these sources contributes by a term $-(\Delta/\eta)\, {\rm arctan}\,\eta$  with the specific $\eta_{\pm}$ depending on the motion of the source, see Figure \ref{fig:bdyquench} for a pictorial representation.

Physically, the two sources appear because the operator inserted at $(t=0,x=x_0)$ creates an entangled pair of wave packets moving in opposite directions (as required by momentum conservation) each with a characteristic size of order $\alpha$. If we further assume that quantum correlations between the two are negligible compared to the quantum correlations of each component with itself whenever the distance between the two is much larger than $\alpha$, we can predict that the entanglement entropy of a region $A$ will be the sum of two contributions, consistent with the separation found in (\ref{deltaS+-}).

Furthermore, this physical picture predicts that, as a function of time, entanglement entropy is maximal when the location of the two endpoints of $\partial A$, $\{x_L,x_R\}$ symmetrically divide each of the traveling wave packets in two. We can calculate the time when this happens.

\begin{itemize}
\item \textit{Operator inserted inside the interval}: In this case we expect that the centers of both right and left moving components $\eta_+$, and $\eta_-$  would reach the endpoints of the entangling region, $x_{R}=x_c+R$ and $x_{L}=x_c-R$, at different times $t_{\pm}>0$. In order to determine these times we need to equate $x(t)=x_0\pm t=x_c\pm R$ and solve for $t$. This gives rise to the times $t_{\pm}=R \pm (x_c-x_0) $.

\item \textit{Operator inserted outside the interval}: Let us assume for concreteness that the operator has been inserted to the left of the interval. In this case, the left moving wave packet will not cross the interval and so its contribution will not reach a maximum. On the other hand the right moving wave packet will reach local maximum at two different times $t_\pm$ given by the solution of
$x(t)=x_0+t=x_c\pm R$, i.e., $t_\pm=R\pm (x_c-x_0)$.

\end{itemize}
Both situations are consistent with our assumptions provided that $\alpha$ is sufficiently small.

If the above picture is correct then one should be able to reproduce the local times derived above in both scenarios using the formula (\ref{deltaS+-}). To do so, notice that the function $-(\Delta/\eta)\, {\rm arctan}\,\eta$ is an even function of $\eta$ and a monotonically increasing function of $|\eta|$, therefore its local maximum values occur when $|\eta|$ is globally maximal. From (\ref{chi}), and for $R-|x_0-x_c|\gg \alpha$, it is clear that this happens at the times
\be\label{times}
\pm t=R-(x_0-x_c) \, , \quad  {\rm and }\quad \pm t=-R-(x_0-x_c) \,.
\ee
Since our setup only considers positive times, then depending on whether the operator is inserted inside or outside the interval, we will have the following values for  $t_{\pm}$:

\begin{itemize}
\item \textit{Operator inserted inside the interval}: In this case $|x_0-x_c|<R$, so $R-(x_0-x_c)>0$ which means that out of the four times given in (\ref{times}) the only positive ones are
$t_+=R+(x_c-x_0)$ at which $\chi_+$ is maximal, and $t_-=R-(x_c-x_0)$ at which $\chi_-$ is maximal.

\item \textit{Operator inserted outside the interval}: In this case $|x_0-x_c|>R$.  Assuming the operator is inserted to the left of the interval $x_0<x_c-R$ then the positive times in this scenario are $t_{+}=R+(x_c-x_0)$ and $t_{-}=R-(x_c-x_0)$ which are both maxima of $\chi_+$.
\end{itemize}
These results are in perfect agreement with the expectations from the interpretation in terms of entangled wave packets traveling at the speed of light.

\subsubsection*{Dynamical contribution}

Let us now consider the dynamical term (\ref{Sdyn-2}). First, we write it in terms of the Euclidean parameters $\{\tau, x_0, x_c, R, \alpha\}$. From (\ref{Suni-2}) and (\ref{Sdyn-2}) it follows that
\bea
\delta S_A^{dyn}=-\frac{\Gamma\(\frac{3}{2}\)\Gamma(2\Delta+1)}{\Gamma(2\Delta+\frac{3}{2})}\[\frac{2\alpha  R }{\sqrt{\[(x_0-x_c)^2-R^2+\alpha^2-\tau^2\]^2+4\tau^2(x_0-x_c)^2}}\]^{4\Delta}\,.
\eea
which, after the analytic continuation to Lorentzian time, $\tau \to it$, becomes
\bea\label{Q-corrs}
\delta S_A^{dyn}=-\frac{\Gamma\(\frac{3}{2}\)\Gamma(2\Delta+1)}{\Gamma(2\Delta+\frac{3}{2})}\[\frac{2\alpha  R }{\sqrt{\[(x_0-x_c)^2-R^2+\alpha^2+t^2\]^2-4t^2(x_0-x_c)^2}}\]^{4\Delta}\,. \label{Q-corrs}
\eea
This expression does not have a similar separation as the one found for the universal piece $\delta S_A^{uni}$. This fact is manifest if we rewrite (\ref{Q-corrs}) as
\bea
\label{realSdyn}
\!\!\!\!\delta S_A^{dyn}=-\frac{\Gamma\(\frac{3}{2}\)\Gamma(2\Delta+1)}{\Gamma(2\Delta+\frac{3}{2})}\!\[\frac{2\alpha  R }{\left|(x_0-x_c-t)^2-R^2+\alpha^2\right|}\]^{2\Delta}
\!\!\[\frac{2\alpha  R }{\left|(x_0-x_c+t)^2-R^2+\alpha^2\right|}\]^{2\Delta}\!\!\!\!\!\!.
\eea
Indeed, this product factorization shows that the full quantum state has non-vanishing entanglement between the two wave packets.

Let us briefly discuss $\delta S_A^{dyn}$ as a function of time. It is maximal when the denominator in (\ref{Q-corrs})
\bea \label{denominator}
\big[ (x_0-x_c)^2-R^2+\alpha^2+t^2 \big]^2-4t^2(x_0-x_c)^2 \, ,
\label{poly}
\eea
is minimal. This function cannot be arbitrarily small as it comes from a sub-leading term in the OPE expansion \eqref{OPE-corr}. The smallest it can be is zero, which happens at the roots of (\ref{denominator}) which are
\bea
t_{\pm, \pm}=\pm (x_0-x_c) \pm \sqrt{R^2-\alpha^2} \, .
\eea
These roots have real, positive values for $R>\alpha$. Hence, consistency with the OPE expansion requires $R<\alpha$. Further, local extrema of \eqref{poly} occur when
\bea
t^*=0, \qquad  t^*_{\pm}=\pm\sqrt{R^2+\(x_0-x_c\)^2-\alpha^2}
\eea
For $t\geq 0$, it is easy to see that the global minima occur at
\begin{itemize}
\item $t^*=0$ for $\alpha^2>R^2+\(x_0-x_c\)^2$
\item  $t^*_+=\sqrt{R^2+\(x_0-x_c\)^2-\alpha^2}$ for $\alpha^2<R^2+\(x_0-x_c\)^2$
\end{itemize}
In the next sections, we will recover terms of this form from the bulk calculation and show that they generically arise from contributions due to bulk entanglement entropy.

\section{Aspects of the gravity duals\label{sec:gravity}}

We will now describe the geometry of quench states considered in the previous section via holography. We will follow the method described in \cite{Nozaki:2013wia}, wherein it was developed for operators of large conformal dimension $\Delta\gg c$. It involves finding the backreacted metric for a localized perturbation which is initially localized near the boundary and falls into the interior of AdS. In the following, we will generalize this calculation to a one-particle excited state of a \textit{light} scalar field coupled to gravity. The lightness of the scalar field implies that the dual operator have small conformal dimension, $\Delta \ll c$.

\subsection{Light operator excited states and bulk backreaction}

We begin by discussing the backreaction of a quantum scalar field on the metric of a pure AdS$_3$ spacetime. We start with the action
\begin{equation}\label{eq:actionscalar}
S=\frac{1}{16\pi G}\int d^3x \, \sqrt{-g} \, \left[ \mathcal{R}-2\Lambda- 8 \pi G \left( \partial_\mu \phi \, \partial^\mu \phi + m^2 \, \phi^2 \right)\right]\,,
\end{equation}
where $G$ is the Newton's constant in 3 dimensions, $\mathcal{R}$ denotes the Ricci scalar and $\Lambda$ is the cosmological constant which is fixed in terms of the AdS radius $
\Lambda = -\frac{1}{L^2}\,$. This theory is dual to a CFT$_2$ with central charge
\bea
c=\frac{3L}{2G}\,,
\eea
which is large provided that $G/L$ is small. Moreover, the mass of the scalar field $m$ is related to the conformal dimension of the dual operator $\mathcal{O}_\Delta$, through
\bea
\Delta=1+\sqrt{1+m^2L^2}\,,
\eea
Since we are interested in the case of light operators, $\Delta \ll c$, we require that $m G \ll 1$.

To find the backreacted metric for the state dual to the local quench, following \cite{Nozaki:2013wia}, we start in global coordinates with the vacuum AdS$_3$ solution
\begin{equation}\label{eq:pureAdSmet}
ds^2 = -\left(1+\frac{r^2}{L^2}\right) d\tau^2 + \frac{dr^2}{\left(1+\frac{r^2}{L^2}\right)} + r^2 \, d\theta^2\,.
\end{equation}
The scalar field can be expanded in terms of modes on this background, each mode labeled by two quantum numbers, corresponding to an expansion in either of the two space coordinates $(r,\theta)$. The lowest energy mode is an $S$-wave, as described in \cite{Maldacena:1998bw}. The wavefunction for this mode is spherically symmetric and is annihilated by the isometries $L_1$ and $\bar{L}_{1}$ of AdS$_3$, i.e., $L_1|\psi\rangle=\bar{L}_{1}|\psi\rangle=0$. This mode defines a one-particle excited state for the scalar field on the pure AdS$_3$ background
\begin{equation}\label{eq:1pexstate}
\ket{\psi} \equiv a_{0,0}^{\dagger} \, \ket{0}\,,
\end{equation}
where $a_{0,0}^\dagger$ denotes the creation operator. Solving for the wavefunction of this mode one can show that
\begin{equation}
\label{eq:swavefunc}
\phi_{0,0} =  \frac{1}{\sqrt{2\pi L} \left(1+\frac{r^2}{L^2}\right)^{\frac{\Delta}{2}}}\,.
\end{equation}
 The normal-ordered stress-energy tensor of the scalar field is
\begin{equation}
\mathcal{T}_{\mu \nu} =\,\, : \partial_\mu \phi \, \partial_\nu \phi - \frac{1}{2}g_{\mu\nu} \left( \partial_\rho \phi \, \partial^\rho \, \phi + m^2 \, \phi^2 \right) \!:\,,
\end{equation}
whose one-point function $
\bra{\psi} \mathcal{T}_{\mu \nu} \ket{\psi}\, $ can be evaluated in the one-particle excited state as defined above. Using spherical symmetry, the off-diagonal components of this one-point function can be shown to vanish. The diagonal components can be evaluated using \eqref{eq:swavefunc} and they are given by \cite{Belin:2018juv}
\begin{align}\label{BulkTmn}
\begin{split}
\bra{\psi} \mathcal{T}_{\tau\tau} \ket{\psi} &= \frac{\Delta(\Delta-1)}{\pi L^3 \left(1+\frac{r^2}{L^2}\right)^{\Delta-1}} \,,\\
\bra{\psi} \mathcal{T}_{rr} \ket{\psi} &=  \frac{\Delta}{\pi L^3 \left(1+\frac{r^2}{L^2}\right)^{\Delta+1}} \,,\\
\bra{\psi} \mathcal{T}_{\theta \theta} \ket{\psi} &=  \frac{\Delta \, r^2 \,  \left[1+\frac{r^2}{L^2}(1-\Delta)\right]}{\pi L^3 \left(1+\frac{r^2}{L^2}\right)^{\Delta+1}}\,.
\end{split}
\end{align}
This stress-energy tensor backreacts on the AdS$_3$ vacuum and the backreacted geometry can be obtained by solving semi-classical Einstein's equations
\begin{equation}
 \mathcal{R}_{\mu \nu} - \frac{1}{2} g_{\mu\nu}\, \mathcal{R}+ \Lambda \, g_{\mu \nu}= 8 \pi G \, \bra{\psi} \mathcal{T}_{\mu \nu} \ket{\psi}\,,
\end{equation}
where $\mathcal{R}_{\mu \nu}$ denotes the Ricci tensor.
Since the source $\bra{\psi} \mathcal{T}_{\mu \nu} \ket{\psi}$ is diagonal, one immediately concludes that the backreacted metric has to be diagonal. So one can propose the following ansatz
\begin{equation}
ds^2=-\left(\frac{r^2}{L^2}+F_1(r)^2\right)d\tau^2+\frac{dr^2}{\frac{r^2}{L^2}+F_2(r)^2} +r^2d\theta^2 \,,
\end{equation}
The functions $F_i(r)$ were first determined in \cite{Belin:2018juv}, and at order $\mathcal{O}(G\Delta/L)$ are given by
\begin{align}
\begin{split}
F_1(r) &= 1-\frac{4 G \Delta}{L}+\mathcal{O}\left(\left(G \Delta/L\right)^2\right) \, ,  \\
F_2(r) &= 1-\frac{4 G \Delta}{L} \left[ 1-\left(1+\frac{r^2}{L^2}\right)^{1-\Delta}\right] +\mathcal{O}\left(\left(G \Delta/L\right)^2\right)\,.
\end{split}
\end{align}
Hence, the backreacted metric at this order is
\begin{equation}\label{backMetric}
 ds^2 = - \left(\frac{r^2}{L^2}+ \left(1-\frac{4 G \Delta}{L}\right)^2 \right)\, d\tau^2 + \frac{dr^2}{\frac{r^2}{L^2} + \left(1-\frac{4 G \Delta}{L} + \frac{4 G \Delta}{L}\left(1+\frac{r^2}{L^2}\right)^{1-\Delta} \right)^2}+ r^2 \, d\theta^2\,.
\end{equation}
The stress-energy tensor, although smooth, becomes sharply peaked around the origin as one increases $\Delta$. In the limit of heavy operators $\Delta\gg c$ or, equivalently, when the mass of the scalar field is large $mG\gg1$, the backreacted background can be shown to approximate to that of a conical defect. This was indeed the case considered
in \cite{Nozaki:2013wia}. However, we will work in the opposite regime of light operators and a non-trivial wavefunction for the scalar field.

\subsection{Local quenches via large diffeomorphisms}
\label{coords}

As discussed in the previous Section, our local quenches are defined on a plane $\mathbb{R} \times \mathbb{R}$. We will take this to be the boundary of \textit{Poincar\'e} AdS$_3$. The local quench can then be described by the motion of a localized perturbation, which is localized near the boundary at time $t=0$ and then falls into the interior of AdS. We will use the backreacted global metric in equation \eqref{backMetric} to compute the perturbed Poincare geometry. As discussed in \cite{Nozaki:2013wia}, the idea is to perform a large diffeomorphism on the circle $\mathbb{R} \times \mathbb{S}$ to obtain the plane $\mathbb{R} \times \mathbb{R}$. The diffeomorphism extends naturally inside the AdS spacetime.

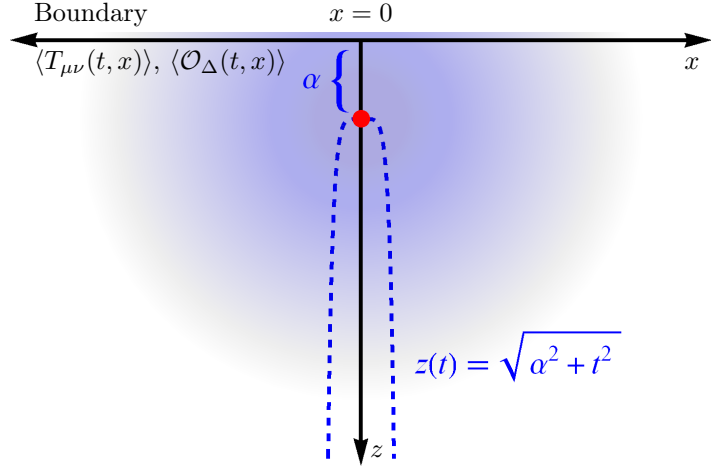
\begin{figure}[t!]
\begin{center}
  \includegraphics[angle=0,width=0.6\textwidth]{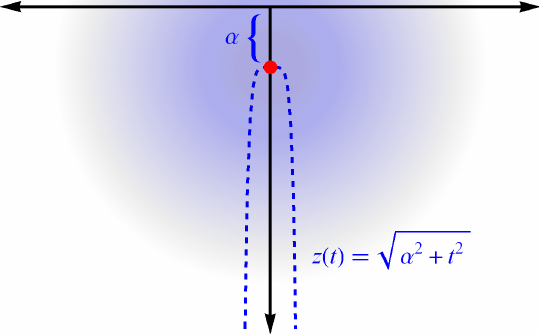}
\begin{picture}(0,0)
\put(-260,150){{\small $\langle T_{\mu\nu}(t,x)\rangle$, $\langle \mathcal{O}_\Delta(t,x)\rangle$}}
\put(-260,168){{\small Boundary}}
\put(-14,150){{\small $x$}}
\put(-148.5,168){{\small $x=0$}}
\put(-133,3){{\small $z$}}
\end{picture}
\end{center}
\vspace{-0.5cm}
\caption{\small Schematic representation of the holographic dual of a local quantum quench. The model consists of a small perturbation that arises by acting locally with an operator $\mathcal{O}_\Delta$ on the vacuum state. The perturbation falls into the AdS interior and excites the metric and other bulk fields. The asymptotic values of the metric and scalar field determine the one point function of the stress-energy tensor $T_{\mu\nu}$ and the scalar operator $\mathcal{O}_\Delta$ in the boundary CFT. For finite $\alpha$, the state at $t=0$ can be prepared by smearing the operator over a region with finite support $\sim\alpha$. This is consistent with the standard notion of UV/IR connection.}
\label{fig:bulkquench}
\end{figure}

The transformation has two parts. First let us recall that AdS can be thought of as an hyperboloid embedded in a higher dimensional Minkowski spacetime with two timelike coordinates. The definition of the global patch and Poincar\'e patch of AdS$_3$ in terms of the $\mathbb{R}^2\times\mathbb{R}^2$ coordinates is the following,
\bea
T 	&=& L\sqrt{1+\frac{r^2}{L^2}}\cos \left(\frac{\tau}{L}\right) 	= \frac{L^2+z^2+x^2-t^2}{2z}\,,\label{GlobPoin1}\\
W 	&=& L\sqrt{1+\frac{r^2}{L^2}}\sin \left(\frac{\tau}{L}\right) 	= \frac{Lt}{z}\,,\\
X &=& r \cos \theta				= \frac{-L^2+z^2+x^2-t^2}{2z}\,,\\
Z 	&=& r \sin\theta				= \frac{L x}{z}\,,\label{GlobPoin4}
\eea
and they satisfy the constraint
\bea
-T^2-W^2+X^2+Z^2 = -L^2\,.
\eea
Moving between the two patches, one finds that a stationary point at $r=0$ in the global patch maps into a non-trivial geodesic in the Poincar\'e patch, with $x=0$ and $z^2-t^2=L^2$. This is problematic, since the particle cannot reach arbitrarily close to the boundary. An easy way to fix this is to consider boost along the $T$ and $X$ directions. This leaves the pure AdS space invariant, but modifies the geodesic to
\bea
\label{trajectory}
x=0\,,\qquad z^2-t^2 = L^2 e^{2\beta} \equiv \alpha^2\,,
\eea
where $\beta$ is a boost parameter. This trajectory captures the desired behavior in the Poincar\'e patch: for small enough $\alpha$ it gets arbitrarily close to the boundary at $t=0$. The backreaction of the perturbation following this geodesic gives us our local quench. For finite $\alpha$, the geodesic does not reach the boundary. This is equivalent to preparing the state at $t=0$ by smearing the operator over a region with finite support of order $\alpha$, consistent with the standard notion of UV/IR connection \cite{Peet:1998wn,Hatta:2010dz,Agon:2014rda}. See Figure \ref{fig:bulkquench} for graphical representation.

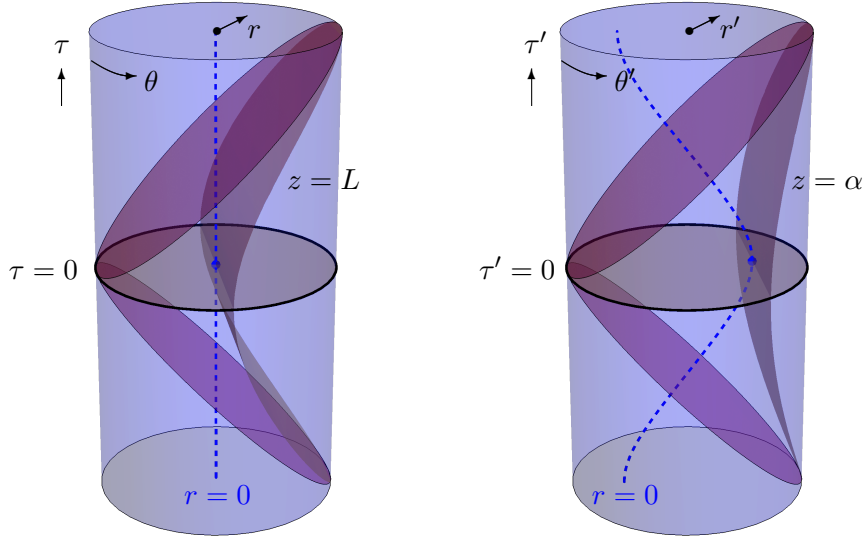
\begin{figure}[t!]
\begin{center}
  \includegraphics[width=4cm]{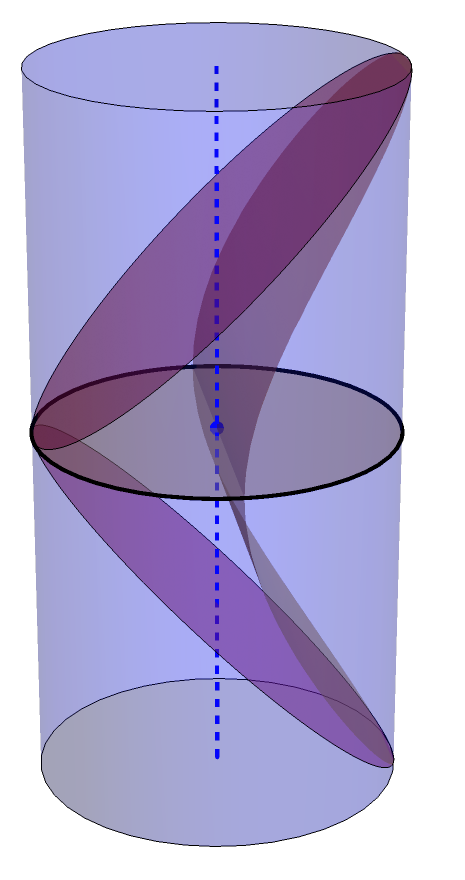}
  \hspace*{2cm}
  \includegraphics[width=4cm]{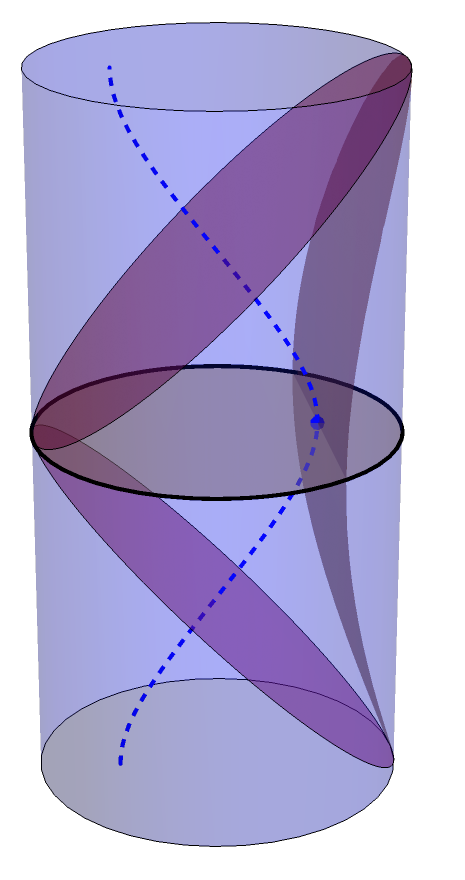}
  \setlength{\unitlength}{1cm}
\begin{picture}(0,0)
\put(-10.7,6.0){\vector(0,1){0.5}}
\put(-10.8,6.7){$\tau$}
\put(-11.4,3.7){$\tau=0$}
\put(-7.7,4.9){$z=L$}
\put(-9.8,6.4){\vector(1,0){0.1}}
\put(-9.6,6.2){$\theta$}
\qbezier(-10.3,6.6)(-10.0,6.4)(-9.8,6.4)
\put(-8.63,7.0){\vector(2,1){0.4}}
\put(-8.23,6.9){$r$}
\put(-8.63,7.0){\circle*{0.1}}
\put(-9.07,0.7){{\color{blue}$r=0$}}
\put(-4.45,6.0){\vector(0,1){0.5}}
\put(-4.55,6.7){$\tau'$}
\put(-3.55,6.4){\vector(1,0){0.1}}
\put(-3.35,6.2){$\theta'$}
\qbezier(-4.05,6.6)(-3.75,6.4)(-3.55,6.4)
\put(-5.15,3.7){$\tau'=0$}
\put(-1,4.9){$z=\alpha$}
\put(-2.35,7.0){\vector(2,1){0.4}}
\put(-1.95,6.9){$r'$}
\put(-2.35,7.0){\circle*{0.1}}
\put(-3.65,0.7){{\color{blue}$r=0$}}
\end{picture}
\end{center}
\vspace{-0.2cm}
\caption{\small Two global coordinate systems and their Poincar\'e patches. In the left figure we have plotted the original global frame, represented with coordinates $(\tau,r,\theta)$, where the particle
lies at the origin $r=0$. Specializing to a Poincar\'e patch maps this geodesic to $z^2=L^2+t^2$, so the minimum approach to the boundary is $z=L$ at $t=0$.
This surface is depicted in brown. In the right figure we have plotted the boosted global frame $(\tau',r',\theta')$ where the particle oscillates according to (\ref{eq:osscil}). Specializing to a Poincar\'e patch of this boosted frame maps the particle's trajectory to $z^2=\alpha^2+t^2$, with $\alpha\equiv L\, e^{\beta}$, so it can now get arbitrarily close to the boundary. The minimum approach is now $z=\alpha$ at $t=0$. This surface is also depicted in brown in the corresponding cylinder.
}
\label{fig:globalfigs}
\end{figure}

As an intermediate step, and for future reference, we give here the explicit transformation between the original global frame $(\tau,r,\theta)$ and a boosted global frame
$(\tau',r',\theta')$:
\bea
T 	&=& L\sqrt{1+\frac{r^2}{L^2}}\cos \left(\frac{\tau}{L}\right) 	= L\sqrt{1+\frac{r'^2}{L^2}}\cos \left(\frac{\tau'}{L}\right)\cosh \beta-r' \cos \theta'\sinh \beta\,,\\
W 	&=& L\sqrt{1+\frac{r^2}{L^2}}\sin \left(\frac{\tau}{L}\right) 	= L\sqrt{1+\frac{r'^2}{L^2}}\sin \left(\frac{\tau'}{L}\right)\,,\\
X &=& r \cos \theta				= r' \cos \theta' \cosh \beta - L\sqrt{1+\frac{r'^2}{L^2}}\cos \left(\frac{\tau'}{L}\right) \sinh \beta	\,,\\
Z 	&=& r \sin\theta				= r' \sin\theta'\,.
\eea
In this boosted frame, the $r=0$ geodesic maps to
\be\label{eq:osscil}
r'^2=\frac{4 L^2 \sinh ^2\beta  \cos ^2\left(\tau'/L\right)}{3+\cosh (2 \beta )-2 \cos \left(2 \tau'/L\right)\sinh ^2(\beta ) }\,,\qquad \sin\theta'=0\,.
\ee
which is periodic in $\tau'$. The geometry in this frame is dual to a CFT state with perpetual collective oscillations, of the kind studied in \cite{Freivogel:2011xc}. The final transformation consists of specializing to a Poincar\'e patch of the boosted global frame. The form of this transformation is the same as given in (\ref{GlobPoin1})-(\ref{GlobPoin4}) but with $(\tau,r,\theta)$ replaced by $(\tau',r',\theta')$. In Figure \ref{fig:globalfigs} we show pictorially the effects of these transformations.

We can also obtain the explicit combined transformation from the original global frame $(\tau,r,\theta)$ to the latter Poincar\'e patch. This map is is given by:
\bea
\label{eq:betaTransformation}
T 	&=& L\sqrt{1+\frac{r^2}{L^2}}\cos \left(\frac{\tau}{L}\right) 	= \frac{L^2e^{\beta}+e^{-\beta}(z^2+x^2-t^2)}{2z}\,,\\
W 	&=& L\sqrt{1+\frac{r^2}{L^2}}\sin \left(\frac{\tau}{L}\right) 	= \frac{Lt}{z}\,,\\
X &=& r \cos \theta				= \frac{-L^2e^{\beta}+e^{-\beta}(z^2+x^2-t^2)}{2z}\,,\\
Z 	&=& r \sin\theta				= \frac{L x}{z}\,.
\eea
Equivalently, inverting these relations, and writing them explicitly in terms of $\alpha$ we get:
\begin{align}
\label{eq:cotrans}
 \tau  &= L\arctan\left(\frac{2 \alpha t}{\alpha^2+x^2+z^2-t^2}\right) \, , \\
r &= \frac{L}{2 \alpha z}\sqrt{\alpha^4+2\alpha^2 \left(x^2-z^2+t^2\right)+\left(x^2+z^2-t^2\right)^2} \, , \\
\theta  &= \arcsin\left(\frac{2 \alpha x}{\sqrt{\alpha^4+2\alpha^2\left(x^2-z^2+t^2\right)+\left(x^2+z^2-t^2\right)^2}}\right) \, .\label{eq:cotrans3}
\end{align}
The full metric after the combined coordinate transformation is straightforward to obtain but is very lengthy and not particularly illuminating. Hence, we will not transcribe it here.

We want to calculate all $\mathcal{O}(G^0)$ corrections to the holographic entanglement entropy due to the presence of scalar field in the bulk. As we will see in section \ref{sec:HEE}, they include a geometric correction, due to the backreaction of the quantum fields on the geometry and a quantum correction due to the bulk entanglement entropy of the quantum fields. Altogether, these two corrections will give all leading $\mathcal{O}(c^0)$ corrections to the local quench in the CFT. One important thing to note is that the bulk entanglement entropy piece is sensitive to the scalar field profile in the bulk. For finite $\Delta$, this is not a problem, since $\phi(t,x,z)$ is smooth everywhere. However, as $\Delta$ increases the profile becomes sharply peaked around the origin, rendering this problem largely degenerate. In fact, the requirement $\Delta \ll c$ is precisely what will enable us to carry out such a calculation without problem.

\subsection{One-point function of local operators}

Before proceeding to the computation of the entanglement entropy, we study here other observables of interest, i.e., the one-point function of local operators dual to light bulk fields. In our theory (\ref{eq:actionscalar}) we have two of these fields: the metric $g_{\mu\nu}$ and the scalar field $\phi$. In Fefferman-Graham (FG) coordinates, we can write the following near-boundary expansions:
\bea\label{eq:FGgauge}
ds^2 = \frac{L^2}{z^2}\left[ dz^2 + (\eta_{\mu\nu} + z^2 \tau_{\mu\nu}+\cdots)dx^{\mu}dx^{\nu} \right],
\eea
\bea\label{eq:FGgaugeS}
\phi=z^{2-\Delta}\phi_{d-\Delta}+z^\Delta\phi_\Delta+\cdots\,.
\eea
In terms of these expansions, and given the normalization of the action (\ref{eq:actionscalar}), the one-point function of the CFT stress-energy tensor and scalar operator are given by \cite{deHaro:2000vlm}:
\bea\label{eq:stressenergy}
\langle T_{\mu\nu} \rangle = \frac{L}{8\pi G} \tau_{\mu\nu}\,,
\eea
\bea\label{eq:scalarop}
\langle \mathcal{O}_\Delta \rangle = 2(\Delta-1)\phi_\Delta\,.
\eea
However, a brief comment is in order. Since we are working in the semiclassical approximation, $\phi$ must be treated as an operator. Hence, in order to use the above formulas, we first need to find the expectation value of $\phi$ in the quantum state $|\psi\rangle=a^\dagger_{0,0}|0\rangle$. This amounts to compute a 3-point function in the bulk, which should vanish for a free theory. Indeed, a quick calculation shows that
\be
\langle\psi|\phi|\psi\rangle=0\,.
\ee
Therefore, it immediately follows that
\be
\langle \mathcal{O}_\Delta \rangle=0\,.
\ee

\begin{figure}[t!]
\begin{center}
  \includegraphics[angle=0,width=0.45\textwidth]{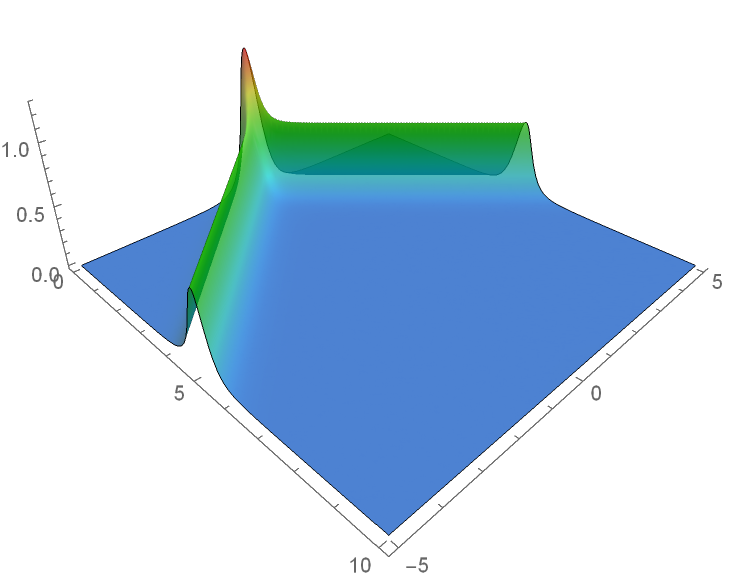}\includegraphics[angle=0,width=0.45\textwidth]{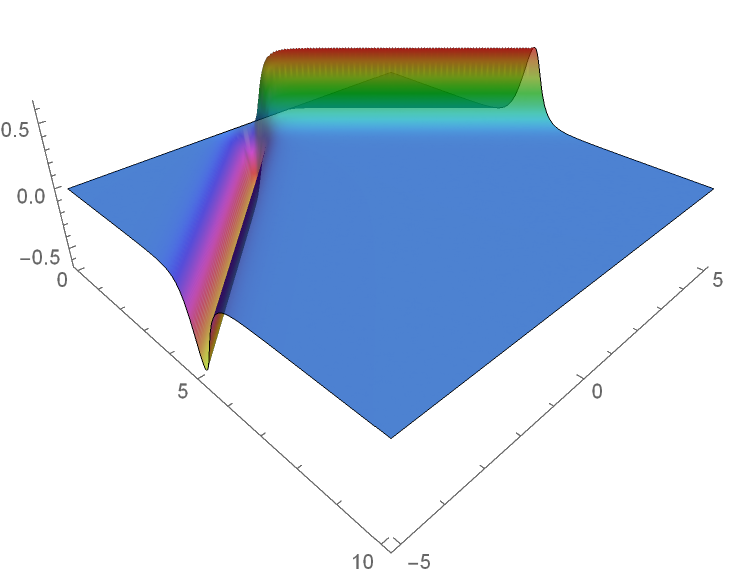}
\begin{picture}(0,0)
\put(-165,45){{\small $t$}}
\put(-36,45){{\small $x$}}
\put(-365,45){{\small $t$}}
\put(-236,45){{\small $x$}}
\put(-391,126){{\small $\tilde{\mathcal{E}}$}}
\put(-192,126){{\small $\tilde{\mathcal{P}}$}}
\end{picture}
\end{center}
\vspace{-0.5cm}
\caption{\small Profiles of the normalized energy density $\tilde{\mathcal{E}}=\mathcal{E}/E$ and momentum density $\tilde{\mathcal{P}}=\mathcal{P}/E$, with $E=\Delta/\alpha$, as a function of $t$ and $x$ for a local quench with $\alpha=1/2$. The parameter $\alpha$ measures the initial spread of the excitations around $x=0$, which is estimated to be of order $O( \alpha)$. The initial perturbation generates two shocks, moving to the left and to the right respectively, that move at the speed of light. The heights of these shocks remain constant due to energy conservation in the CFT.}\label{fig:stress}
\end{figure}
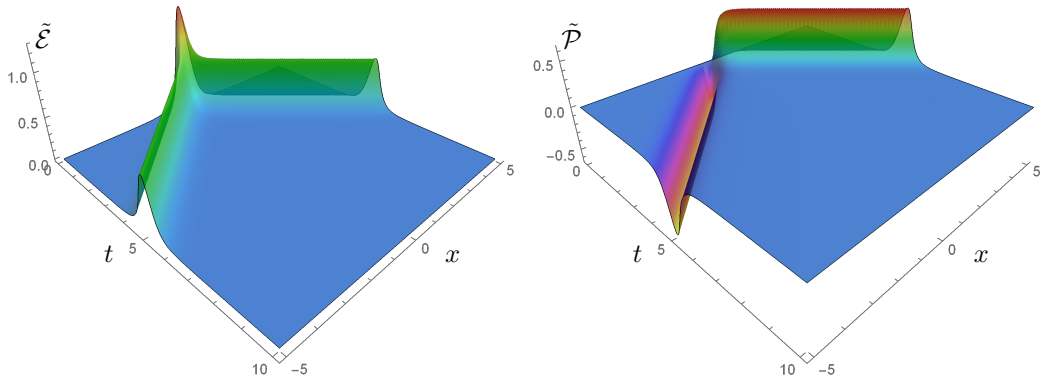

Next, for the computation of  the stress-energy tensor of the boundary CFT we must write the bulk metric in the Fefferman-Graham gauge (\ref{eq:FGgauge}). In order to do so we need a second coordinate transformation, which can be obtained perturbatively as
\bea
\label{FGtrans}
z \to z' = z[ 1 + z^2 f(t,x) + \cdots ]\,,
\eea
The function $f(t,x)$ which satisfies the conditions of the FG gauge is found to be
\bea
f(t,z) = -\frac{8G\Delta}{L}\frac{\alpha^2}{\alpha^4+2\alpha^2(t^2+x^2)+(t^2-x^2)^2}
\eea
With this change of coordinates, we can make use of (\ref{eq:stressenergy}) to obtain the stress-energy tensor in the CFT:
\bea\label{eq:Tmunu}
\langle T_{\mu\nu}\rangle=\frac{2 \alpha ^2 \Delta }{\pi}\left(
                                                                                                            \begin{array}{cc}
                                                                                                              \displaystyle\frac{(t^2 + x^2 + \alpha^2)^2 + 4 t^2 x^2}{ \left[(x^2-t^2-\alpha^2)^2+4 \alpha ^2 x^2\right]^2} & \displaystyle\frac{-4 t x (t^2 + x^2 + \alpha^2)}{ \left[(x^2-t^2-\alpha^2)^2+4 \alpha ^2 x^2\right]^2} \\
                                                                                                              \displaystyle\frac{-4 t x (t^2 + x^2 +\alpha^2)}{ \left[(x^2-t^2-\alpha^2)^2+4 \alpha ^2 x^2\right]^2} & \displaystyle\frac{(t^2 + x^2 + \alpha^2)^2 + 4 t^2 x^2}{ \left[(x^2-t^2-\alpha^2)^2+4 \alpha ^2 x^2\right]^2} \\
                                                                                                            \end{array}
                                                                                                          \right)\,.
\eea
From these expressions one can obtain quantities of interest, such as the energy density $\mathcal{E}=\langle T^{tt}\rangle$ and momentum density $\mathcal{P}=\langle T^{tx}\rangle$, which by symmetry equals the energy flux $\langle T^{xt}\rangle$. The pressure in this case also equals the energy density, since $\langle T^{xx}\rangle=\langle T^{tt}\rangle$, as expected for a conformal theory in 2-dimensions. We now make a couple of comments. First notice that both the traceless condition and stress-energy conservation are satisfied,
\bea
\langle T^{\mu}_{\,\,\,\mu} \rangle=0\,,\qquad\nabla_\mu \langle T^{\mu\nu}\rangle=0\,.
\eea
Second, the total energy is constant, and in agreement with the expectation for a one-particle state, for an insertion of a primary of dimension $\Delta$
\bea
\label{energy}
E= \int dx \,\mathcal{E} =  \frac{\Delta}{\alpha}= \frac{\Delta}{L }e^{-\beta}\,.
\eea
The extra term $e^{-\beta}$ here accounts for the boost factor. The total momentum vanishes, $p=0$, because the excitations generated by the quench move both to the left and to the right. In order to better understand this time dependence, we plot in figure \ref{fig:stress} the two non-trivial components of the stress-energy tensor, namely the energy density $\mathcal{E}$ and the momentum density $\mathcal{P}$. As expected, we observe profiles that are peaked on the light-cone $-t^2+x^2=0$, which can be understood as shock waves that move at the speed of light \cite{Horowitz:1999gf} due to the initial excitation at $t=x=0$.

In the CFT, these shock waves arise from the action of a primary operator $\mathcal{O}_\Delta$ on the vacuum \cite{Nozaki:2014hna,He:2014mwa,Nozaki:2014uaa,Caputa:2014vaa}, as discussed near equation \eqref{eq:normstate}. In fact, we have deliberately identified the parameter $\alpha$ in the bulk trajectory \eqref{trajectory} with the UV regulator in the quenched state (\ref{eq:normstate}). For finite values of $\alpha$, the parameter gives the smearing of the operator $\cO_\Delta$ around $x=0$. Only in the limit $\alpha \to 0$, the bulk trajectory touches the boundary and we get an exactly local quench. In this limits, the energy \eqref{energy} post-quench diverges, and so will energy density as seen in Figure \ref{fig:stress}. Analytically, the density in this limit looks like
\bea\label{energydelta}
\lim_{\alpha\to0}\mathcal{E}=\frac{E}{2}\left(\delta(t+x)+\delta(t-x)\right)\,.
\eea
The equal constant $E/2$ upfront implies that the heights of the two peaks remain constant, a consequence of energy conservation in the CFT\footnote{In higher dimensions, one would expect a spherical shock with amplitude proportional to $\mathcal{E}\sim \delta(t-r)/r^{d-2}$.}.

Our results for the stress-energy tensor are in agreement with the results of \cite{Nozaki:2013wia}, even though they worked in the regime of heavy operators. This is because the value of the conformal dimension $\Delta$ only appears as a multiplicative factor in the stress-energy tensor. In the bulk, this is manifested in the fact that the backreacted metric of the one-particle state looks identical to that of a conical defect for an observer at infinity \cite{Belin:2018juv}.

\section{Holographic entanglement entropy\label{sec:HEE}}

On the gravity side, order $1/N$ corrections to entanglement entropy can be computed using the FLM prescription \cite{Faulkner:2013ana,Engelhardt:2014gca}, which states that
\begin{equation}
\label{FLM}
S_A =\underset{\gamma_A^{\text{ext}}}{\text{min}}\, \underset{\gamma_A}{\text{ext}} \,  \[\frac{\text{Area}(\gamma_A)}{4 G}\]+S_{\text{bulk}}(\Sigma_A) +\cdots\,.
\end{equation}
In this formula, $\gamma_A$ is a codimension-2 bulk surface anchored at the boundary, with $\partial \gamma_A =\partial A$, and $S_{\text{bulk}}$ is the entanglement entropy of bulk fields across the extremal surface $\gamma_A^\text{ext}$, in a Cauchy slice $\Sigma\supset A\cup\gamma_A$. For perturbative excited states over the vacuum, $\Sigma$ can be taken to be a constant-$t$ slice (or a boosted version of it, for intervals in generic time-slices).

The leading order term in (\ref{FLM}) can be calculated from the standard RT/HRT prescription \cite{Ryu:2006bv,Hubeny:2007xt}, using only the extremal area in the unperturbed geometry. At order $\mathcal{O}(1)$ there are two contributions: one due to the correction to the area term in the backreacted geometry, and another due to the entanglement entropy of bulk fields in the given quantum state. The latter contribution can be computed in the unperturbed geometry because the bulk entanglement in the perturbed geometry would be further suppressed in $1/N$ and would only appear at higher order in the expansion. The dots in the above equation represent such higher order contributions and could be computed in the framework of quantum extremal surfaces \cite{Engelhardt:2014gca}.

In this section, we will compute these $\mathcal{O}(1)$ contributions to entanglement entropy in the perturbed geometry constructed in section \ref{sec:gravity}. We will study them separately in sections \ref{sec:geomC} and \ref{sec:quanC} respectively.

\subsection{Geometric corrections to entanglement entropy\label{sec:geomC}}

Here we will discuss some $\CO(1)$ corrections to the holographic entanglement entropy of a single interval in an excited state following a local quench. These corrections come purely because the bulk metric that corresponds to the excited state is different from the one that corresponds to the pure state in the CFT. Hence we denote these corrections as ``geometric corrections''. We relegate the discussion of the $\CO(1)$ corrections due to bulk entanglement entropy to section \ref{sec:quanC}.

Consider an arbitrary perturbation over pure AdS due to matter fields, such that the metric takes the form\footnote{The true expansion parameter should be dimensionless, and can depend on the particular type of perturbation. In our case it is given by $G\Delta/L$.}
\begin{equation}\label{eq:metexpansion}
g_{\mu\nu}=g_{\mu\nu}^{(0)}+g_{\mu\nu}^{(1)}+\mathcal{O}(G^2)\,,
\end{equation}
The corrections to entanglement entropy due to the change in the geometry can be computed from the area term in (\ref{FLM}). The change in geometry (\ref{eq:metexpansion}) induces a linear variation in area as follows
\begin{equation}
\text{Area}(\gamma_A)=\text{Area}^{(0)}(\gamma_A)+\delta\text{Area}(\gamma_A)\,.
\end{equation}
The leading order correction in the metric is of order $\mathcal{O}(G)$ so it is clear that the $\delta\text{Area}$ term contributes at order $\mathcal{O}(1)$ to the entanglement entropy. The leading term and first order variation of the area are found to be
\begin{equation}\label{Area0}
\text{Area}^{(0)}(\gamma_A)=\int d^{d-1}\xi\, \sqrt{h^{(0)}}\,,
\end{equation}
and
\begin{equation}\label{deltaArea}
\delta\text{Area}(\gamma_A)=\frac{1}{2}\int d^{d-1}\xi\, \sqrt{h^{(0)}}\,\text{Tr}[h^{(1)}(h^{(0)})^{-1}]\,,
\end{equation}
respectively, where
\begin{equation}
h_{\alpha\beta}=h_{\alpha\beta}^{(0)}+h_{\alpha\beta}^{(1)}+\mathcal{O}(G^2)
\end{equation}
is the induced metric on the extremal surface, with
\begin{equation}
h_{\alpha\beta}^{(0)}=\frac{\partial X^\mu}{\partial\xi^\alpha}\frac{\partial X^\nu}{\partial \xi^\beta}g_{\mu\nu}^{(0)}\,, \qquad h_{\alpha\beta}^{(1)}=\frac{\partial X^\mu}{\partial\xi^\alpha}\frac{\partial X^\nu}{\partial \xi^\beta}g_{\mu\nu}^{(1)}\,,
\end{equation}
and $\xi^\alpha$ are coordinates parametrizing the surface. An important point here is that, at this order in the perturbation,
the embedding functions $X^\mu(\xi)$ can be taken to be the same as in empty AdS. This means that we do not need to know the precise shape of $\gamma_A$ in the perturbed geometry to evaluate (\ref{deltaArea}). This simple but useful observation can be nicely illustrated by making use of the variational principle \cite{Kundu:2016cgh,Lokhande:2017jik}, and holds true regardless of the expansion parameter.

In the presence of a 1-particle excited state of a light scalar field \eqref{eq:1pexstate} in AdS$_3$, the backreacted geometry in global coordinates is given by \eqref{backMetric}.\footnote{When the scalar field is heavy, the backreaction is instead given by a conical defect geometry. The geometric corrections to entanglement entropy in this case were calculated in \cite{Nozaki:2013wia}. We will focus here on the case where the scalar field has a small mass, or equivalently, is dual to a light operator in the CFT, i.e. $\Delta\ll c$.}
We can expand this metric as in (\ref{eq:metexpansion}), to obtain
\begin{equation}
g_{\mu\nu}^{(0)}=\left(
                   \begin{array}{ccc}
                     -\left(1+\frac{r^2}{L^2}\right) & 0 & 0 \\
                     0 & \frac{1}{\left(1+\frac{r^2}{L^2}\right)} & 0 \\
                     0 & 0 & r^2 \\
                   \end{array}
                 \right)\,,\qquad g_{\mu\nu}^{(1)}=\frac{8G\Delta}{L}\left(
                   \begin{array}{ccc}
                     1 & 0 & 0 \\
                     0 & \frac{1-\left(1+\frac{r^2}{L^2}\right)^{1-\Delta}}{\left(1+\frac{r^2}{L^2}\right)^2} & 0 \\
                     0 & 0 & 0 \\
                   \end{array}
                 \right)\,.
\end{equation}
Next, we need to implement the bulk diffeomorphism (\ref{eq:cotrans})-(\ref{eq:cotrans3}) to arrive to a Poincar\'e wedge where (\ref{backMetric}) is viewed as a local quench. The resulting metric is lengthy so we will not transcribe it here. Next, we would like to compute (\ref{Area0}) and (\ref{deltaArea}) in this new frame. However, since areas are invariant under coordinate transformations, we can work directly in the original global coordinate system but taking care of properly transforming the embedding functions $X^\mu(\xi)$. In the Poincar\'e patch of AdS$_3$, the metric at order zero is given by
\begin{equation}
  ds^2=\frac{L^2}{z^2}\left(-dt^2+dx^2+dz^2\right)\,.
\end{equation}
In these coordinates, the extremal surfaces that we are interested in are given by
\be
t=\text{constant}\,,\qquad (x-x_c)^2+z^2=R^2\,.
\ee
These are semicircles at a constant-$t$ slice, with radius $R$ and centered at $x=x_c$. The two endpoints of these geodesics are $x_{\pm}=x_c\pm R$, so they naturally span boundary intervals of length
\begin{equation}
\ell=2 R\,.
 \end{equation}
Importantly, note that the local quench considered here is due to the insertion of a primary operator at $x=0$, whereas the interval of the entangling region is centered at $x=x_c$. The special case $x_c=0$ naturally has more symmetry than the generic case with $x_c\neq0$. In the following we will specialize to the more symmetric case with $x_c=0$, since the calculations will be simpler, but at a later stage we will study the most general case.

\paragraph{Centered Intervals:\label{symintervals}} For intervals centered at the origin ($x_c=0$) we choose to parametrize the geodesic in terms of $\xi=z\in(0,R)$, i.e., with $X^{\mu}=\{t(z),x(z),z\}$, where
\begin{equation}\label{eq:embedPoinc}
t(z)=t=\text{constant}\,, \qquad x(z)=\pm\sqrt{R^2-z^2}\,.
\end{equation}
Now we use the transformations (\ref{eq:cotrans})-(\ref{eq:cotrans3}) to obtain the embeddings in the global coordinate system. In terms of the parameter $\xi=z$, $X^{\mu}=\{\tau(z),r(z),\theta(z)\}$, where
\begin{align}
\tau(z)  &= L\arctan\left(\frac{2 \alpha t}{\alpha^2+x(z)^2+z^2-t^2}\right) \, , \label{eq:emb1}\\
r(z) &= \frac{L}{2 \alpha z}\sqrt{\alpha^4+2\alpha^2 \left(x(z)^2-z^2+t^2\right)+\left(x(z)^2+z^2-t^2\right)^2} \, , \\
\theta(z)  &= \arcsin\left(\frac{2 \alpha x(z)}{\sqrt{\alpha^4+2\alpha^2\left(x(z)^2-z^2+t^2\right)+\left(x(z)^2+z^2-t^2\right)^2}}\right) \label{eq:emb3}\, ,
\end{align}
and $x(z)$ is given in (\ref{eq:embedPoinc}).

We can compute the leading term and first variation of entanglement entropy by evaluating the embedding functions (\ref{eq:emb1})-(\ref{eq:emb3}) in (\ref{Area0}) and (\ref{deltaArea}), respectively, and using the RT/HRT formula. Since we are considering the $x_c=0$ case, the symmetry of the problem allows us to take one branch of $x(z)$, say the positive one, and multiply the resulting integrals by a factor of two. After some manipulations we arrive at
\begin{equation}
S^{(0)}_A=\frac{L R}{2G} \int_{\epsilon}^{R} \!\!\frac{dz}{z \sqrt{R^2-z^2}}\,,
\end{equation}
and
\begin{equation}
\delta S^\text{geom}_A\big|_{x_c=0}=\frac{2\Delta}{R} \int_{0}^{R} dz\,\frac{z\sqrt{R^2-z^2}}{a(t)^2-z^2}\left[1-\left(\frac{z}{a(t)}\right)^{2(\Delta-1)}\right]\,,
\end{equation}
where we have defined the function
\begin{equation}
\label{def:aoft}
a(t)\equiv\sqrt{R^2+\frac{(\alpha^2+t^2-R^2)^2}{4\alpha^2}}\geq R\,.
\end{equation}
The first integral gives rise to the standard result for the entanglement entropy in the vacuum of a 2D CFT,
\begin{equation}\label{vacummEE0}
S^{(0)}_A=\frac{L}{2G}\log\left(\frac{2R}{\epsilon}\right)=\frac{c}{3}\log\left(\frac{\ell}{\epsilon}\right)\,.
\end{equation}
The second integral is a bit more involved, but can be explicitly performed to obtain
\begin{equation}\label{eq:finalcentered0}
\delta S^\text{geom}_A\big|_{x_c=0}=2\Delta\left[1-\arcsin\big(\tfrac{R}{a(t)}\big)\sqrt{\tfrac{a(t)^2}{R^2}-1}\right]
-\frac{\Gamma(\tfrac{3}{2})\Gamma(\Delta+1)R^{2\Delta}}{\Gamma(\Delta+\tfrac{3}{2})a(t)^{2\Delta}}\,\!_2F_1\left[1,\Delta,\Delta+\tfrac{3}{2},\tfrac{R^2}{a(t)^2}\right].
\end{equation}
This expression can be massaged into a more familiar form, by defining a function $\vartheta(t)$ according to
\begin{equation}
\frac{R}{a(t)}=\sin\left(\frac{\vartheta(t)}{2}\right)\,,
\end{equation}
or equivalently,
\begin{equation}\label{eq:vartheta}
\vartheta(t)=2\arcsin\left(\frac{2\alpha R}{\sqrt{(\alpha^2+t^2-R^2)^2+4\alpha^2R^2}}\right)\,.
\end{equation}
With this definition, and using the following hypergeometric identity
\bea
{}_2F_1(a,b,c,z)=(1-z)^{-b} {}_2F_1\left[c-a,b,c,-\tfrac{z}{1-z}\right]\,,
\eea
we arrive at
\begin{align}
\begin{split}
\label{eq:finalcentered}
\delta S^\text{geom}_A\big|_{x_c=0} &= \Delta\left[2-\vartheta(t)\cot\big(\tfrac{\vartheta(t)}{2}\big)\right] \\
&\quad -\frac{\Gamma(\tfrac{3}{2})\Gamma(\Delta+1)\tan^{2\Delta}\!\big(\frac{\vartheta(t)}{2}\big)}{\Gamma(\Delta+\tfrac{3}{2})}
 \,\!_2F_1\left[\Delta+\tfrac{1}{2} ,\Delta ,\Delta+\tfrac{3}{2},-\tan^2\big(\tfrac{\vartheta(t)}{2}\big) \right].
 \end{split}
\end{align}
Upon identifying $\vartheta(t)\leftrightarrow\theta$, this formula coincides exactly with the result of \cite{Belin:2018juv} for the geometric corrections to the entanglement entropy of a 1-particle excited state of a light scalar field in global AdS. Notice that this is expected, since the length of the geodesic is invariant under general coordinate transformations. Indeed, in the Poincar\'e coordinates, a constant time-slice interval with endpoints at $x=\pm\ell/2$ is mapped through (\ref{eq:cotrans})-(\ref{eq:cotrans3}) to an interval in the global coordinates at a constant-$\tau$ slice with opening angle $\delta\theta=\vartheta(t)$ given by (\ref{eq:vartheta}). Reference \cite{Belin:2018juv} considered an interval with opening angle $\delta \theta$, so the agreement of the two results is not suprising. On the other hand, non-centered intervals in Poincar\'e coordinates (with $x_c\neq0$) map to intervals in global coordinates that are tilted in the time direction (with $\delta \tau\neq0$). These were not considered originally in \cite{Belin:2018juv}.

Before moving to the most general case of non-centered intervals, let us briefly analyze our final result of geometric correction for centered intervals. First, notice that when $\Delta$ is an integer (\ref{eq:finalcentered}) takes a much simpler form. For example, for a marginal operator dual to a massless scalar field ($\Delta=2$), we have
\begin{equation}\label{deltaS2:cen}
\delta S_A^\text{geom}\big|_{x_c=0,\,\Delta=2}=\frac{16\alpha^2R^2}{3}\frac{1}{(\alpha^2+t^2-R^2)^2+4\alpha^2R^2}\,.
\end{equation}
For other integer values of $\Delta$, the hypergeometric function simplifies to a rational function as well, but the final expressions are longer as we increase the value of $\Delta$.

Second, note that the first part of (\ref{eq:finalcentered}) matches exactly with the universal geometric term (\ref{deltaS+-}) obtained in section \ref{sec:CFTEE}. It turns out that this piece can be easily extracted from the modular Hamiltonian in the CFT, as we will show in section \ref{sec:1stlaw}.

Lastly, for future reference, we discuss the explicit expansion of the final result (\ref{eq:finalcentered}) for small intervals. In this limit we get two distinct contributions
\be\label{smallRexpsC}
\delta S^\text{geom}_A\big|_{x_c=0}=\frac{2\Delta}{3}\left(\frac{2\alpha R}{t^2+\alpha^2}\right)^{2}\left[1+\mathcal{O}(R^2)\right]
-\frac{\Gamma(\tfrac{3}{2})\Gamma(\Delta+1)}{\Gamma(\Delta+\frac{3}{2})}\left(\frac{2\alpha R}{t^2+\alpha^2}\right)^{2\Delta}\left[1+\mathcal{O}(R^2)\right]\,.
\ee
The first series arises from expanding the first part of (\ref{eq:finalcentered}) while the second one from the hypergeometric function.
The second series might seem puzzling at this point. Recall that the full CFT result at order $\mathcal{O}(1)$, obtained in section \ref{sec:CFTEE}, does not contain terms proportional to $R^{2\Delta+2i}$ with $i\in\mathbb{N}$. Later in section \ref{sec:quanC} we will see that the second series precisely cancels out with the $\mathcal{O}(1)$ corrections coming from the bulk modular Hamiltonian. This remarkable matching is in fact expected and follows directly from the exact relation between the CFT and bulk modular Hamiltonians. We will come back to this point in section \ref{sec:quanC}.

\begin{figure}[t!]
\begin{center}
\includegraphics[scale=0.6]{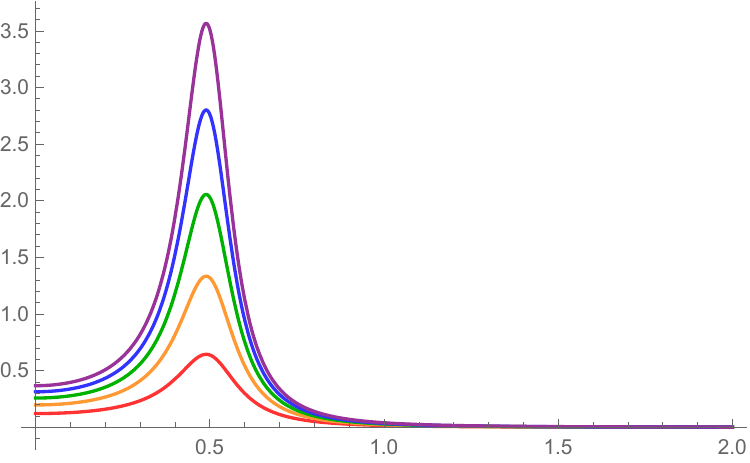}$\quad$\includegraphics[scale=0.6]{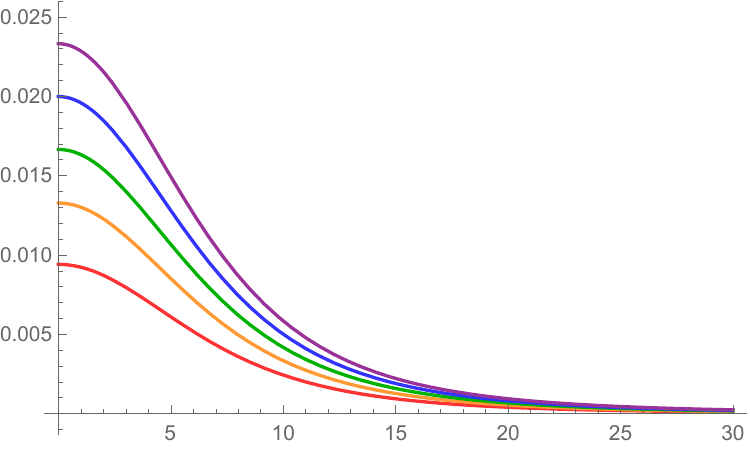}
\begin{picture}(0,0)
\put(-204,138){{\small $\delta S_A^{\text{geom}}$}}
\put(-18,32){{\small $t/\ell$}}
\put(30,138){{\small $\delta S_A^{\text{geom}}$}}
\put(211,32){{\small $t/\ell$}}
\end{picture}
\vspace{-0.7cm}
\caption{Time evolution of the geometric correction to entanglement entropy $\delta S_A$ for centered intervals, with $\Delta=3/2,2,5/2,3,7/2$ corresponding to the red, orange, green, blue and purple curves, respectively. In the left panel we fixed $\alpha/\ell=10^{-1}$, while in the right panel we chose $\alpha/\ell=10$.
}
\label{fig:traceDis2}
\end{center}
\end{figure}
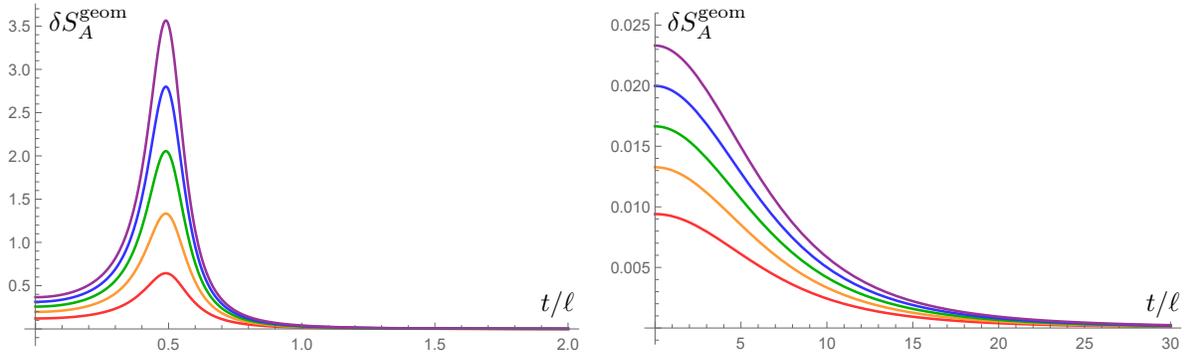
In Figure \ref{fig:traceDis2} we show various plots of the geometric correction \eqref{eq:finalcentered} for centered intervals in different physical scenarios. In the left panel, we plot entanglement entropy for several values of $\Delta$, in a situation where $\alpha/\ell \ll 1$, so the quench is localized and the perturbation is sharply peaked. The entanglement itself peaks at $t=R=\ell/2$, which shows that the excitations created by the quench carry entanglement at the speed of light $v_E=1$. This behavior resembles the seminal result for evolution of entanglement entropy after global quenches in 2d CFTs \cite{Calabrese:2005in}, which can be interpreted in terms of EPR pairs. In holographic CFTs, the entanglement still propagates at the speed of light, however, the quasi-particle picture is replaced by the \emph{entanglement tsunami} interpretation \cite{AbajoArrastia:2010yt,Albash:2010mv,Liu:2013iza,Liu:2013qca}. The difference between global quenches and our setup is that the energy injected in a local quench eventually disperses to infinity, so the entanglement entropy drops to zero as the state relaxes to the vacuum state. This sharp transition has a clear interpretation from the bulk perspective: it happens exactly at the moment that the bulk particle crosses the entangling surface $\gamma_A$. The entanglement entropy peaks at this moment because the particle has largest relative backreaction on the area of the entangling surface. At late times though, the particle falls into the deep infrared, eventually escaping the Poincar\'e patch at $t\to\infty$.

In the right panel we show the opposite situation, where $\alpha/\ell \gg 1$. In this case the quench is smoothed over a region much bigger than the entangling interval and the bulk particle is always outside and far from the entanglement wedge. The evolution of entanglement entropy in this case is monotonically decreasing in time, since the particle is always moving away from the surface $\gamma_A$, hence reducing its effect over time. Remarkably, its full behavior strongly resembles the late-time evolution of entanglement entropy in an expanding boost-invariant plasma, studied in \cite{Pedraza:2014moa,DiNunno:2017obv}, specially for large $\Delta$. This can be explained by the fact that, at leading order in hydrodynamics, the bulk dual to the latter system is given precisely by a boosted black brane that moves into the radial direction $z$. Indeed, in the limit $\ell/\alpha \ll 1$, the backreacted metric near $\gamma_A$ (close to the boundary) enjoys also an approximate planar symmetry. As a consequence of these two facts, the entanglement entropy obeys the `first law'-like relation $\delta S_A=\delta E_A/T_A$ \cite{Bhattacharya:2012mi,Allahbakhshi:2013rda} even though the state is time-dependent \cite{Lokhande:2017jik}. Moreover, such a relation can be easily derived from the first-law of entanglement $\delta S_A=\delta \langle H_A\rangle$ provided that the time-evolution is sufficiently slow and the region is small. We will discuss this point more in detail in section \ref{sec:1stlaw}.

\paragraph{Non-centered Intervals:\label{symintervals}} For generic intervals  with center $x_c\neq0$, we parametrize the RT surface using a variable $\xi\in(0,1)$ such that $X^{\mu}=\{t(\xi),x(\xi),z(\xi)\}$, where
\begin{equation}
\label{eq:embPoincB}
t(\xi)=t=\text{constant}\,, \qquad x(\xi)=(x_c-R)+2R\xi\,,\qquad z(\xi)=2R\sqrt{(1-\xi)\xi}\,.
\end{equation}
As one varies $\xi$ from 0 to 1, $x(\xi)$ varies linearly between $x_{-}=x_c-R$ and $x_{+}=x_c+R$. We again use the transformations (\ref{eq:cotrans})-(\ref{eq:cotrans3}) to obtain the embeddings in the global coordinates, in terms of this parameter. It is $X^{\mu}=\{\tau(\xi),r(\xi),\theta(\xi)\}$ and takes the form
\begin{align}
\tau(\xi)  &= L\arctan\left(\frac{2 \alpha t}{\alpha^2+x(\xi)^2+z(\xi)^2-t^2}\right) \, , \label{eq:emb1B}\\
r(\xi) &= \frac{L}{2 \alpha z}\sqrt{\alpha^4+2\alpha^2 \left(x(\xi)^2-z(\xi)^2+t^2\right)+\left(x(\xi)^2+z(\xi)^2-t^2\right)^2} \, , \\
\theta(\xi)  &= \arcsin\left(\frac{2 \alpha x(\xi)}{\sqrt{\alpha^4+2\alpha^2\left(x(\xi)^2-z(\xi)^2+t^2\right)+\left(x(\xi)^2+z(\xi)^2-t^2\right)^2}}\right) \label{eq:emb3B}\, ,
\end{align}
and $x(\xi)$ and $z(\xi)$ are given in (\ref{eq:embPoincB}).

Next, we compute the leading term and first variation of entanglement entropy by evaluating the embedding functions (\ref{eq:emb1B})-(\ref{eq:emb3B}) in (\ref{Area0}) and (\ref{deltaArea}), respectively, and using the RT/HRT formula. The leading order term yields the same result as for the centered interval (\ref{vacummEE0}), i.e., the value of entanglement entropy in the vacuum, as expected\footnote{To arrive to this expression we have used the relation between the $x-$ and $z-$cutoffs, $\delta$ and $\epsilon$, which can be obtained by expanding the embedding (\ref{eq:embPoincB}) near the boundary: $
\delta=\epsilon^2/(2R)^2$.}
\begin{equation}
S^{(0)}_A=\frac{L}{8G} \int_{\delta}^{1-\delta} \!\!\!\!\!\!\frac{d\xi}{\xi(1-\xi)}=\frac{L}{2G}\log\left(\frac{2R}{\epsilon}\right)=\frac{c}{3}\log\left(\frac{\ell}{\epsilon}\right)\,.
\end{equation}
The first variation of entanglement entropy yields an integral of the form
\be\label{intSgeneral}
\delta S^\text{geom}_A=\frac{\Delta}{L}\int_0^1 \frac{N(\xi)}{D(\xi)}d\xi\,,
\ee
where we have defined
\be
\begin{split}
\label{NandD}
&N(\xi)\equiv \tau'(\xi)^2+\left[\frac{1-\left(1+ r(\xi)^2/L^2\right)^{1-\Delta}}{\left(1+r(\xi)^2/L^2 \right)^2}\right]r'(\xi)^2\,,\\
&D(\xi)\equiv \sqrt{-\left(1+\tfrac{r(\xi)^2}{L^2}\right)\tau'(\xi)^2+\frac{r'(\xi)^2}{\left(1+r(\xi)^2/L^2 \right)}+r(\xi)^2 \, \theta'(\xi)^2}\,.
\end{split}
\ee
The full integrand as a function of $\xi$ is lengthy so we will not transcribe it here.

The next step is to perform the integral. This can be done analytically for integer values of $\Delta$ but the final expressions are cumbersome and not very enlightening. To give a flavor for the kind of expressions one obtains, the explicit result for an operator dual to a massless scalar field ($\Delta=2$) is given by
\begin{align}
\label{eq:noncfinal}
\begin{split}
\delta S_A^\text{geom} \big|_{\Delta=2} &=\mathcal{F}(t,R,x_c,\alpha)+\mathcal{G}(t,R,x_c,\alpha)\,\times\\
&\qquad \left[\text{arccot}\left(\frac{2 \alpha  t}{\alpha ^2+(x_c+R)^2-t^2}\right)-\text{arccot}\left(\frac{2 \alpha  t}{\alpha^2+(x_c-R)^2-t^2}\right)\right],
\end{split}
\end{align}

where $\mathcal{F}(t,R,x_c,\alpha)$ and $\mathcal{G}(t,R,x_c,\alpha)$ are rational functions given by
\be
\mathcal{F}(t,R,x_c,\alpha) =-\frac{[(t+R)^2+\alpha^2][(t-R)^2+\alpha^2]}{8t^2x_c^2}+\frac{R^2+9t^2-\alpha^2}{4t^2}-\frac{x_c^2}{8t^2}\,,
\ee
\be
\begin{split}
&\mathcal{G}(t,R,x_c,\alpha)=\frac{[R^4+2R^2(\alpha^2-t^2)+(\alpha^2+t^2)^2]^2}{64R t^3\alpha x_c^3}-\frac{R^6+R^4(3t^2+\alpha^2)}{16R t^3\alpha x_c}\\
&\quad\,\,+\frac{R^2(9t^4+2t^2\alpha^2+\alpha^4)-(t^2+\alpha^2)^2(4t^2-\alpha^2)}{16R t^3\alpha x_c}+\frac{R^4x_c}{32R t^3\alpha} +\frac{x_c^5}{64R t^3 \alpha}  \\
&\quad\,\,+\frac{[2(R^2+9t^2)(R^2-\alpha^2)+19t^4+3\alpha^4]x_c}{32R t^3\alpha} -\frac{(R^2+5t^2-\alpha^2)x_c^3}{16R t^3\alpha} \, \,  \, ,
\end{split}
\ee
respectively. This expression looks singular in the limit of centered intervals $x_c\to0$. However, although each individual piece has terms that diverge in this limit, they cancel out amongst each other upon resummation. Therefore, expanding for small\footnote{Here we mean $x_c$  small with respect to all other scales, i.e., $x_c/t\ll 1$, $x_c/R \ll 1$, and $R/\alpha\ll 1$, with \emph{all} other dimensionless ratios arbitrary.} $x_c$ we obtain
\be
\delta S_A^\text{geom} \big|_{\Delta=2}=\delta S^\text{geom}_A \big|_{x_c=0,\,\Delta=2}+\mathcal{O}(x_c^2)+\cdots\,,
\ee
where $\delta S^\text{geom}_A|_{x_c=0,\,\Delta=2}$ is given in (\ref{deltaS2:cen}). Similarly, the geometric correction for the non-centred intervals in the limits $t\to0$ and $R\to0$ is given by
\begin{align}
\begin{split}
\delta S_A^\text{geom} \big|_{\Delta=2} &= \frac{16\alpha^2R^2}{3}\frac{1}{[(x_c+R)^2+\alpha^2][(x_c-R)^2+\alpha^2]}+\mathcal{O}(t^2)+\cdots\,,
\end{split} \\
\begin{split}
\label{eq:smallL}
\delta S_A^\text{geom} \big|_{\Delta=2} &= \frac{16\alpha ^2 R^2}{3}\frac{ (t^2+x_c^2+\alpha ^2)^2+4 t^2 x_c^2}{ [(t^2-x_c^2+\alpha ^2)^2+4 \alpha ^2 x_c^2]^2}+\mathcal{O}(R^4)+\cdots\,,
\end{split}
\end{align}
respectively, which can be easily verified to be well-behaved.

Now for arbitrary $\Delta$, the geometric correction for the non-centered intervals can be obtained by expanding the integrand of (\ref{intSgeneral}) in powers of $R$ and performing the individual integrals. This gives
\begin{align}
\begin{split}
\label{smallRexpsNC}
\delta S^\text{geom}_A  &= \frac{8 \Delta \alpha ^2 R^2}{3} \frac{(t^2+x_c^2+\alpha ^2)^2+4 t^2 x_c^2}{[(t^2-x_c^2+\alpha ^2)^2+4 \alpha ^2 x_c^2]^2}\Big[1+\sum_{i=1}^{\infty}\mathcal{P}_i(t,x_c,\alpha)R^{2i}\Big] \\
&\quad -\frac{\Gamma(\tfrac{3}{2}) \Gamma (\Delta +1)}{\Gamma(\Delta +\tfrac{3}{2})}\frac{ (2 \alpha R)^{2 \Delta }}{ [(t^2-x_c^2+\alpha ^2)^2+4 \alpha ^2 x_c^2]^{\Delta }}\Big[1+\sum_{i=1}^{\infty}\mathcal{Q}_i(t,x_c,\alpha)R^{2i}\Big]\,,
\end{split}
\end{align}
which agrees with leading term of the $\Delta=2$ result (\ref{eq:smallL}) and generalizes the expansion found for centered intervals (\ref{smallRexpsC}) to non-centered intervals. As we will show later in section \ref{sec:1stlaw}, the complete series in the first term above agrees with the result obtained from the first law of entanglement entropy \textit{in the CFT}, $\delta S_A=\delta \langle H_A\rangle$. Further, the first term of that series captures the `first law'-like relation $\delta S_A=\delta E_A/T_A$, valid only for small intervals. Lastly, resumming the series in the first term gives rise to the universal term (\ref{deltaS+-}) obtained from the CFT calculation in section \ref{sec:CFTEE}.

The series in the second term in \eqref{smallRexpsNC} turns out to cancel out with the result coming from the first law of entanglement entropy \textit{in the bulk}, $\delta S_A=\delta \langle H_{\text{bulk}}\rangle$, as will be shown in section \ref{sec:quanC}. This explains the absence of terms proportional to $R^{2\Delta+2i}$ $(i\in\mathbb{N})$, in the final result for the entanglement entropy at order $\mathcal{O}(1)$, as is the case for the CFT calculation of section \ref{sec:CFTEE}.

Finally, it is worth noticing that the two series in (\ref{smallRexpsNC}) can
be identified with specific contributions coming from the integral in (\ref{intSgeneral}). More specifically, we can split the numerator of (\ref{intSgeneral}) such that
\be
\delta S^\text{geom}_A = \frac{\Delta}{L}\int_0^1 \frac{N_\mathcal{P}(\xi)}{D(\xi)}d\xi+\frac{\Delta}{L}\int_0^1 \frac{N_\mathcal{Q}(\xi)}{D(\xi)}d\xi\,,
\ee
where
\be\label{def:NPQ}
N_\mathcal{P}(\xi)\equiv \tau'(\xi)^2+\frac{r'(\xi)^2}{\left(1+\tfrac{r(\xi)^2}{L^2}\right)^2}\,,\qquad N_\mathcal{Q}(\xi)\equiv -\frac{r'(\xi)^2}{\left(1+\tfrac{r(\xi)^2}{L^2}\right)^{\Delta+1}}\,,
\ee
and $D(\xi)$ is given in \eqref{NandD}. From these definitions, together with the explicit form of the embedding functions (\ref{eq:emb1B})-(\ref{eq:emb3B}), we see that these terms have expansions
\be\label{eq:contP}
\delta S_A^\mathcal{P}\equiv\frac{\Delta}{L}\int_0^1 \frac{N_\mathcal{P}(\xi)}{D(\xi)}d\xi=\frac{8 \Delta \alpha ^2 R^2}{3} \frac{(t^2+x_c^2+\alpha ^2)^2+4 t^2 x_c^2}{[(t^2-x_c^2+\alpha ^2)^2+4 \alpha ^2 x_c^2]^2}\Big[1+\sum_{i=1}^{\infty}\mathcal{P}_i(t,x_c,\alpha)R^{2i}\Big],
\ee
\be\label{eq:contQ}
\delta S_A^\mathcal{Q}\equiv\frac{\Delta}{L}\int_0^1 \frac{N_\mathcal{Q}(\xi)}{D(\xi)}d\xi=\frac{-\Gamma(\tfrac{3}{2}) \Gamma (\Delta +1)}{\Gamma(\Delta +\tfrac{3}{2})}\frac{ (2 \alpha R)^{2 \Delta }}{ [(t^2-x_c^2+\alpha ^2)^2+4 \alpha ^2 x_c^2]^{\Delta }}\Big[1+\sum_{i=1}^{\infty}\mathcal{Q}_i(t,x_c,\alpha)R^{2i}\Big].
\ee
This separation is motivated by how the two terms depend on $\Delta$. While both sides of (\ref{eq:contP}) depend only linearly in $\Delta$, (\ref{eq:contQ}) involves a nontrivial power of $\Delta$ that arises from the denomiator of $N_{\mathcal{Q}}$. This non-trivial dependence is also manifested in the small interval approximation which gives terms like $ R^{2\Delta+2 i}$. As we will see, the quantities $ \delta S_A^\mathcal{P} $ and $\delta S_A^\mathcal{Q}$ will prove useful in sections \ref{sec:1stlaw} and \ref{sec:quanCmH} respectively.

\begin{figure}[t!]
\begin{center}
\includegraphics[scale=0.6]{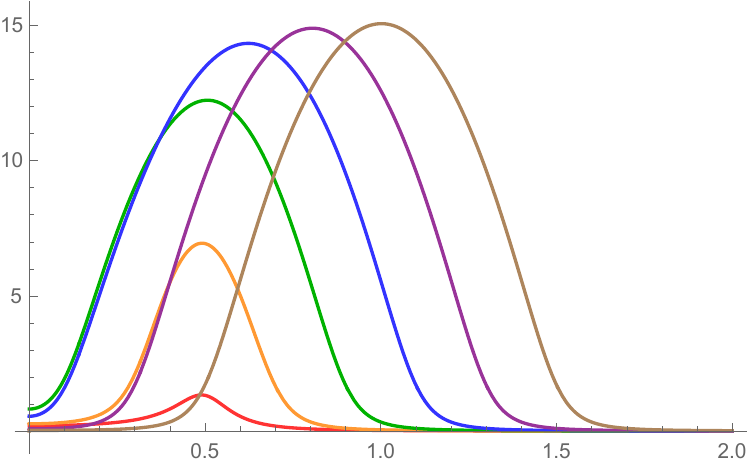}$\quad$\includegraphics[scale=0.6]{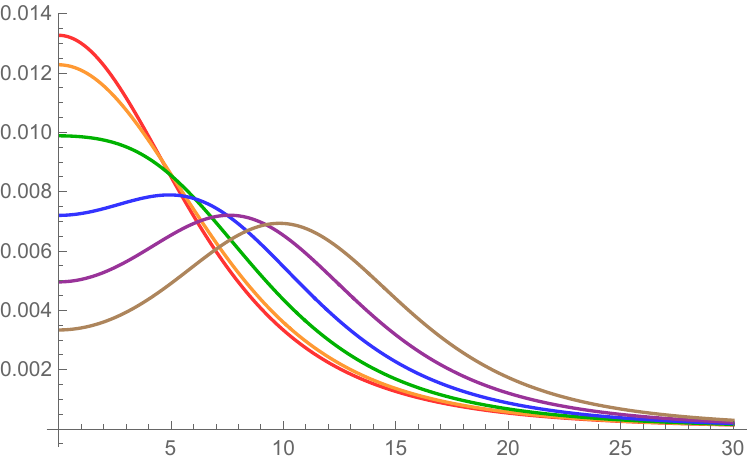}
\begin{picture}(0,0)
\put(-206,138){{\small $\delta S_A^\text{geom}$}}
\put(-18,32){{\small $t/\ell$}}
\put(32,138){{\small $\delta S_A^\text{geom}$}}
\put(211,32){{\small $t/\ell$}}
\end{picture}
\vspace{-0.7cm}
\caption{Time evolution of the geometric correction to entanglement entropy $\delta S_A$ for non-centered intervals. All plots correspond to a quench by an operator with $\Delta=2$. In the left panel we fixed $\alpha/\ell=10^{-1}$ and pick $x_c/\ell=0,1/5,2/5,3/5,4/5,1$ corresponding to the red, orange, green, blue, purple and brown curves, respectively. In the right panel we fixed $\alpha/\ell=10$ and pick $x_c/\ell=0,2,4,6,8,10$ corresponding to the red, orange, green, blue, purple and brown curves, respectively.
}
\label{fig:noncenter}
\end{center}
\end{figure}
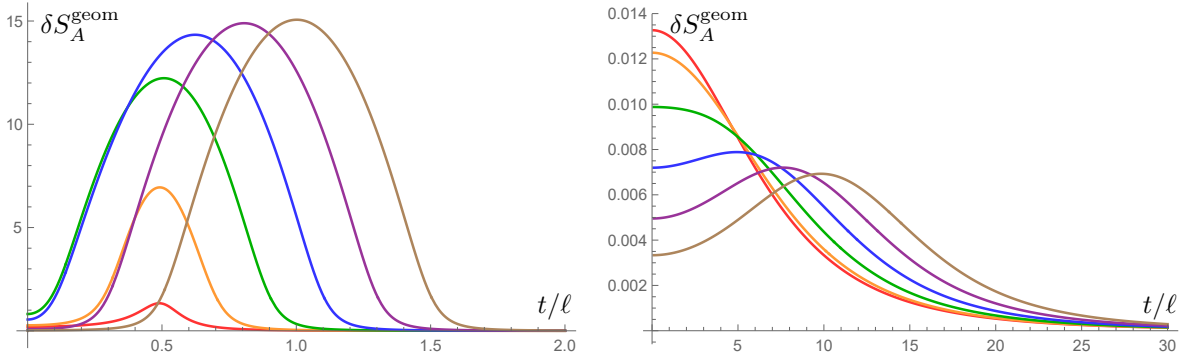
Lastly, we study the time evolution of the geometric correction for the non-centered intervals. Figure \ref{fig:noncenter} shows plots for different physical scenarios. In the left panel, we consider the situation $\alpha/\ell \ll 1$, so the quench is localized and the perturbation is sharply peaked around $x=0$. We see two distinct behaviors: for $0\leq x_c/\ell \lesssim \tfrac{1}{2}$ the peak of the entanglement entropy is located at $t\sim R$, but its value increases with $x_c/\ell$. In this situation the excitations are created within $A$, and they take on average $t\sim R$ to exit region. However, as $x_c/\ell$ is increased, the perturbations are created closer and closer to the boundary of $A$, contributing more to the short-range entanglement across $\partial A$. For $\tfrac{1}{2}\lesssim x_c/\ell$, the peak happens at $t\sim x_c$ while its value slightly increases as $x_c/R$ is increased. The quench in this case is localized around $x=0$, which happens to be outside $A$, and the excitations take on average $t\sim x_c$ to localize within the entangling region. We can attribute the latter behavior to the fact that $\alpha/\ell$ is in fact finite, so for $x_c/\ell$ just above $\tfrac{1}{2}$, a small portion of the quench profile is still supported within $A$. All these results are consistent with entanglement propagation at the speed of light $v_E=1$, be it via quasi-particles \cite{Calabrese:2005in} or wave packets  described in section \ref{sec:realtime}.

In the right panel, we show plots for the case $\alpha/\ell \gg 1$, so the quench is supported over a region $\alpha$ much bigger than the entangling interval. From the bulk perspective, the behavior follows because the particle is always falling in the deep IR, far from the entanglement wedge. We observe two distinct behaviors: for $0\leq x_c/\alpha\lesssim \tfrac{1}{2}$ the peak is still located at $t=0$, while its value decreases as $x_c/\ell$ is increased. In this range of $x_c/\alpha$, the $A$ lies in the region where the quench is supported, however as $x_c/\ell$ increases, less and less energy is initially contained within $A$. For $\tfrac{1}{2}\lesssim x_c/\alpha$, most of the quench energy is initially supported outside $A$ so the peak moves and is now located at $t\sim x_c$, while slowly decreasing in amplitude. Similar to the case of centered intervals, the late-time evolution of the geometric correction for the non-centered intervals resemble that in an expanding boost-invariant plasma \cite{Pedraza:2014moa,DiNunno:2017obv}. In particular, the non-monotonic behavior as we vary $x_c$ is reminiscent of the non-monotonic behavior of the correlators with respect to the rapidity \cite{Pedraza:2014moa,DiNunno:2017obv}. The comments made in section \ref{sec:1stlaw} about emergent planar symmetry and `first law'-like behavior also apply here.

\subsubsection{First law of entanglement and CFT modular Hamiltonian\label{sec:1stlaw}}

For small perturbations of a reference state, $\rho=\rho^{(0)}+ \delta\rho$, entanglement entropy satisfies a first law relation,
\be\label{eq:1stlaw}
\delta S_A=\delta\langle H_A\rangle\,,
\ee
where $H_A$ is the so-called modular Hamiltonian. To see this, recall that by definition this operator is related to the reduced density matrix $\rho_A=\text{tr}_{A^c}[\rho]$ through
\be
\rho_A=\frac{e^{-H_A}}{\text{tr}[e^{-H_A}]}\,.
\ee
The small perturbation of the full state translates generically into a small perturbation of the reduced density matrix $\rho_A=\rho_A^{(0)}+\delta\rho_A$. Hence, to linear order in this perturbation, the variation of entanglement entropy $S_A=-\text{tr}[-\rho_A\log \rho_A]$ is given by
\begin{eqnarray}
\delta S_A &=& -\text{tr}\left[ \delta\rho_A \log \rho_A \right]-\text{tr} \left[\rho_A\, \rho_A^{-1}\delta\rho_A \right]\,,\nonumber\\
\label{derfirst}
&=& \text{tr}\left[\delta\rho_A\, H_A \right]-\text{tr} \left[ \delta\rho_A \right]\,.
\end{eqnarray}
The last term in \eqref{derfirst} is identically zero, since the trace of the reduced density matrix is one by definition. Hence, the leading order variation of the entanglement entropy is given by (\ref{eq:1stlaw}), as advertised. However, there are very few cases for which $H_A$ is known explicitly. The most famous example is the case where $A$ is half-space, say $x_1>0$, and $\rho$ corresponds to the vacuum state. In this case \cite{Bisognano:1975ih,Unruh:1976db}
\begin{equation}\label{modular1}
H_A=2\pi\int_A x_1 \, T_{00}(t,\vec{x}) \, d^{d-1}x\,.
\end{equation}
For generic CFTs, this setup can be conformally mapped to the case where
$A$ is a ball of radius $R$, centered at $x_c$, in which case \cite{Hislop:1981uh,Casini:2011kv}
\begin{equation}\label{modular2}
  H_A=2\pi\int_A  \frac{R^2-(\vec{x}-\vec{x}_c)^2}{2R} T_{00}(t,\vec{x})\,d^{d-1}x\,.
\end{equation}

Now, local quenches are \emph{not} perturbatively close to the vacuum. Although the energy injected is small,  some of the eigenvalues of the full density matrix $\rho$ will drastically differ from those of the vacuum $\rho^{(0)}$ due to the sharply peaked perturbation. However, in some limiting cases, the reduced density matrix of a subsystem may still satisfy $\rho_A=\rho_A^{(0)}+\delta\rho_A$. This happens when there is an additional small parameter to carry out the expansion, e.g., when the size of the subsystem $A$ is small in comparison to other scales of the state. If this is true, then the expectation value of the energy density operator would be approximately constant inside region $A$, $\langle T_{00}(t,\vec{x})\rangle|_{A}\simeq\mathcal{E}(t)$, so (\ref{eq:1stlaw}) becomes
\begin{equation}
\delta S_A=2\pi\mathcal{E}(t)\hspace{0.1em}\Omega_{d-2}\int_0^R\frac{R^2-r^2}{2R}r^{d-2}dr=\frac{2\pi\mathcal{E}(t)\hspace{0.1em}\Omega_{d-2}R^d}{d^2-1}\,.
\end{equation}
Here, $\Omega_{d-2}=2\pi^{\frac{d-1}{2}}/\Gamma[\frac{d-1}{2}]$ is the surface area of a $(d-2)$-dimensional unit sphere. Defining $\delta E_A$ as the energy inside region $A$,
\begin{equation}
\delta E_A=\mathcal{E}(t)\hspace{0.1em} V_A\,,\qquad V_A\equiv\frac{\Omega_{d-2}}{d-1}R^{d-1}\,,
\end{equation}
where $V_A$ is the volume of region $A$, it follows that
\begin{equation}\label{firstlawexc}
\delta S_A=\frac{\delta E_A}{T_A}\,,\qquad T_A\equiv\frac{d+1}{2\pi R}\,,
\end{equation}
where $T_A$ is known as the entanglement temperature. This `first law'-like relation holds true for arbitrary static states provided that $\mathcal{E}\hspace{0.1em} R^d\ll1$ \cite{Bhattacharya:2012mi,Allahbakhshi:2013rda}. For time-dependent states, equation (\ref{firstlawexc}) is still expected to be valid if additional conditions are satisfied: in addition to $\mathcal{E}\hspace{0.1em} R^d\ll1$, the size of the region must be smaller than all characteristic \emph{time scales} of the state, i.e. $\dot{\mathcal{E}}(t)R^{d+1}\ll1$, $\ddot{\mathcal{E}}(t)R^{d+2}\ll1$ and so on \cite{Lokhande:2017jik}. In the following, we will specialize to the case of 2d CFTs and take $R=\ell/2$ as the half-length of the interval.

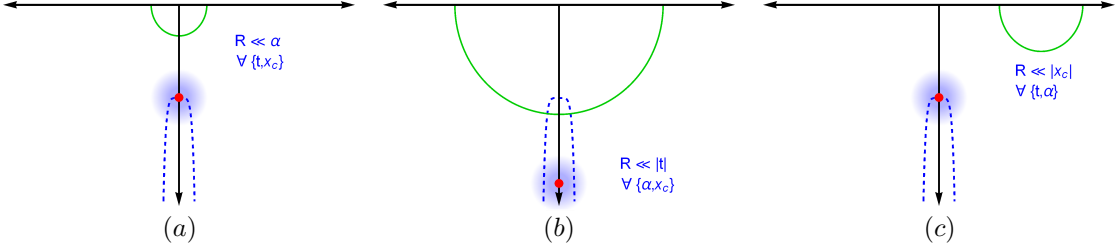
\begin{figure}[t!]
\begin{center}
  \includegraphics[angle=0,width=0.3\textwidth]{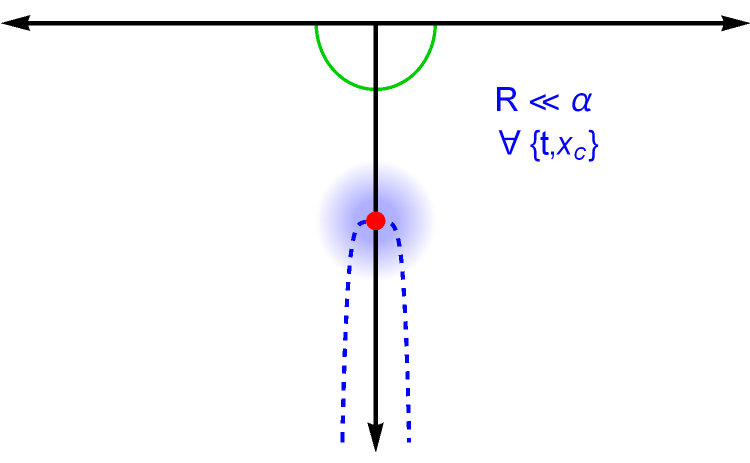}$\quad$\includegraphics[angle=0,width=0.3\textwidth]{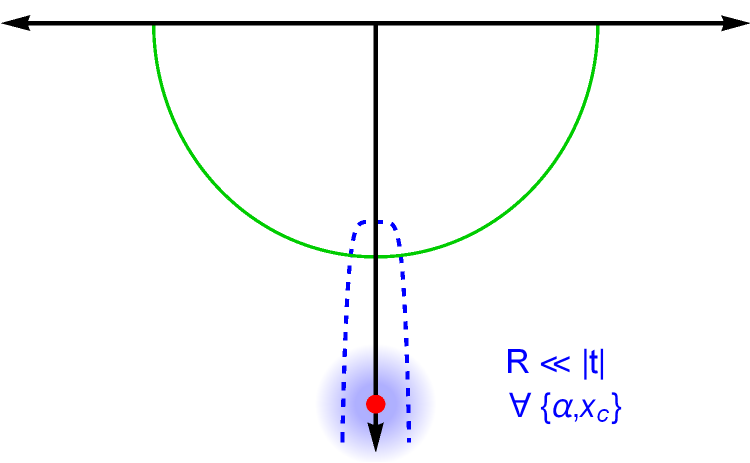}$\quad$\includegraphics[angle=0,width=0.3\textwidth]{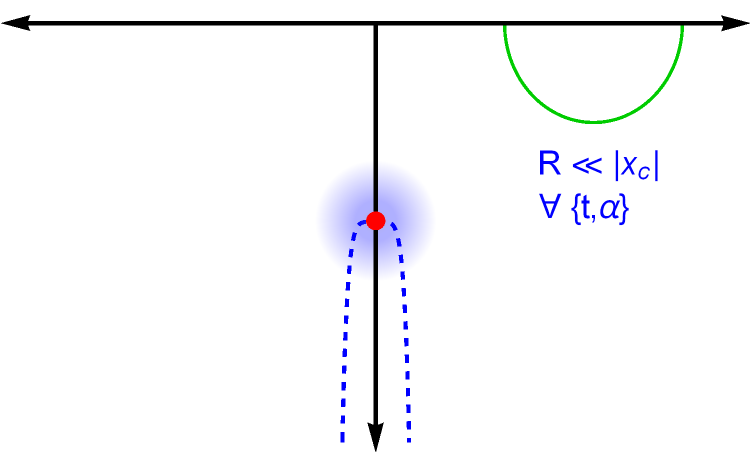}\\
\begin{picture}(0,0)
\put(-150,5){{\small $(a)$}}
\put(-6,5){{\small $(b)$}}
\put(138,5){{\small $(c)$}}
\end{picture}
\end{center}
\vspace{-0.3cm}
\caption{\small Bulk illustration of the cases where the `first law'-like relation (\ref{firstlawexc}) is expected to hold: $(a)$ $R\ll\alpha$ and $\forall\,\{t,x_c\}$, $(b)$ $R\ll t$ and $\forall\,\{\alpha,x_c\}$ or $(c)$ $R\ll x_c$ and $\forall\,\{t,\alpha\}$. For all of them, one can check that $\langle T_{00}(t,\vec{x})\rangle|_{A}\simeq\mathcal{E}(t)$ is indeed a good approximation so (\ref{firstlawexc}) follows directly from (\ref{eq:1stlaw}).}
\label{fig:bulkfirstlaw}
\end{figure}

In our setup of a local quench in 2D CFT, we can distinguish between three scenarios: $(a)$ $R\ll\alpha$ and $\forall\,\{t,x_c\}$, $(b)$ $R\ll t$ and $\forall\,\{\alpha,x_c\}$ or $(c)$ $R\ll x_c$ and $\forall\,\{t,\alpha\}$. These three cases are depicted in Figure \ref{fig:bulkfirstlaw}. The common feature among them is that the bulk region where the backrection is large is parametrically far away from the entanglement wedge of region $A$. Since the gravitational potential of a point particle decays with the distance, the small parameter in each case guarantees that the metric around the entanglement surface (i.e., near the boundary) is $i)$ perturbatively small in such parameter and $ii)$ approximately spatially homogeneous. The former condition implies that $\mathcal{E}\hspace{0.1em} R^2\ll1$, while the latter one ensures that $\langle T_{00}(t,x)\rangle|_{A}\simeq\mathcal{E}(t)$, as required for the derivation of (\ref{firstlawexc}). Now, for the state that we are considering, all the time variations are smooth provided that $\alpha$ is finite. It is only in the strict limit $\alpha\to0$ that $\mathcal{E}(t)$ approaches the sum of two delta functions (\ref{energydelta}). However, in this limit the two conditions above are not satisfied so (\ref{firstlawexc}) is not expected to hold anyway. For any finite value of $\alpha$, we can chose a sufficiently small $R$ such that the extra conditions are satisfied: $\dot{\mathcal{E}}(t)R^{3}\ll1$, $\ddot{\mathcal{E}}(t)R^{4}\ll1$, and so on, so that we can expect (\ref{firstlawexc}) to be valid in a corner of the space of parameters.

Let us check in detail the above claims. At this point it will be useful to recall the analytic expressions for the stress-energy tensor in the CFT (\ref{eq:Tmunu}) and entanglement entropy for centered and non-centered intervals, (\ref{eq:finalcentered}) and (\ref{eq:noncfinal}), respectively. As explained above, either if we are in situation $(a)$, $(b)$ or $(c)$, we can assume that the energy density $\left\langle T_{00}(t,x)\right\rangle$ is approximately constant in the region $A$, namely for $x\in(x_c-R,x_c+R)$. For simplicity, we will evaluate it at the center of the interval,
  \be
  \left\langle T_{00}(t,x)\right\rangle\simeq\left\langle T_{00}(t,x_c)\right\rangle=\mathcal{E}(t)=
\frac{2 \alpha ^2 \Delta }{\pi}\frac{(\alpha^2 + t^2+x_c^2)^2 + 4 t^2 x_c^2}{ \left[(\alpha^2+t^2-x_c^2)^2+4 \alpha ^2 x_c^2\right]^2}\,.
  \ee
Performing the integral of the kernel over the region $A$, the first law (\ref{eq:1stlaw}) then implies that
\be\label{eq:1stLresult}
\delta S_A=\frac{8\Delta\alpha ^2 R^2}{3}\frac{ (\alpha ^2+t^2+x_c^2)^2+4 t^2 x_c^2}{ [(\alpha ^2+t^2-x_c^2)^2+4 \alpha ^2 x_c^2]^2}=\frac{\delta E_A}{T_A}\,,\qquad R^2\ll1/\mathcal{E}(t)\,.
\ee
where
\be
\delta E_A=V_A\hspace{0.1em} \mathcal{E}(t)=2R\hspace{0.1em} \mathcal{E}(t)\,,\qquad T_A=\frac{3}{2\pi R}\,.
\ee
Observe that this expression matches the leading order term of the entanglement entropy that we obtained from the direct computation in (\ref{smallRexpsC}) and (\ref{smallRexpsNC}), for centered and generic intervals, respectively. This concludes our check.

Going beyond the leading order in the size of the region, we still expect the first law (\ref{eq:1stlaw}) to give the full result at linear order in the density matrix, $\mathcal{O}(\delta\rho)$, but receive corrections at higher orders in $\delta\rho$. Since we know the exact result for expectation value of the energy density operator (\ref{eq:Tmunu}), it is easy to compute this contribution. The easiest way to perform the integral is to change variables, $x\to \zeta=x-x_c+t$ so that the integrand becomes:
\be
\delta S_A^{(\delta\rho)}=2\alpha^2\Delta\int_{t-R}^{t+R}\!\!\frac{R^2-(\zeta-t)^2}{2R}\!\left[\frac{1}{(\alpha^2+(\zeta+x_c)^2)^2}+\frac{1}{(\alpha^2+(\zeta+x_c-2t)^2)^2}\right]\!d\zeta\,.
\ee
These two integrals can be written in terms of the $\arctan(z)$ function. Using the sum identity $\arctan(z_1)\pm\arctan(z_2)\!=\!\arctan(\frac{z_1\pm z_2}{1\mp z_1 z_2})$ we can massage the full result into the following form:
\be\label{eq:deltaSfirstL}
\delta S_A^{(\delta\rho)}=\Delta\left[2-\(\frac{1}{\eta_{+}} \arctan\eta_{+}+\frac{1}{\eta_{-}} \arctan\eta_{-}\)\right]\,, \qquad \eta_\pm\equiv \frac{2 \alpha  R}{R^2-(x_c\pm t)^2-\alpha ^2}\,.
\ee
This expression matches exactly the universal term (\ref{deltaS+-}) obtained from the CFT calculation in section \ref{sec:CFTEE}. Expanding the above for small $R$, we obtain a complete match with the full first series of (\ref{smallRexpsC}) and (\ref{smallRexpsNC}), which can be checked order by order in the small $R$ expansion. These series can alternatively be written in a compact form by considering the integral expression for $\delta S_A^{\mathcal{P}}$ presented in (\ref{eq:contP}), thus also matching the universal term
\be
\delta S_A^{uni}=\delta S_A^{\mathcal{P}}\,.
\ee

\subsection{Corrections due to bulk entanglement entropy\label{sec:quanC}}

The second class of $\mathcal{O}(1)$ corrections to entanglement entropy are due to the entanglement of bulk fields in the given quantum state. The bulk is state specified by the density matrix $\rho_{\text{bulk}}$, defined on a Cauchy slice $\Sigma\supset A\cup\gamma_A$. Defining $\Sigma_A$ as the codimension-one region bounded by $\gamma_A$ and $A$, and $\Sigma_{A^c}=\Sigma\setminus \Sigma_{A}$, one then performs the trace over $\Sigma_{A^c}$ and compute the von Neumann entropy associated to the reduced density matrix $\rho_{\Sigma_A}=\Tr_{\Sigma_{A^c}}(\rho_{\text{bulk}})$, $S_{\text{bulk}}=-\Tr (\rho_{\Sigma_A} \log \rho_{\Sigma_A})$.
In practise, this can be accomplished by implementing the replica trick in the bulk, as was done in \cite{Belin:2018juv}.

Inspired by the CFT discussion around equation \eqref{Renyi}, we will want compute the difference between the entanglement in the one-particle excited state $\ket{\psi}$, defined in (\ref{eq:1pexstate}), and the vacuum state $\ket{0}$: $\delta S_{\text{bulk}}=S_{\text{bulk}}^{\ket{\psi}}-S_{\text{bulk}}^{\ket{0}}$. This quantity is free of UV divergences.
Given the crucial success of the first law of entanglement in the CFT, not only in reproducing the leading behavior of the geometric correction to entanglement entropy in the limit of small intervals but also in recovering the full universal piece of the entanglement entropy, it will prove useful to directly apply the first law of entanglement in the bulk to isolate the contribution coming from the entanglement of bulk fields. This first law stipulates that
\be
\label{bulkfirstLaw}
\delta S_{\text{bulk}}=\langle H_{\text{bulk}}\rangle\,,
\ee
where $H_{\text{bulk}}$ is the bulk modular Hamiltonian. We emphasize that equation \eqref{bulkfirstLaw} is not expected to hold generally for the states we consider here, but only at leading order in $\delta \rho_{\text{bulk}}=\rho_{\text{bulk}}^{\ket{\psi}}-\rho_{\text{bulk}}^{\ket{0}}$. The arguments follow closely to those for the first law in the CFT, as explained in section \ref{sec:1stlaw}. Nevertheless, this should suffice to capture the difference between the universal term (\ref{deltaS+-}) expected from CFT considerations and the full result for the geometric corrections, obtained in section \ref{sec:geomC}. In particular, we should be able to extract an infinite series of terms proportional to $R^{2\Delta+2i}$ $(i\in\mathbb{N})$ that are needed to cancel the second series of (\ref{smallRexpsC}) and (\ref{smallRexpsNC}). This expectation is based on the exact relation between the bulk modular Hamiltonian $H_{\text{bulk}}$ and the CFT modular Hamiltonian $H_A$ \cite{Jafferis:2015del}
\be\label{modularHs}
H_A=\frac{\hat{A}}{4G}+H_{\text{bulk}}\, .
\ee
In the following section we will use the first law (\ref{bulkfirstLaw}) to compute the contribution of the bulk entanglement entropy at linear order in $\delta \rho_\text{bulk}$, which should enable us to verify the above claims. We relegate the study of the next order corrections to section \ref{sec:replica} and the interpretation in the context of the black hole information problem to section \ref{bulkInterp}.

\subsubsection{Linear order corrections and bulk modular Hamiltonian\label{sec:quanCmH}}

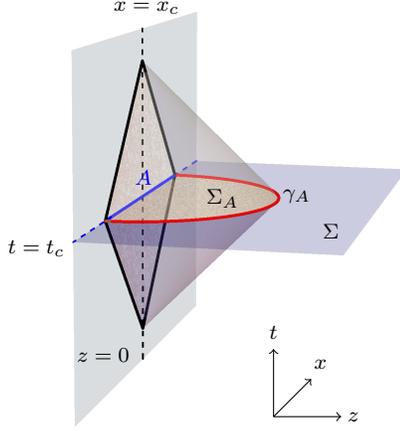
\begin{figure}[t!]
 \centering
 \begin{center}
 $\qquad\quad$
   \includegraphics[width=0.3\linewidth]{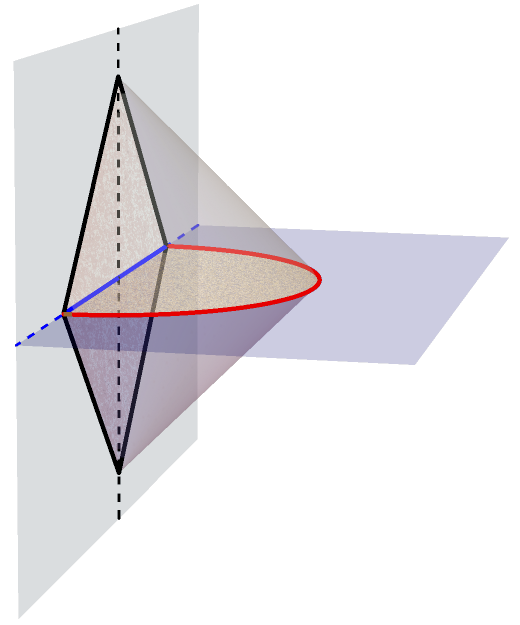}
\begin{picture}(0,0)
\put(-53,88){{\scriptsize $\gamma_A$}}
\put(-109,93){{\color{blue}{\scriptsize $A$}}}
\put(-117,159){{\scriptsize $x=x_c$}}
\put(-131,26){{\scriptsize $z=0$}}
\put(-157,67){{\scriptsize $t=t_c$}}
\put(-82,86){{\scriptsize $\Sigma_A$}}
\put(-38,73){{\scriptsize $\Sigma$}}
\end{picture}
\begin{tikzpicture}[x  = {(0.9cm,0cm)},
                    y  = {(0.5cm,0.5cm)},
                    z  = {(0cm,0.9cm)},
                    scale = 1,
                    color = {lightgray}]
\hspace{-2.3cm}
\begin{scope}[canvas is xy plane at z=3]
  \draw[black,->] (0,0) -- (1,0) node (x-axis) [right] {{\scriptsize$\!z$}};
  \draw[black,->] (0,0) -- (0,1) node (y-axis) [above] {{\scriptsize$\,\,\,\,\,x$}};
\end{scope}
\begin{scope}[canvas is yz plane at x=0]
  \draw[black,->] (0,3) -- (0,4) node (z-axis) [above] {{\scriptsize$t$}};
\end{scope}

\end{tikzpicture}
\end{center}
 \vspace{-3mm}
 \caption{Calculation of the quantum correction to the entanglement entropy associated to a region $A=\{x_c-R\leq x\leq x_c+R,\, t=t_c\}$. Given a quantum state on $\Sigma=\{M:\,t=t_c\}$, specified by its density matrix $\rho_\Sigma$, we define $\Sigma_A\subset\Sigma$ as the codimension-one bulk region within $\gamma_A$ and $A$, and $\Sigma_{A^c}=\Sigma\setminus \Sigma_{A}$. The entanglement entropy in the bulk is computed by tracing over the degrees of freedom in $\Sigma_{A^c}$ and then finding the von Neumann entropy associated to the reduced state $\rho_{\Sigma_A}$.}
 \label{fig:ewedge}
\end{figure}

To calculate the contribution at linear order in the density matrix $\delta \rho_\text{bulk}$, we will need an explicit expression for the bulk modular Hamiltonian in Poincar\'e coordinates:
\be\label{bulkMH}
H_{\text{bulk}}=\int_{\Sigma_A}\sqrt{g_{\Sigma_A}} d\Sigma_A N^\mu\xi_A^\nu T_{\mu\nu}\,,
\ee
where $N^\mu$ is the unit normal vector associated to $\Sigma_A$, $\xi^\nu$ is the Killing vector that generates the entanglement wedge (normalized such that the surface gravity at the bifurcate horizon $\gamma_A$ is $\kappa=2\pi$), and $T_{\mu\nu}$ is the bulk stress energy tensor. For simplicity, we will take $\Sigma$ to be constant-$t$ slice, as depicted in Figure \ref{fig:ewedge}.
This implies that
\be
\sqrt{g_{\Sigma_A}}d\Sigma_A=\frac{L^2}{z^2}dxdz\,, \quad N^{\mu}=\frac{z}{L}\delta^{\mu}_0\,.
\ee
The Killing vector $\xi_A$ takes the following form in the Poincar\'e patch \cite{Faulkner:2013ica}
\be
\label{killvec}
\xi_A=-\frac{2\pi}{R}(t-t_c)[z\partial_z+(x-x_c)\partial_x]+\frac{\pi}{R}[R^2-z^2-(t-t_c)^2-(x-x_c)^2]\partial_t\,,
\ee
which is already correctly normalized since the surface gravity at the horizon yields $\kappa=2\pi$.
The expression \eqref{killvec} simplifies further for a constant $t=t_c$ slice and we get
\be
\xi_A|_{\Sigma_A}=\frac{\pi}{R}[R^2-z^2-(x-x_c)^2]\partial_t\,.
\ee
Defining the coordinates $(\rho,\phi)$ such that $x=x_c+\rho \cos \phi$, $z=\rho \sin\phi$ ($0\leq\rho\leq R$, $0\leq \phi\leq\pi$), we arrive to the following expression for the change of entanglement entropy in the bulk, to linear order in the density matrix
\begin{align}
\begin{split}
\label{deltaSMH}
\delta S_{\text{bulk}}^{(\delta\rho)}&=2\pi L \int_{\Sigma_A}  \frac{dxdz}{z} \, \frac{(R^2-x^2-z^2)}{2R}\langle T_{00}(t,x,z)\rangle\, ,  \\
&=2\pi L\int_0^R d\rho\int_0^\pi d\phi \,\csc\phi\,\frac{R^2-\rho^2}{2R}\langle T_{00}(t,\rho,\phi)\rangle\,.
\end{split}
\end{align}
This expression resembles the change of entanglement in flat space, but with an extra factor of $L/z$ that arises from the volume form in AdS.

We now need the expectation value of the energy density operator $T_{00}$ in the Poincar\'e frame to evaluate \eqref{deltaSMH}. In the original global coordinates, we have explicit expressions for the components of the stress-energy tensor in (\ref{BulkTmn}), while the coordinate transformation from global to Poincar\'e  coordinates $x^\mu=(\tau,r,\theta) \to x^{\mu'}=(t,x,z)$ is given by equations (\ref{eq:cotrans})-(\ref{eq:cotrans3}). The standard transformation rule gives
\be
\label{T00bulk}
T_{\mu'\nu'}=\frac{\partial x^{\mu}}{\partial x^{\mu'}}\frac{\partial x^{\nu}}{\partial x^{\nu'}}T_{\mu\nu} \,
\implies T_{00}=\left(\frac{\partial \tau}{\partial t}\right)^2T_{\tau\tau}+\left(\frac{\partial r}{\partial t}\right)^2T_{rr}+\left(\frac{\partial \theta}{\partial t}\right)^2T_{\theta\theta}\,.
\ee
The explicit expression is lengthy and not particularly illuminating, so we will not display it here. For our purposes it will suffice to analyze the small interval limit of (\ref{deltaSMH}). Expanding the evaluated form of (\ref{T00bulk}) for small $\rho$, we obtain a leading order term proportional to $T_{00}\sim \mathcal{O}(\rho^{2\Delta-2})$ and further corrections suppressed by higher powers of $\rho$. Plugging this expansion into the equation \eqref{deltaSMH} leads to
\bea
\!\!\!\!\!\!\!\!\delta S_{\text{bulk}}^{(\delta\rho)}&=&\frac{(2\alpha)^{2 \Delta }\Delta(\Delta -1)}{R[(t^2-x_c^2+\alpha ^2)^2+4 \alpha ^2 x_c^2]^{\Delta }}\int_0^\pi (\sin \phi)^{2 \Delta -3} d\phi\int_0^R \rho^{2\Delta-2}(R^2-\rho^2)d\rho+\cdots ,\nonumber\\
&=&\frac{ (2\alpha)^{2 \Delta }\Delta(\Delta -1)}{R[(t^2-x_c^2+\alpha ^2)^2+4 \alpha ^2 x_c^2]^{\Delta }}\left(\frac{\sqrt{\pi}\Gamma(\Delta-1)}{\Gamma(\Delta-\frac{1}{2})}\right)\left(\frac{R^{2\Delta+1}}{2(\Delta-\frac{1}{2})(\Delta+\frac{1}{2})}\right)+\cdots,\nonumber\\
&=&\frac{\Gamma(\tfrac{3}{2})  \Gamma(\Delta+1)}{\Gamma (\Delta+\tfrac{3}{2} )}\frac{(2 \alpha R)^{2\Delta} }{[(t^2-x_c^2+\alpha ^2)^2+4 \alpha ^2 x_c^2]^{\Delta }}+\cdots,\label{deltaSMH-limit}
\eea
where the dots denote higher order terms in $R$.
Remarkably, the leading order result in (\ref{deltaSMH-limit}) exactly cancels the terms of order $\mathcal{O}(R^{2\Delta})$ in the expansions (\ref{smallRexpsC}) and (\ref{smallRexpsNC}) for centered and general intervals, respectively. In fact, one can check order-by-order that the full result coming from (\ref{deltaSMH}) yields
\be
\label{linbulkEE}
\delta S_{\text{bulk}}^{(\delta\rho)}=\frac{\Gamma(\tfrac{3}{2})  \Gamma(\Delta+1)}{\Gamma (\Delta+\tfrac{3}{2} )}\frac{(2 \alpha R)^{2\Delta} }{[(t^2-x_c^2+\alpha ^2)^2+4 \alpha ^2 x_c^2]^{\Delta }}\Big[1+\sum_{i=1}^{\infty}\mathcal{Q}_i(t,x_c,\alpha)R^{2i}\Big]\,,
\ee
and cancels the whole series $\mathcal{O}(R^{2\Delta+2i})$, $i\in \mathbb{N}$,  in \eqref{smallRexpsNC}. This is expected from field theory considerations, and the exact relation between bulk and CFT modular Hamiltonians (\ref{modularHs}). For $x_c=0$ this series can be resummed as
\be
\delta S_{\text{bulk}}^{(\delta\rho)}\big|_{x_c=0}=\frac{\Gamma(\tfrac{3}{2})\Gamma(\Delta+1)R^{2\Delta}}{\Gamma(\Delta+\tfrac{3}{2})a(t)^{2\Delta}}\,\!_2F_1\left[1,\Delta,\Delta+\tfrac{3}{2},\tfrac{R^2}{a(t)^2}\right]\,,
\ee
with $a(t)$ given in (\ref{def:aoft}). For $x_c\neq0$ we do not know the most general form but we can formally write it in a compact integral form as
\be
\delta S_{\text{bulk}}^{(\delta\rho)}= -\delta S_A^{\mathcal{Q}}\,,
\ee
where $\delta S_A^{\mathcal{Q}}$ is given in (\ref{eq:contQ}). We can obtain explicit expressions for integer $\Delta$, by considering the difference between the universal term (\ref{deltaS+-}) and the result coming from the area term (\ref{intSgeneral}) (e.g. (\ref{eq:noncfinal}) for $\Delta=2$), however the expressions are lengthy and we do not show them.

\subsubsection{Quadratic corrections and bulk replica trick\label{sec:replica}}

We now go further and compute the vacuum subtracted bulk entanglement entropy to higher orders in $\delta \rho_{\text{bulk}}$. One way to do this is by implementing the replica trick in the bulk
\bea\label{replica-dn}
\delta S_{\text{bulk}}=\lim_{n\to 1}\partial_{n}\[ \log \(\frac{\tr{\rho_\psi^n}}{\tr{\rho_0^n}}\)\],
\eea
where $\rho_\psi=\tr_{\Sigma_{A^c}}\ket{\psi}\bra{\psi}$ and $\rho_0=\tr_{\Sigma_{A^c}}\ket{0}\bra{0}$ are the reduced density matrices in the one-particle excited state and the vacuum respectively. We start by expanding the operator\footnote{The second order term has an interesting similarity with the sum (\ref{O2O2}) if one relates each $\cO^2$ insertion on a given sheet with a $\delta \rho$ factor and the index $m$ with the distance in number of sheets between such operators. In particular, the factor of $n/2$ is explained in the exact same way as in (\ref{O2O2}): the replica symmetry ensures that for fixed $m$ one has $n$ equal contributions related by an overall translation, while the extra $1/2$ is introduced to avoid the double counting of equivalent configurations.}
\bea \label{rho-n}
\tr{\rho_\psi^n}=\tr{\(\rho_0 +\delta \rho\)^n}=\tr{\rho_0^n}+n\,\tr{\rho_0^{n-1}\delta \rho} +\frac{n}{2}\,\sum_{m=0}^{n-2}\tr \(\, {\delta\rho\, \rho_0^{m}\, \delta\rho\, \rho_0^{n-2-m}}\)+\cO(\delta \rho^3)\,.
\eea
Plugging this expansion back in (\ref{replica-dn}), we arrive to an expansion of $\delta S_{\text{bulk}}$ at different orders in $\delta\rho$, of the form
\be
\delta S_{\text{bulk}}=\delta S_{\text{bulk}}^{(\delta\rho)}+\delta S_{\text{bulk}}^{(\delta\rho^2)}+\cdots\,.
\ee
The first term in this expansion gives exactly the contribution of the bulk modular Hamiltonian, computed in section \ref{sec:quanCmH}. Here we are interested in the second order contribution, which can be obtained by isolating the piece
\bea\label{rhon2order}
\!\!\!\log \(\frac{\tr{\rho^n}}{\tr{\rho_0^n}}\)\Bigg|_{\cO(\delta \rho^2)}\!\!\!=\frac{n}{2}\,\frac{ \tr \(\, { \rho_0^{n-2} \delta\rho\,\, \delta\rho_n\, }\)}{ \tr{\rho_0^n}}-\frac{n^2}{2}\(\frac{\tr{\rho_0^{n-1}\delta \rho}}{ \tr{\rho_0^n}} \)^2\!\!,\quad\delta \rho_{n}\equiv \sum_{m=0}^{n-2} \rho_0^m\, \delta \rho \, \rho_0^{-m}.
\eea
From (\ref{rhon2order}) and (\ref{replica-dn}), we obtain that the second order contribution is
\bea\label{2ndSrho}
\delta S^{(\delta\rho^2)}_{\text{bulk}}=-\frac12 \tr\(\delta \rho\, \delta \rho'_1 \rho_0^{-1}\)\,,
\eea
where $ \delta \rho'_1 \equiv \partial_n \(\delta \rho_n\)|_{n=1}$ and we have used some formal properties of $\delta \rho_n$.\footnote{We assume that $\delta\rho_n$ has a proper analytic continuation for $n\approx 1$. In that case one can show that
$$
\tr \delta\rho_n=\(n-1\)\tr{\delta \rho}=0, \quad {\rm and } \quad \delta \rho_1\equiv \lim_{n\to 1}  \delta \rho_n=0\,.
$$
The first property follows from the cyclicity of the trace and the normalization of both, $\rho$ and $\rho_0$. The second is a consistency condition of the $n \to 1$ limit of (\ref{rho-n}). Namely, since every order contribution to the left hand side of (\ref{rho-n}) must vanish in the $n\to 1$ limit, this requires that
$$
\lim_{n\to 1}\frac{n}{2}\tr\, \( {\delta\rho\, \delta\rho_n\, \rho_0^{n-2}}\)=\frac{1}{2}\tr\, \( {\delta\rho\, \delta\rho_1\, \rho_0^{-1}}\)=0\,.
$$
This must hold for arbitrary $\delta \rho$ and $\rho_0$ and therefore $\delta \rho_1=0$ must hold as an operator equation.}

The calculation of this term is more naturally performed in the Hilbert space of the Rindler observer,
which leads to a relatively straightforward answer in terms of the Bogoliubov coefficients that relates the global and Rindler modes \cite{Belin:2018juv}.
We relegate the computation of the Bogoliubov coefficients for our combined transformation (global $\to$ Poincar\'e $\to$ Rindler), given by (\ref{eq:cotrans})-(\ref{eq:cotrans3}) and (\ref{poincaretobtz-2a})-(\ref{poincaretobtz-2c}), to Appendix \ref{AppendixA}. Here we will  merely point out that, in the small $R$ limit, the Bogoliubov coefficients in our case differ only to those in \cite{Belin:2018juv} by a constant phase and hence, the resulting integral coincides with the one in \cite{Belin:2018juv}, up to the identification of the corresponding small parameters (\ref{etaR}). With this in mind, we obtain
\bea\label{2ndSRindler}
\delta S^{(\delta\rho^2)}_{\text{bulk}}=-\[\frac{2 \alpha R }{\sqrt{4\alpha^2x_c^2+\left(\alpha^2+t_c^2-x_c^2\right)^2}}\]^{4\Delta}\!\!\!\!\! \int\limits_{\,\,\omega_{1,2}>0} \!\!\!\!\! d\omega_1d\omega_2dk_1dk_2 \,\, G(\omega_1,\omega_2;k_1,k_2)\,,
\eea
where
\bea
G(\omega_1,\omega_2;k_1,k_2)=\frac{2^{4\Delta} }{32 \pi^4}\(\frac{\pi (\omega_1+\omega_2)}{\sinh \pi (\omega_1+\omega_2)} +\frac{\pi (\omega_1-\omega_2)}{\sinh \pi (\omega_1-\omega_2)}  \)F(\omega_1,k_1)F(\omega_2,k_2)\,,
\eea
and
\bea
\qquad F(\omega,k)=\frac{\left|\Gamma\(\frac\Delta2+i\frac{\omega+k}{2}\)\right|^2\left|\Gamma\(\frac\Delta2+i\frac{\omega-k}{2}\)\right|^2}{|\Gamma(\Delta)|^2}\,.
\eea
As in \cite{Belin:2018juv}, we could not get a closed expression for the integral but by numerical evaluation we can confirm that it is consistent with the expected result,
\bea
\label{quadbulkEE}
\delta S^{(\delta\rho^2)}_{\text{bulk}}=-\frac{\Gamma(\tfrac{3}{2})\Gamma(2\Delta+1)}{\Gamma(2\Delta+\tfrac{3}{2})}\[\frac{2 \alpha R }{\sqrt{4\alpha^2x_c^2+\left(\alpha^2+t_c^2-x_c^2\right)^2}}\]^{4\Delta}\,,
\eea
which matches with our formula for the dynamical contribution $\delta S_A^{dyn}$ obtained from the CFT analysis (\ref{Q-corrs}). This completes our check of the FLM formula in the dynamical setting of local quenches. We emphasize, again, that our result for $\delta S^{(\delta\rho^2)}_{\text{bulk}}$ (and similarly for $\delta S_A^{dyn}$) is only valid at the leading order in the small $R$ expansion, while our result for $\delta S^{(\delta\rho)}_{\text{bulk}}$ is valid for any $R$.


\subsubsection{Interpretation as the entropy of Hawking radiation
\label{bulkInterp}}

We now study the behavior of $\delta S_{\text{bulk}}$ and interpret the results. In our interpretation, two transformations play an important role: the large gauge transformation in equations \eqref{eq:cotrans}-\eqref{eq:cotrans3} and the CHM map that takes a Poincare patch to a Rindler patch. The CHM \cite{Casini:2011kv} map is given by
\begin{eqnarray}
\label{poincaretobtz-2}
t&=&t_c+\frac{R\sqrt{\mathfrak{r}^2-1}\sinh\bt}{\mathfrak{r}\cosh\by+\sqrt{\mathfrak{r}^2-1}\cosh\bt}~,\label{poincaretobtz-2a}\\
x&=&x_c+\frac{R\mathfrak{r}\sinh\by}{\mathfrak{r}\cosh\by+\sqrt{\mathfrak{r}^2-1}\cosh\bt}~,\\
z&=&\frac{R}{\mathfrak{r}\cosh\by+\sqrt{\mathfrak{r}^2-1}\cosh\bt}\,. \label{poincaretobtz-2c}
\end{eqnarray}
where $\mathfrak{r}\in(1,\infty)$, $\mathfrak{t}\in(-\infty,\infty)$ and $\by\in(-\infty,\infty)$. In this coordinate system, the bulk metric takes the form of a planar BTZ geometry
\begin{equation}
\label{btzmetric}
ds^2=L^2\left(-(\mathfrak{r}^2-1)d\bt^2+\mathfrak{r}^2d\by^2+\frac{d\mathfrak{r}^2}{\mathfrak{r}^2-1}\right)\,,
\end{equation}
It is useful to study the effect of the composite map created from transformations \eqref{eq:cotrans}-\eqref{eq:cotrans3} and the CHM map \eqref{poincaretobtz-2}-\eqref{poincaretobtz-2c}.   When we start in global coordinates with a static, single-particle wavefunction for the bulk excitation, this map takes it to a multi-particle time-dependent wavefunction. This is because of a Bogoliubov transformation implicit in it, see Appendix \ref{AppendixA}. We thus get a thermally-populated state of the scalar field in the planar BTZ geometry \eqref{btzmetric}. We interpret these excitations as Hawking radiation outside the BTZ black brane. Consequently, when viewed from the Rindler frame, $S_{\text{bulk}}$ computes the entanglement entropy of bulk fields with the interior of the brane traced over. In black hole context, this is often referred to as the \emph{entropy of the Hawking radiation}.

The dependence of the multi-particle wavefunction on Rindler time $\mathbf{t}$ has an interesting consequence. As a result, the boundary condition for the scalar field in the CHM frame is time-dependent. This allows a \textit{leakage} of the wavefunction from the wedge described by \eqref{btzmetric} to its complement. We interpret the complement as a \textit{bath} that \textit{absorbs} the Hawking radiation from outside the black brane \eqref{btzmetric}. This should be contrasted with the setup in \cite{Almheiri:2019hni} (and related works) where a non-gravitating region needs to be coupled to the black hole to absorb the Hawking radiation. The global frame point of view discussed later further supports these claims.

\begin{figure}[!htb]
\begin{center}
  \includegraphics[width=4cm]{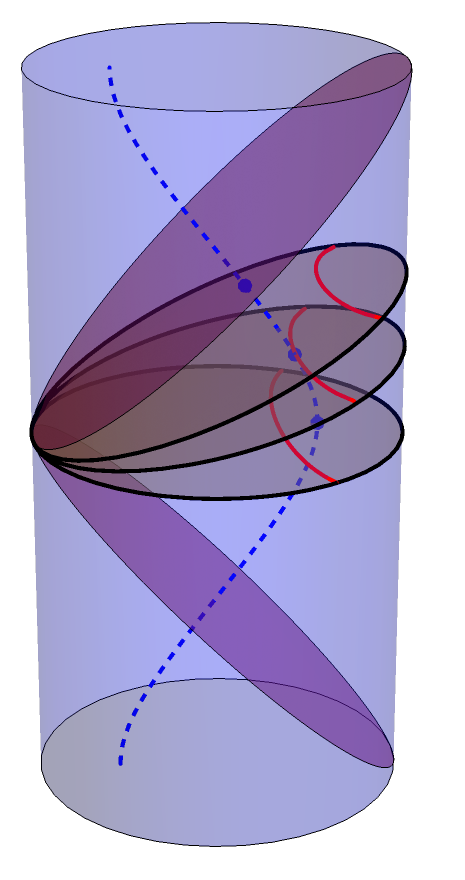}
  \hspace*{2cm}
  \includegraphics[width=4cm]{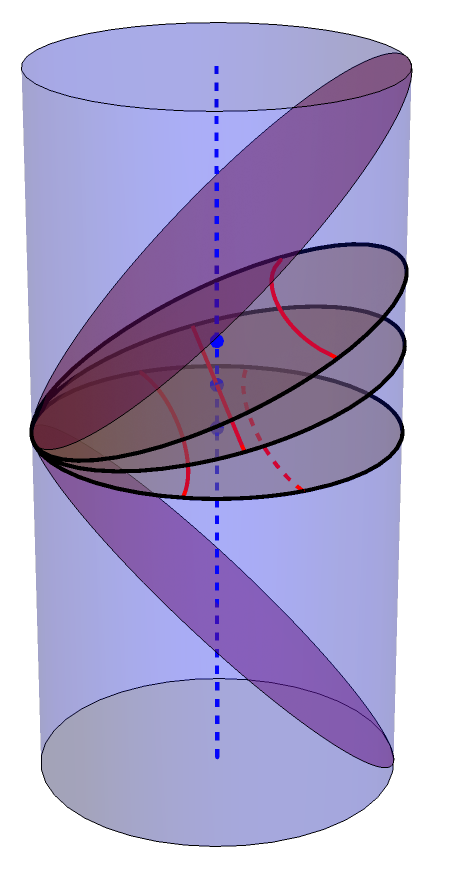}
  \setlength{\unitlength}{1cm}
\begin{picture}(0,0)
\put(-10.7,6.0){\vector(0,1){0.5}}
\put(-10.8,6.7){$\tau'$}
\put(-9.8,6.4){\vector(1,0){0.1}}
\put(-9.6,6.2){$\theta'$}
\qbezier(-10.3,6.6)(-10.0,6.4)(-9.8,6.4)
\put(-8.63,7.0){\vector(2,1){0.4}}
\put(-8.23,6.9){$r'$}
\put(-8.63,7.0){\circle*{0.1}}
\put(-6.95,3.8){$\delta\theta'|_{t=0}$}
\put(-6.95,4.55){$\delta\theta'|_{t=t_{\text{Page}}}$}
\put(-6.95,5.25){$\delta\theta'|_{t>t_{\text{Page}}}$}
\put(-4.45,6.0){\vector(0,1){0.5}}
\put(-4.55,6.7){$\tau$}
\put(-3.55,6.4){\vector(1,0){0.1}}
\put(-3.35,6.2){$\theta$}
\qbezier(-4.05,6.6)(-3.75,6.4)(-3.55,6.4)
\put(-2.39,7.0){\vector(2,1){0.4}}
\put(-1.99,6.9){$r$}
\put(-2.39,7.0){\circle*{0.1}}
\put(-0.71,3.8){$\delta\theta|_{t=0}$}
\put(-0.71,4.55){$\delta\theta|_{t=t_{\text{Page}}}$}
\put(-0.71,5.25){$\delta\theta|_{t>t_{\text{Page}}}$}
\end{picture}
\end{center}
\vspace{-0.2cm}
\caption{
\footnotesize Evolution of the size of the RT surface (i.e. black brane horizon from the perspective of the Rindler observer) for centered intervals from the point of view of the global frame. Left: we plot the situation as seen from the boosted global coordinates. Right: we plot the situation in the original global frame. In both cases we observe that the size of $\gamma_A$ (hence the size of the black brane) decreases monotonically as the Poincar\'e time evolves from $t=0$ to $t\to\infty$. This implies that, for centered intervals, our setup can indeed be interpreted as a toy model of black hole evaporation. We remark that $\delta S_{\text{bulk}}^{(\delta\rho)}$ indeed behaves as a Page curve for $t\in[0,\infty]$, reaching a peak at $t=t_{\text{Page}}$ and then decreasing as the state evolves in time. In our setup, the Page time $t_{\text{Page}}$ can be identified as the time at which the particle crosses the RT surface $\gamma_A$.
}
\label{fig:globalfigs2}
\end{figure}
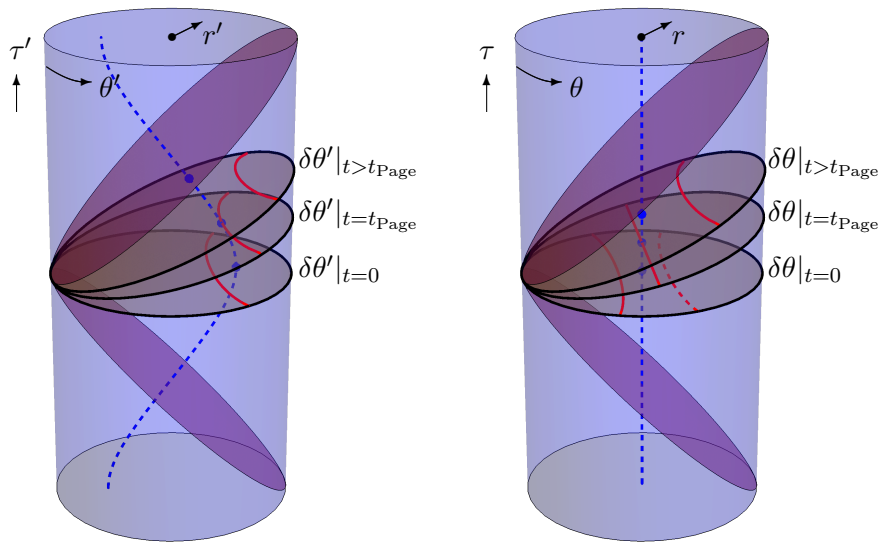

In fact, if we go to the global patch, we can make precise the sense in which our model is a toy model of black hole evaporation and also see the absorption of Hawking radiation in a different way. First notice that the black brane (\ref{btzmetric}) is actually time dependent. There are different Rindler wedges associated to the region $A$ at different time slices. In the Poincar\'e frame, these Rindler wedges are adapted to a region of fixed length $\ell$. However, they have different sizes from the point of view of the global frame. For centered intervals, $x_c=0$, a careful calculation shows that
\bea
\delta \theta'&\equiv&|\theta'_2-\theta'_1|= \begin{cases}
  \displaystyle 2\pi - \left|2 \arctan\left(\tfrac{2L R}{R^2-L^2-t^2}\right)\right| \, , & \displaystyle \{R>L,\,\,\,0<t<\sqrt{R^2-L^2}\} \ ,\\[3ex]
  \displaystyle  \left|2 \arctan\left(\tfrac{2LR}{R^2-L^2-t^2}\right)\right| \, , & \displaystyle \text{otherwise} \ .
 \end{cases}\\
 \delta \tau'&\equiv&|\tau'_2-\tau'_1|=0\,,
\eea
in the boosted global frame, or
\bea
\label{deltatheta}
\delta \theta&\equiv&|\theta_2-\theta_1|=\begin{cases}
  \displaystyle 2\pi - \left|2 \arctan\left(\tfrac{2\alpha R}{R^2-\alpha^2-t^2}\right)\right| \, , & \displaystyle 0<t<t_{\text{Page}} \ ,\\[3ex]
  \displaystyle  \left|2 \arctan\left(\tfrac{2\alpha R}{R^2-\alpha^2-t^2}\right)\right| \, , & \displaystyle t>t_{\text{Page}} \ .
 \end{cases}\\
 \delta \tau&\equiv&|\tau_2-\tau_1|=0\,,
\eea
in the original global frame. The reason that $\delta \theta'$ and $\delta \theta$ have two branches is due to the periodicity of
the two coordinate systems, $\theta'\sim\theta'+2\pi$ and $\theta\sim\theta+2\pi$. In both cases we show that the size of the RT surface in fact decreases in time as $t$ goes from $t=0$ to $t\to\infty$, as depicted in Figure \ref{fig:globalfigs2}. In the left figure, we show an example as seen in the boosted global frame with $R<L$ so we only have the second branch of $\delta\theta'$ for $t\in[0,\infty]$. In this case we have $\delta\theta'|_{t=0}<\pi$ and decreases monotonically as $t$ evolves, with $\delta\theta'|_{t\to\infty}\to0$. In the right figure, we show an
example as seen in the original global frame. In this case, one has to always include both branches of $\delta\theta$. In fact, the branches are switched at exactly the Page time since then $\delta\theta|_{t=t_{\text{Page}}}=\pi$. This implies that $\delta\theta|_{t<t_{\text{Page}}}>\pi$ and $\delta\theta|_{t>t_{\text{Page}}}<\pi$, as can be confirmed from (\ref{deltatheta}). With the inclusion of both branches, $\delta\theta$ decreases monotonically for $t\in[0,\infty]$ also in this frame. Hence in either boosted or original global frame, the size of the black brane always decreases in time for centered intervals. We interpret this as a toy model of an \textit{eternally evaporating} black hole. 

We can also see the absorption of the Hawking radiation mentioned earlier while working in the original global frame, where the wavefunction for the bulk excitation is static and spherically-symmetric. It is useful to identify the region which maps to the exterior of the BTZ black brane under the aforementioned composite map. In the global frame, the overlap of the wavefunction with this region is proportional to the probability of the particle being in the exterior of the black brane. However, the size of that region decreases monotonically as plotted in Figure \ref{fig:globalfigs2}. This decreasing probability fits in with our interpretation of the \textit{leakage} of Hawking radiation. Conversely, as the size of the complementary region increases, the probability of the particle to be in the complement increases with time. This further justifies our interpretation that the interior of the black brane serves the purpose of a bath. 

\begin{figure}[t!]
\begin{center}
\includegraphics[scale=0.6]{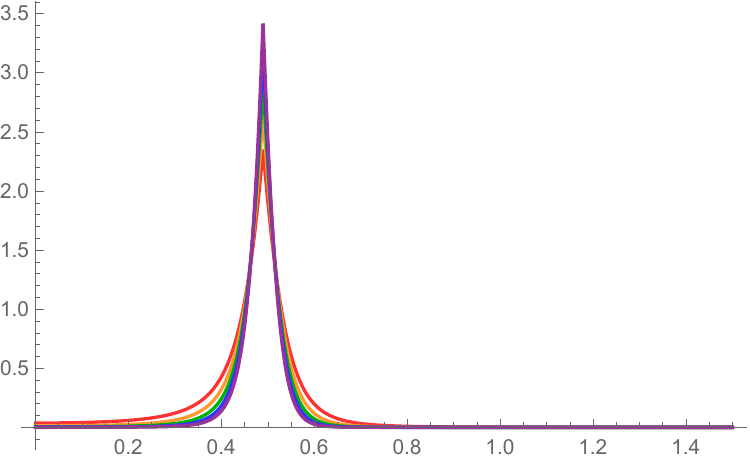}$\quad$\includegraphics[scale=0.6]{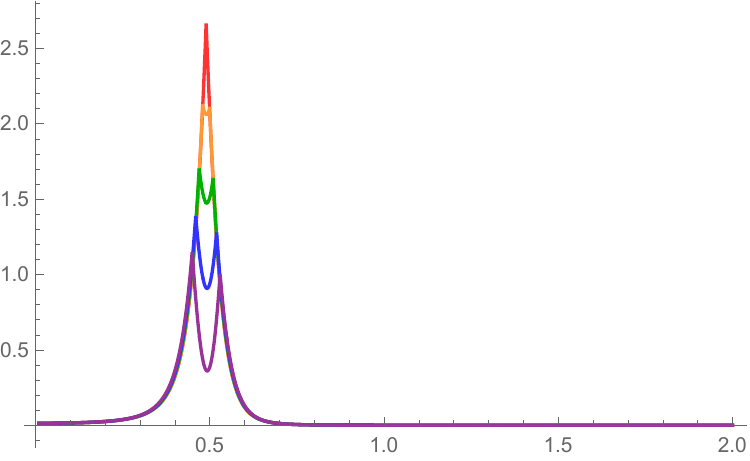}
\begin{picture}(0,0)
\put(-206,138){{\small $\delta S_{\text{bulk}}^{(\delta\rho)}$}}
\put(-18,32){{\small $t/\ell$}}
\put(22,138){{\small $\delta S_{\text{bulk}}^{(\delta\rho)}$}}
\put(211,32){{\small $t/\ell$}}
\end{picture}
\vspace{-0.7cm}
\caption{
\footnotesize
Time evolution of the bulk entanglement entropy contribution at linear order in the perturbation, $\delta S_{\text{bulk}}^{(\delta\rho)}$, for cases that satisfy the condition (\ref{cond:localthq}). The latter condition is imposed such that the particle is initially inside the entanglement wedge so for a Rindler observer the state can be viewed as a local perturbation of a black brane geometry. In the left panel we have plotted the case of centered intervals, so we have set $x_c=0$, and we have varied $\Delta=3/2,2,5/2,3,7/2$, corresponding to the red, orange, green, blue and purple curves, respectively. In the right panel we have plotted the case of non-centered intervals, specializing to $\Delta=2$ and varying $x_c/\ell=0,10^{-1},2\times 10^{-1},3\times 10^{-1},4\times 10^{-1}$, corresponding to the red, orange, green, blue and purple  curves, respectively. For concreteness we have fixed $\alpha/\ell=10^{-1}$ in all the plots and, accordingly, chosen values for $x_c/\ell$ such that (\ref{cond:localthq}) is satisfied.
}
\label{fig:bulkEE}
\end{center}
\end{figure}
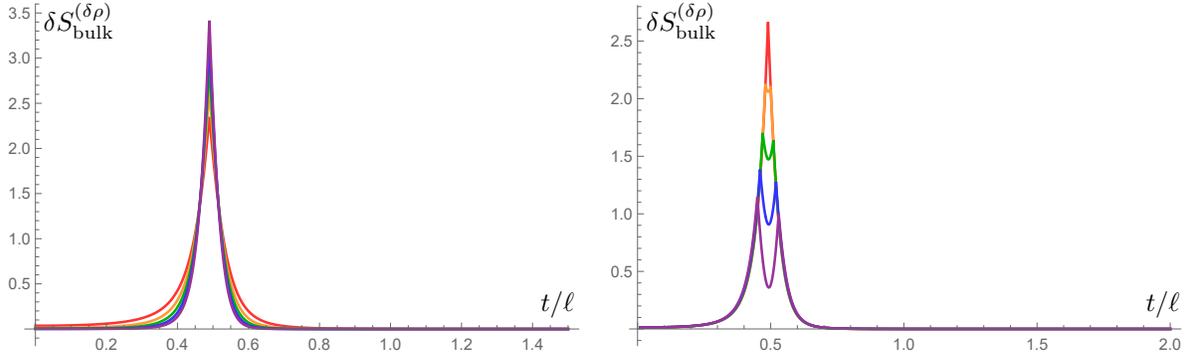

In our toy model of black hole evaporation, we can ask how the entropy of the radiation evolves with time.  Let us consider the case where the multi-particle wavefunction is initially inside the entanglement wedge (i.e. exterior of the brane). This is given by the condition
\begin{equation}
\label{cond:localthq}
\alpha<\sqrt{R^2-x_c^2}\,.
\end{equation}
In this case, the state as seen by the Rindler observer is a thermal perturbation of the black brane geometry, which is a quench of a \textit{thermal state} in the dual CFT. The entropy of Hawking radiation is given by the bulk entanglement entropy $\delta S_\text{bulk}$. For concreteness, we will only consider the expressions for $\delta S_{\text{bulk}}^{(\delta\rho)}$, which are valid for any $R$ (recall that our result for $\delta S_{\text{bulk}}^{(\delta\rho^2)}$ requires $R\ll\alpha$, so it is not valid in the regime where (\ref{cond:localthq}) holds true). In Figure \ref{fig:bulkEE} we plot $\delta S_{\text{bulk}}^{(\delta\rho)}$ for different scenarios when (\ref{cond:localthq}) is satisfied. In the left panel, we plot $\delta S_{\text{bulk}}^{(\delta\rho)}$ for the case of centered intervals, $x_c=0$, and different values of $\Delta$. Strikingly, we observe that the bulk entanglement at this order follows the expected behavior for a Page curve, i.e., increasing up to a time $t=t_{\text{Page}}$ and then decreasing as $t\to\infty$. The Page time $t_{\text{Page}}$ in our setup can be identified as the time at which the particle exits the wedge,
\be
t_{\text{Page}}=\sqrt{R^2-\alpha ^2}\,.
\ee
We can perform an analysis similar to that in equation \eqref{deltatheta} for non-centered intervals. It is straightforward so we will omit the specifics here and directly state our results. In this case, the size of the black brane as viewed from the global perspective changes non-monotonically as we vary the Poincar\'e time from $t=0$ to $t\to\infty$. We take this to mean that for $x_c\neq0$ our setup cannot be interpreted as a toy model of black hole evaporation. The plot in the right panel of Figure \ref{fig:bulkEE} confirms our justification. In this case, the evolution of  $S_{\text{bulk}}^{(\delta\rho)}$ does not follow a standard Page curve, but instead, develop two peaks at different times $t_{1,2}>0$.

Before closing this section, we comment on the implications of the above analysis for the \textit{information problem}. Since our state is pure, the von Neumann entropy of a full time slice vanishes, $S_{\Sigma}=0$ and the Araki-Lieb inequality implies
\be
\label{araki-lieb}
 |S_{\Sigma_A}-S_{\Sigma_{A^c}} |\leq S_{\Sigma_A\cup\Sigma_{A^c}}=S_{\Sigma}=0\quad \implies \quad S_{\Sigma_A}=S_{\Sigma_{A^c}}\,.
\ee
For excited states, we can subtract the vacuum contributions to obtain
\be\label{purity}
\delta S_{\Sigma_A}=\delta S_{\Sigma_{A^c}}\,.
\ee
When applied to our discussion of decreasing entanglement wedges in the original global frame, we can easily explain the Page curve in the left panel of Figure \ref{fig:bulkEE} in terms of the purity of the state. Recall that $\delta\theta|_{t<t_{\text{Page}}}>\pi$ while $\delta\theta|_{t>t_{\text{Page}}}<\pi$. If we now replace $\delta S_{bulk}^{(\delta \rho)}$ for $t<t_\text{Page}$ by the entanglement of the complementary region $\Sigma_A^c$ using (\ref{purity}), the increasing and decreasing parts of the Page curve immediately follow from the monotonicity of $\delta S_{bulk}^{(\delta \rho)}$ for $\delta\theta\in[0,\pi]$ (see red dashed line in the right plot of Figure \ref{fig:globalfigs2}). Although this result is only manifest in the unboosted global frame, it must be true in general, even in the Poincar\'e frame. We thus conclude that, at least in our toy model for black hole evaporation, \emph{the behavior of the Page curve can be directly associated with the purity of the global state}.

Interestingly, the above result follows from a semiclassical calculation in the bulk, without the requirement of the so-called island contribution. This is due to the fact that our quantum state is \emph{simple} enough so that there are no competing saddles as time evolves. Nevertheless, the mere fact that we are finding a Page curve consistent with unitarity in the framework of semiclassical gravity implies that, in more realistic models, information loss or the lack thereof should \emph{not} be interpreted as an artifact of the semiclassical approximation.

\section{Conclusions and outlook\label{sec:conclusions}}

In this paper, we studied the evolution of entanglement entropy after a local quench in two-dimensional conformal field theories with holographic duals both from the CFT perspective and using its gravitational description. Our CFT computation was carried out using the replica trick. This entailed the calculation of a $2n$-point correlator on a topologically non-trivial multi-sheeted Riemann surface. The computation was considerably simplified by virtue of the enhanced conformal symmetry special to two dimensions which allowed a map that relates the $2n$-point correlator on the aforementioned Riemann surface to a $2n$-point correlator on the complex plane. As a consequence, a separation of the entanglement entropy into two pieces (universal and dynamical) emerged naturally. We provided insights into their nature from the CFT perspective. In the bulk, the aforementioned calculation amounted to compute two separate contributions, the area of an extremal surface in the backreacted geometry and the bulk entanglement entropy associated to the bulk fields in the excited state. We carried out explicit computations of both contributions and showed that the sum matched the CFT result, thus, providing a non-trivial check of the FLM proposal in a fully dynamical situation.

The calculation of the \emph{bulk entanglement entropy} piece was particularly interesting when interpreted in the context of the \emph{black hole information paradox}. Due to the purity of the bulk quantum state, the bulk entanglement entropy of a subregion in a compact space must follow a Page curve as a function of size. Interestingly, under various coordinate transformations we could relate the bulk entanglement entropy of a fixed region in the quench state with the bulk entanglement entropy of a region that decreases with time from the point of view of the global state. For fixed time, this could be further associated to the entropy of bulk fields on a planar BTZ black hole geometry. Thus, one could relate the Page curve that follows from purity of the bulk quantum state with the Page curve associated to the \emph{entropy of Hawking radiation} in an eternally evaporating black hole background. Is in this sense that our setup could be interpreted as a simple \emph{toy model for unitary black hole evaporation}. We emphasize that, even though our bulk calculation was semiclassical, we did \emph{not} require the so-called island contribution to restore unitarity. This was the case because our particular quantum state was \emph{simple}, leading to only one saddle point dominating the path integral throughout the evolution. By \textit{simple}, we mean states with an order $\mathcal{O}(1)$ number of excitations in the bulk which change the total energy and entropy by an order $\mathcal{O}(1)$ amount. For more generic complex states this is not necessarily the case, and one must consider multiple saddle points that eventually exchange dominance. It is nevertheless interesting to note that unitarity \emph{can} be preserved within the semiclassical approximation, at least in simple setups as the one considered here. This implies that information loss or the lack thereof is, in fact, \emph{not} an artifact of the approximation.

There are many open problems and generalizations, some of which we discuss now:
\begin{itemize}

\item { \it{Higher dimensions:}} Many of the steps used in the bulk computations can be generalized for boundary spherical regions or bulk hemispheres on higher dimensional AdS spaces. However, our CFT computation relies heavily on the enhanced conformal symmetry special to two dimensions. To generalize our analysis to higher dimensions, one could imagine instead using higher dimensional twists operators in the CFT, as the ones introduced in \cite{Cardy:2013nua} to mimic the non-trivial geometry of the replica manifold. Further approximations might be required to gain analytic control. A natural would be the small radius limit of the entangling sphere which was successfully used in the analysis of \cite{Cardy:2013nua} and which nevertheless could provide a non-trivial match with the bulk computations.

\item {\it Relative entropy and mutual information:} Relative entropy and mutual information are interesting quantities from quantum information theory perspective and have been widely studied in recent works. There exist replica like techniques for the computation of these quantities in QFT as well as concrete proposals for their holographic duals \cite{Jafferis:2015del}. Therefore, it would be very interesting to extend our framework to the computations of these quantities in our set up. For instance, the mutual information for disjoint regions in excited states was computed recently in \cite{Ugajin:2016opf}, and relative entropy for similar states was discussed in \cite{Sarosi:2016atx}, for general CFTs. See \cite{Asplund:2013zba} for a recent dynamical analysis of mutual information.

\item {\it More general local quenches:} One can consider local quantum quenches of a different type compared to the ones studied in this paper. For example, to consider operators with spin. Recently the backreaction of operators with spin was considered in \cite{Belin:2019mlt}, which generalized the work of \cite{Belin:2018juv} to more general one-particle states. One can imagine performing the same set of bulk transformations that we considered here to obtain quench states in these scenarios. Another option would be to consider a reference state that is not the vacuum. For instance, if the unperturbed state is taken to be a thermal state, then the gravity dual would be a localized perturbation of a black hole geometry \cite{Caputa:2014eta}. This setup is interesting because from the entanglement entropy calculation one could obtain the leading quantum corrections to the butterfly velocity $v_B$, e.g. by extending the work of \cite{Mezei:2016wfz} to order $\mathcal{O}(1)$. Finally, one could consider \textit{bilocal} quenches as in \cite{Arefeva:2017pho,Caputa:2019avh}. These are obtained by inserting a local operator at two different points in space and evolving the state with the Hamiltonian. In particular, computing quantum corrections in these scenarios could shed light on the nature of gravitational interactions beyond the classical regime.

\item {\it Quantum extremal surfaces:} Generalizing beyond the leading $\cO(1)$ corrections to holographic entanglement entropy, there is a proposal for an all-order resummed result in terms of the so-called quantum extremal surfaces \cite{Engelhardt:2014gca}.  The aforementioned surfaces are obtained as the extremal surface with respect to the generalized entropy as opposed to the area term, as in the FLM prescription. Elevating our bulk calculation to this next level of complexity requires the understanding of the bulk entanglement entropy for general regions or arbitrary small perturbations of the semicircle. One would also need to find the various maps of interest for slightly deformed regions. An interesting work along these lines is the perturbative analysis in \cite{Faulkner:2016mzt} for the modular Hamiltonian of deformed spheres and its generalizations for the bulk modular Hamiltonian following \cite{Faulkner:2017vdd}. It would be interesting to work out this example in detail.

\end{itemize}

We hope to come back to some of these problems in the near future.

\section*{Acknowledgements}

We would like to thank Alexandre Belin, Pawel Caputa and Matthew Headrick for useful discussions and comments on the manuscript. We are specially grateful to Alexandre Belin for pointing out an error in the first version of this paper. CAA is supported by the National Science Foundation (NSF) Grant No. PHY-1915093. CAA further acknowledges support from the $\Delta$-ITP visiting program and thank the Institute for Theoretical Physics at the University of Amsterdam for the welcoming atmosphere that ignited this collaboration. SFL is partially supported by the United States Department of Energy QuantISED program, under contract DE-SC0019517. JFP is supported by the Simons Foundation through the \emph{It from Qubit} collaboration.

\appendix

\section{Details of the CFT calculation \label{firstappendix}}

In this appendix we will present details of specific intermediate-step calculations of the entanglement entropy in the CFT.

\subsection{Universal contribution\label{app:univ}}

The argument in the limit of (\ref{Suni-z}) is well defined for $n\in\mathbb{Z}$ provided $n\geq1$, so the $n\to 1$ limit must be taken from above. In the following, we will write down an expansion of the logarithm in (\ref{Suni-z}) for small $(n-1)>0$ and keep only the first order term. This will in fact give the exact answer once the above limit is taken. We will repeatedly use the following linear approximations: $e^x\approx 1+x$, and $\log(1+x)\approx x$.

First, consider the factor
\bea\label{log1}
\log \left[ \frac{\(|z_1||z_2|\)^{2n(1-n)h}}{n^{4nh}} \right]
&=&-n(n-1)h\log\(z_1 \bz_1 z_2 \bz_2 \) -4nh\log(n)\,, \nonumber \\
&\approx &-(n-1)h\[\log(z_1)+\log(\bar{z}_1)+\log(z_2)+\log(\bar{z}_2)+4\]\,.
\eea
Here we have rewritten $\log(n)=\log\(1+(n-1)\)\approx(n-1)$.  The second factor,
\bea\label{log11}
\log \left[ \(\frac{|z_1^n-z_2^n|}{|z_1-z_2|}\)^{4nh}  \right]=2nh \left[\log\( \frac{z_1^n-z_2^n}{z_1-z_2}\)+ \log\( \frac{\bz_1^n-\bz_2^n}{\bz_1-\bz_2}\)\right],
\eea
can be studied in a similar way. For small $(n-1)$ we have
\bea
z_1^n=z_1 z_1^{n-1}=z_1e^{(n-1)\log(z_1)}\approx z_1 \[1+(n-1)\log(z_1)\]\,,
\eea
so:
\bea
\frac{z_1^n-z_2^n}{z_1-z_2} \approx 1+(n-1)\[\frac{z_1\, \log(z_1)-z_2\, \log(z_2)}{z_1-z_2}\].
\eea
Plugging this expression as well as its complex conjugate into (\ref{log11}) leads to
\bea\label{log2}
\!\!\!\!\!\!\!\log \left[ \(\frac{|z_1^n-z_2^n|}{|z_1-z_2|}\)^{4nh}  \right]\approx 2(n-1)h \left[\frac{z_1\, \log(z_1)-z_2\, \log(z_2)}{z_1-z_2}+\frac{\bar{z}_1\, \log(\bar{z}_1)-\bar{z}_2\, \log(\bar{z}_2)}{\bar{z}_1-\bar{z}_2} \right].
\eea
Combining (\ref{log1}) and (\ref{log2}) and replacing the result into (\ref{Suni-z}) allows us to compute the limit which, as mentioned previously, is simply given by the linear coefficient of the expansion in powers of $(n-1)$. This yields equation (\ref{Suni-1}), as advertised in the main body of the paper.

\subsection{Dynamical contribution\label{app:dyn}}

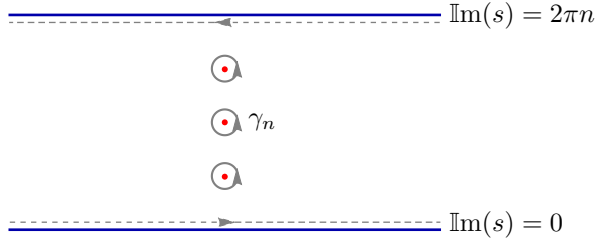
\begin{figure}[t!]
\begin{center}
\hspace{-1.5cm}\includegraphics[scale=0.5]{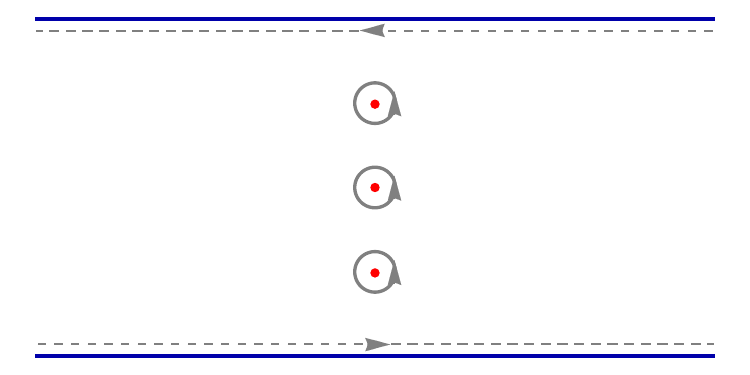}
\begin{picture}(0,0)
\put(-85,44){{\small $\gamma_n$}}
\put(-9,4){{\small $\operatorname{\mathbb{I}m}(s)=0$}}
\put(-9,83){{\small $\operatorname{\mathbb{I}m}(s)=2\pi n$}}
\end{picture}
\end{center}
\vspace{-0.4cm}
\caption{Integration contour (depicted in solid gray) used to evaluate the complex integral in (\ref{eq:contInt}). Assuming that integrand vanishes for
$\operatorname{\mathbb{R}e}(s) \to \pm\infty$, we can then deform the contour $\gamma_n$ to just be the dashed gray lines at $\operatorname{\mathbb{I}m}(s)=\delta$ and at
$\operatorname{\mathbb{I}m}(s)=2\pi n-\delta$ with $\delta>0$. For illustrative purposes, we have picked the value $n=4$ to make this figure.}
\label{fig:contour}
\end{figure}

In this appendix we will perform the sum (\ref{O2O2}) using the technique developed in \cite{Agon:2015ftl}.
We start by rewriting the sum in (\ref{O2O2}) as
\begin{equation}
\label{eq:approxsum}
 \sum_{k=1}^{n-1}\frac{1}{\left| \sin\(\frac{\pi k}{n} \)\right|^{4\Delta}}  = \sum\limits_{k=1}^{n-1} G_n^2(2 \pi k)\,,\qquad  G_n(\tau) \equiv \frac{1}{\left| \sin\(\frac{\tau}{2n} \)\right|^{2\Delta}}\,.
\end{equation}
We analytically continue $G_n(\tau)$  to the complex time plane $G_n(\tau)\to G_n(-is)$ with $s\in \mathbb{C}$ and $0<\operatorname{\mathbb{I}m}(s)<2\pi n$.
By an application of residue theorem, we can then write the sum as
\begin{equation}\label{eq:contInt}
\sum\limits_{k=1}^{n-1} G_n^2(2 \pi k) \cong \int_{\gamma_n} \frac{ds}{2 \pi i} \frac{G_n^2(-i s)}{e^s-1}\,,
\end{equation}
with the contour $\gamma_n$ chosen as depicted in Figure \ref{fig:contour}.
However, this contour integral does not give zero when we set $n=1$. This is why we have the approximate equality in equation \eqref{eq:contInt}. To fix this, we can add a term to the integral
\begin{equation}
\sum\limits_{k=1}^{n-1} G_n^2(2 \pi k) = \int_{\gamma_n} \frac{ds}{2 \pi i} \frac{ G_n^2(-i s)}{e^s-1} -  \int_{\gamma_n}  \frac{ds}{2 \pi i} \frac{G_n^2(-i s)}{e^{s/n}-1}\,.
\end{equation}
This integral can be repackaged using the kernel function
\begin{equation}
k_n(s) \equiv \frac{ 1}{e^s-1} -  \frac{1}{e^{s/n}-1}\,,
\end{equation}
so that
\begin{equation}
\sum\limits_{k=1}^{n-1} G_n^2(2 \pi k) = \int_{\gamma_n} \frac{ds}{2 \pi i}  G_n^2(-i s)k_n(s)\,.
\end{equation}
This integral can be evaluated provided that $i)$ $\operatorname{\mathbb{R}e}(\Delta) > -1/2$ and that $ii)$ $G_n^2(-i s) k_n(s) \to 0$ as $\operatorname{\mathbb{R}e}(s)\to\pm\infty$.
We always assume the former, while the latter can be explicitly checked to be true given the definition of $G_n(s)$. Using the vanishing of the integrand for
$\operatorname{\mathbb{R}e}(s) \to \pm\infty$, we can then deform the contour $\gamma_n$ to just be the two dashed lines at $\operatorname{\mathbb{I}m}(s)=\delta$ and at
$\operatorname{\mathbb{I}m}(s)=2\pi n-\delta$ with $\delta>0$, depicted in the Figure \ref{fig:contour}. Once we do this, and using the periodicity $G^2_n(s)=G^2_n(s+2\pi i n)$, the sum becomes
\begin{equation}\label{eq:divide}
\sum\limits_{k=1}^{n-1} G_n^2(2 \pi k) =  \int_{-\infty}^{\infty}  \frac{ds}{2\pi i} \bigg[G_n^2(-i s + \delta) k_n(s+i \delta) - G_n^2(-i s-\delta)k_n(s-i \delta) \bigg]\,.
\end{equation}
This integral vanishes for $n=1$ since, by construction, $k_1(s)=0$. However, we need the coefficient of the term proportional to $(n-1)$ in the $n\to1$ expansion. In order to extract this coefficient, we divide (\ref{eq:divide}) by $(n-1)$
and take the $n \to 1$ limit on both sides,
\begin{align}
\begin{split}
\label{eq:sumlimit}
\lim\limits_{n \to 1} \frac{1}{n-1}  \sum\limits_{k=1}^{n-1} G_n^2(2 \pi k) =
  \int_{-\infty}^{\infty} \frac{ds}{2\pi i} \bigg[G_1^2(-i s + \delta) \hat{k}(s+i \delta) - G_1^2(-i s-\delta) \hat{k}(s-i \delta) \bigg]\,,
\end{split}
\end{align}
where
\begin{equation}
\hat{k}(s) \equiv -\frac{s}{4 \sinh^2(s/2)}\,.
\end{equation}
Finally, we deform the integration contour to $\operatorname{\mathbb{I}m}(s)=\pi$ in the first term and to $\operatorname{\mathbb{I}m}(s)=-\pi$ in the second term,
so that the $\delta\to0$ limit is non-singular. By doing so, we obtain an expression that can be readily integrated
\begin{equation}
\label{eq:sumcont}
\lim\limits_{n \to 1} \frac{1}{n-1}  \sum\limits_{k=1}^{n-1} G_n^2(2 \pi k) = \frac{1}{4}\int_{-\infty}^{\infty}\frac{ds}{\cosh(s/2)^{4 \Delta+2}}= \frac{\Gamma(\tfrac{3}{2})\Gamma(2 \Delta+1) }{ \Gamma(2 \Delta+\tfrac{3}{2}) }\,.
\end{equation}
This leads to the result reported in (\ref{sum}).

\section{Bogoliubov coefficients and bulk entanglement \label{AppendixA}}
The scalar field has two different mode expansions depending on the choice of coordinates: in global coordinates we can write
\bea\label{global-modes}
\phi(\tau, r, \varphi)=\sum_{m,n}\(a_{m,n} e^{-i\Omega_{m,n}\tau} f_{m,n}(r,\varphi) +a^{\dagger}_{m,n} e^{i\Omega_{m,n}\tau} f^*_{m,n}(r,\varphi)\),
\eea
while in Rindler coordinates we have that
\bea\label{Rindler-modes}
\phi(\mathfrak{r}, \bt,\by)=\sum_{I \in L,R}\int_{\!\omega>0}\!\!  \frac{ d\omega d k}{(2\pi)^2}\( e^{-i\omega \bt} b_{\omega,k,I} g_{\omega,k,I}(\mathfrak{r},\by) +e^{i\omega \bt} b^{\dagger}_{\omega,k,I} g^*_{\omega,k,I}(\mathfrak{r},\by)  \)\,.
\eea
The creation and annihilation operators in the two frames are related by the Bogoliubov coefficients $\alpha,\beta$, e.g.,
\be
a_{m,n}=\sum_{I,\omega,k}\left(\alpha_{m,n;\omega,k,I}b_{\omega,k,I}+\alpha^*_{m,n;\omega,k,I}b^\dag_{\omega,k,I}\right)\,,
\ee
and similarly for $a^{\dag}_{m,n}$. We will compute the asymptotic form of these coefficients, focussing on the large-$r$ limit. As we will see below, this will indeed suffice to extract
the leading result for $\delta S_{\text{bulk}}^{(\delta\rho^2)}$ in the small $R$ limit.

Before going further, we need to determine how does the state $\ket{\psi}=a_{0,0}^\dagger\ket{0}$ look in Rindler coordinates. This was worked out in detail in \cite{Belin:2018juv}, so we will only quote the result here
\bea
\ket{\psi}=a_{0,0}^\dagger\ket{0}=\sum_{\omega,k}\( (1-e^{-2\pi \omega})\alpha^*_{\omega,k,R}b^\dagger_{\omega,k,R}+(1-e^{2\pi \omega})\beta_{\omega,k,R}b_{\omega,k,R} \,. \)\ket{0}
\eea
Notice that this relation only involves Rindler modes associated to the right wedge and so it is very convenient for the computation of the reduced density matrix obtained after tracing out the Hilbert space of the left modes $\mathcal{H}_R$, $\rho_L=\tr_{\mathcal{H}_R}\ket{\psi}\bra{\psi}$. Since the global vacuum reduced to the right wedge in the Rindler description is given by a thermal bath in Rindler coordinates then $\rho$ can be interpreted as a perturbation to this thermal bath. The entanglement entropy is thus the thermal entropy of this perturbed thermal state. The authors of \cite{Belin:2018juv} went further and compute a general expression for $\delta S_{\text{bulk}}^{(\delta\rho^2)}$ solely in terms of the $\alpha_{\omega,k,R}$ and $\beta_{\omega,k,R}$. The result was given in equation (5.24) of that paper. However, the
expression is very long so we will not transcribe it here.

The general form of the Bogoliubov coefficients was derived in \cite{Belin:2018juv} and applies in the exact same form for our case:
\bea\label{alphawk}
\alpha_{\omega,k,R}&=&\frac{1}{N_{\omega,k}}\int_{-\infty}^{\infty} d\bt d\by  e^{-i\omega \bt +i k \by} \frac{e^{i h\, \tilde{ \tau}(\bt,\by)}}{\sqrt{2\pi } \tilde{ r}^h(\bt,\by)}\,, \\ \label{betawk}
\beta_{\omega,k,R}&=&-\frac{1}{N_{\omega,k}}\int_{-\infty}^{\infty} d\bt d\by e^{-i\omega \bt +i k \by} \frac{e^{-i h\,  \tilde{\tau}(\bt,\by)}}{\sqrt{2\pi } \tilde{ r}^h(\bt,\by)}\,,
\eea
where
\bea
N_{\omega,k}=\frac{\left| \Gamma\[\frac{\Delta}{2}+i\frac{(\omega+k)}{2} \]\right|\left| \Gamma\[\frac{\Delta}{2}+i\frac{(\omega-k)}{2} \]\right|}{\sqrt{2\omega} |\Gamma(\Delta)| |\Gamma(i\omega)|}\,.
\eea
The only difference lies on the explicit functions relating the asymptotic global time $\tilde{\tau}(\bt,\by)$, and normalized global radial coordinate $ \tilde{ r}(\bt,\by)$ with the asymptotic Rindler coordinates $\{\bt,\by\}$, defined as
\bea
\tilde{\tau}(\bt,\by)\equiv \lim_{\mathfrak{r}\to \infty} \tau(\mathfrak{r},\bt,\by) ,\quad {\rm and}\quad \tilde{r}(\bt,\by)\equiv \lim_{\mathfrak{r}\to \infty} \frac{r(\mathfrak{r},\bt,\by)}{L\mathfrak{r}}\,.
\eea
In our setup, the relation between global and Rindler coordinates is given by the composed trasformation (\ref{eq:cotrans})-(\ref{eq:cotrans3}) and (\ref{poincaretobtz-2a})-(\ref{poincaretobtz-2c}). This is in fact more general than the transformation used in \cite{Belin:2018juv}, which specialized to a Rindler wedge for a boundary region at constant $\tau$. From our transformation, it follows that
\bea
 \tilde{\tau}(\bt,\by) &=&L\arctan\left(\frac{2 \alpha t_c }{\alpha^2+x_c^2-t_c^2}\right)+\cO(R)\,, \\
 \tilde{r}(\bt,\by) &=
&\frac{1}{2R\alpha }\sqrt{4\alpha^2x_c^2+\left(-\alpha^2+x_c^2-t_c^2\right)^2}(\cosh\by+\cosh\bt)+\cO(1)\,,
\eea
where we have kept only the leading terms in the small $R$ regime. Plugging these into the formulas for the Bogoliubov coefficients (\ref{alphawk}), (\ref{betawk}) and carrying out the integrals leads to\footnote{Notice that the authors of \cite{Belin:2018juv} computed the Bogoliubov coefficients without the aforementioned approximation. As mentioned there the exact result is important in order to have the Bogoliubov coefficients to satisfy consistency conditions such as the normalization and completeness relations. Nevertheless for the purpose of computing $\delta S_{\text{bulk}}^{(\delta\rho^2)}$ it is sufficient to keep only the leading small $R$ results even at the level of the integrands. The reason is that while the normalization condition involves slowly convergent integrals in Fourier space, the integrals involved in the computation of $\delta S_{\text{bulk}}^{(\delta\rho^2)}$ converge fast.
}
\bea\label{Bogo-coeff}
\!\!\!\!\!\!\!\!\alpha_{\omega,k,R}&=&\frac{2^{\Delta}e^{i\phi_0}}{N_{\omega,k}}\!\!\[\frac{2 \alpha R }{\sqrt{4\alpha^2x_c^2+\left(\alpha^2+t_c^2-x_c^2\right)^2}}\]^{\!\Delta}\!\!\frac{\left| \Gamma\[\frac{\Delta}{2}+i\frac{(\omega+k)}{2} \]\right|^2\left| \Gamma\[\frac{\Delta}{2}+i\frac{(\omega-k)}{2} \]\right|^2}{\sqrt{8\pi }|\Gamma[\Delta]|^2},\\
\!\!\!\!\!\!\!\!\beta_{\omega,k,R}&=&-\frac{2^{\Delta}e^{-i\phi_0}}{N_{\omega,k}}\!\!\[\frac{2 \alpha R }{\sqrt{4\alpha^2x_c^2+\left(\alpha^2+t_c^2-x_c^2\right)^2}}\]^{\!\Delta}\!\!\frac{\left| \Gamma\[\frac{\Delta}{2}+i\frac{(\omega+k)}{2} \]\right|^2\left| \Gamma\[\frac{\Delta}{2}+i\frac{(\omega-k)}{2} \]\right|^2}{\sqrt{8\pi }|\Gamma[\Delta]|^2} .
\eea
Up to the constant phases $\pm \phi_0$,
\bea
\phi_0=\frac{1}{2} \Delta \tilde{\tau}(\bt,\by)=\frac{\Delta L}{2}\arctan\left(\frac{2 \alpha t_c }{\alpha^2+x_c^2-t_c^2}\right)\,,
\eea
the Bogoliubov coefficients coincide with the ones from \cite{Belin:2018juv} in the small $R$ regime, provided that we identify the parameter
\bea\label{etaR}
\cosh\eta \iff \frac{1}{2R\alpha }\sqrt{4\alpha^2x_c^2+\left(-\alpha^2+x_c^2-t_c^2\right)^2}\,.
\eea
Since the constant phases $\pm \phi_0$ drop off from physical quantities the rest of the analysis is the same as the one presented in \cite{Belin:2018juv}, keeping in mind the substitution of the factor (\ref{etaR}) everywhere. With this change, equation (5.25) of \cite{Belin:2018juv} becomes (\ref{2ndSRindler}).

\bibliographystyle{ucsd}
\bibliography{refs}

\end{document}